\documentclass[aps,prd,onecolumn,showpacs,superscriptaddress,nofootinbib,preprintnumbers]{revtex4-2}

\usepackage{graphicx}
\usepackage{color}
\usepackage{amsmath}
\usepackage{amssymb}
\usepackage{enumerate}
\usepackage{subcaption}   
\usepackage{caption}
\usepackage[colorlinks=true,citecolor=blue,urlcolor=blue]{hyperref}

\graphicspath{ {Figures/} }

\begin{document}

\title{Reconstructing Dark Matter Mass and Discriminating \\
Standard and Non-Standard WIMP–Nucleus Interactions with Paleo-Detectors}

\author{Dionysios~P.~Theodosopoulos}
\email{d.theodosopoulos@utexas.edu}
\affiliation{Department of Physics, The University of Texas at Austin, Austin, TX 78712, USA}

\author{Katherine~Freese}%
 \email{ktfreese@utexas.edu}
\affiliation{Department of Physics, The University of Texas at Austin, Austin, TX 78712, USA}
\affiliation{The Oskar Klein Centre, Department of Physics, Stockholm University, AlbaNova, SE-106 91 Stockholm, Sweden}
\affiliation{Nordita, Stockholm University and KTH Royal Institute of Technology, Hannes Alfvéns väg 12, SE-106 91 Stockholm, Sweden}

\author{Chris~Kelso}
\email{ckelso@unf.edu}
\affiliation{
 Department of Physics and Astronomy, University of North Florida, 1 UNF Dr, Jacksonville, FL 32224, USA
}%

\author{Patrick~Stengel}
\email{patrick.stengel@ijs.si}
\affiliation{
 Jo\v{z}ef Stefan Institute, Jamova 39, 1000 Ljubljana, Slovenia
}%

\begin{abstract}
Paleo-detectors record and retain crystal damage in ancient minerals from nuclear recoils induced by dark matter scattering over geological timescales. Previous studies have shown that paleo-detectors can provide sensitivity to a variety of dark matter scenarios which is complementary to conventional direct-detection experiments. In this paper, we complete the first detailed study of how well paleo-detectors can reconstruct dark matter parameters or distinguish between different types of dark matter interactions with nuclei in the presence of a dark matter signal, considering both elastic and inelastic dark matter-nucleus scattering. For representative nuclear recoil track read-out scenarios, we demonstrate that weakly interacting massive particle (WIMP) dark matter masses can be reconstructed for a variety of Non-Relativistic Effective Field Theory (NREFT) interactions between WIMPs and nuclei. In particular, paleo-detectors are projected to be capable of reconstructing WIMP masses $\lesssim 10 \, \mathrm{GeV}/c^2$, a regime that is challenging for conventional direct-detection experiments; 
further, we find that paleo-detectors could reconstruct WIMP masses up to
$1\ \mathrm{TeV}/c^2$ for hypothetical signals within their accessible parameter space,
extending the mass range over which reconstruction is possible by up to a factor of $\sim 2$ compared with analogous studies of conventional direct-detection experiments.
In addition, we demonstrate that paleo-detectors could discriminate between canonical spin-independent or spin-dependent NREFT interactions and non-canonical interactions which can depend on the relative velocity or momentum transferred between the WIMP and nucleus. 
Specifically, at WIMP masses $\gtrsim 10 \, \mathrm{GeV/c^2}$, we project that canonical NREFT interactions can be excluded by paleo-detectors in the cases of nearly all non-canonical interactions without measurement of nuclear recoil direction, which conventional experiments typically require in addition to measurement of nuclear recoil energy.
\end{abstract}

\maketitle

\section{Introduction}

Despite decades of experimental searches with continuously improving sensitivity to rare dark matter interactions, the particle nature of dark matter remains one of the most challenging open questions in high energy and astroparticle physics. Direct-detection experiments are designed to probe the interactions of Weakly Interacting Massive Particle (WIMP) dark matter with atomic nuclei~\cite{Drukier:1984vhf,Goodman:1984dc,Drukier:1986tm}. In particular, nuclear recoils induced by WIMP scattering can be detected through a combination of phonons, ionization and scintillation light produced within instrumented volumes of various target materials~\cite{LZ:2022lsv,DarkSide-50:2022qzh,XENON:2023cxc,PandaX:2024qfu}. Next generation direct-detection experiments aim to improve sensitivity to lower energy nuclear recoils induced by lighter WIMPs, with masses $m_\chi \lesssim 10 \, \mathrm{GeV/c^2}$, using a variety of techniques to lower energy thresholds for recoil detection, while the sensitivity to heavier WIMPs, with masses $m_\chi \gtrsim 10 \, \mathrm{GeV/c^2}$, is envisioned to improve by instrumenting ever larger volumes of target material~\cite{SuperCDMS:2016wui,DarkSide-20k:2017zyg,XLZD:2024nsu}.

As a complement to current and future direct-detection experiments, paleo-detectors record and retain damage to mineral crystals caused by dark matter interactions over geological timescales, up to $\mathcal{O}(1)\, $Gyr. The paleo-detector concept has been the subject of increasing theoretical and experimental work~\cite{Baum:2023cct,Baum:2024eyr,Hirose:2025jht}, following initial searches for damage tracks from nuclear recoils induced by WIMP interactions in ancient mica several decades ago~\cite{PhysRevLett.74.4133,Collar:1995aw,Snowden-Ifft:1996dug}. While most paleo-detector studies have focused on two interactions, the standard elastic spin-independent (SI) and spin-dependent (SD) interactions of WIMPs with nuclei~\cite{Baum:2018tfw,Drukier:2018pdy,Edwards:2018hcf,SinghSidhu:2019znk,Ebadi:2021cte,Acevedo:2021tbl,Baum:2021jak,Fung:2025cub}, in a recent paper (Ref.~\cite{Theodosopoulos:2026ehn} hereafter referred to as TFKS) we investigated the sensitivity of paleo-detectors to both elastic and inelastic WIMP-nucleus interactions within a more general Non-Relativistic Effective Field Theory (NREFT) framework~\cite{Fan:2010gt,Fitzpatrick:2012ix}. 
Due to the combination of nano-scale spatial resolution of advanced read-out techniques allowing for sensitivity to low energy nuclear recoils, as well as large exposure 
that, depending on the readout technique and
sample mass, can exceed that of conventional direct-detection experiments
by factors of up to $\sim10^{3}$--$10^{4}$,
we demonstrated in TFKS that paleo-detectors could be sensitive to couplings of NREFT operators over an order of magnitude smaller than currently constrained by conventional direct-detection experiments \cite{XENON:2017fdd,PandaX-II:2018woa,SuperCDMS:2022crd,LZ:2023lvz}.

Building on the sensitivity projections of TFKS, we first investigate the ability of paleo-detectors to reconstruct the WIMP mass in the case of a dark matter detection. We consider hypothetical signals
generated by a representative set of NREFT operators and determine the range of WIMP masses that would remain compatible with the observed track-length spectrum at  the 2$\sigma$ confidence level. An earlier study of parameter reconstruction with paleo-detectors was presented in Ref.~\cite{Edwards:2018hcf}, which considered standard elastic SI scattering and used the Euclideanized-signal method. The present work extends that analysis to
both elastic and inelastic scattering through a broad set of NREFT operators and employs a profile-likelihood-ratio approach. We find that paleo-detectors could reconstruct WIMP masses up to
$1\ \mathrm{TeV}/c^2$ for hypothetical signals within their accessible parameter space,
extending the mass range over which reconstruction is possible by up to a factor of $\sim 2$ compared with analogous studies of conventional direct-detection experiments~\cite{Green:2007rb,Green:2008rd,Edwards:2018lsl,Bozorgnia:2018jep}. 
Furthermore, paleo-detectors enable mass reconstruction in regions that are difficult or inaccessible for conventional experiments, including elastic
scattering with WIMP masses $\lesssim 10\ \mathrm{GeV}/c^2$ and inelastic scattering with a representative mass splitting
$\delta_m=50\ \mathrm{keV}/c^2$ over the mass range $30$--$400\ \mathrm{GeV}/c^2$.

We then study whether track-length spectral information can be used to identify the type of WIMP--nucleon interaction responsible for a hypothetical detected signal. In particular, we ask whether paleo-detectors can distinguish signals generated by non-standard NREFT operators from the
standard SI or SD interaction hypotheses using only the measured track-length spectrum, without nuclear recoil directionality. 
We find that paleo-detectors can reject the standard SI or SD hypotheses for signals generated by a broad class of non-standard operators, especially for WIMP masses above a few tens of $\mathrm{GeV}/c^2$ in a low-resolution read-out scenario (n.b. lower resolution allows for higher exposure). 
This provides a complementary discrimination strategy to conventional direct-detection experiments, for which previous studies have shown that distinguishing NREFT operators typically requires directional information in addition to the nuclear recoil energy~\cite{Kavanagh:2015jma,Fieguth:2018vob}.

The remaining sections of this paper are outlined as follows. The NREFT framework for WIMP--nucleus interactions, the corresponding signals in paleo-detectors and the relevant backgrounds are briefly described in respective Secs.~\ref{nreft},~\ref{DMpaleo}, and~\ref{background}, summarizing more detailed discussions in TFKS. In Sec.~\ref{mass_reconstruction}, we project how well paleo-detectors can reconstruct the WIMP mass assuming a signal from a variety of NREFT interactions. In Sec.~\ref{comparison}, we project the capacity of paleo-detectors to discriminate between standard and non-standard NREFT operators. We conclude with a summary and discussion of potential future work in Sec.~\ref{conclusion}. In the main text we focus on the seven NREFT operators with spin-independent interactions, whereas in the Appendix we show the results of a similar analysis for spin-dependent interactions.

\section{Non-Relativistic Effective Field Theory} \label{nreft}

A complete formulation of the Non-Relativistic Effective Field Theory (NREFT) describing WIMP–nucleon elastic interactions at non-relativistic energies was developed in Refs.~\cite{Fan:2010gt,Fitzpatrick:2012ix}. This framework provides a model-independent basis for parametrizing elastic WIMP–nucleon scattering in terms of effective four-fermion operators. Because direct-detection experiments probe momentum transfers well below the QCD scale, the NREFT formalism captures all possible interactions consistent with Galilean invariance.

In this section, we briefly review the NREFT operators governing elastic WIMP--nucleon scattering and then describe the modifications required to extend the formalism to inelastic interactions. Throughout this work, the NREFT framework is used to study how different operators imprint distinct signatures in the track-length spectrum, with track length serving as a proxy for recoil energy, to reconstruct dark matter mass, and to quantify degeneracies between interaction hypotheses.

\subsection{Elastic WIMP--Nucleon Interactions}

In the NREFT framework, the interaction between a dark matter particle $\chi$ and a nucleon $N$ is expressed in terms of operators constructed from the following four Hermitian building blocks:
\begin{equation}
     i\frac{\vec{q}}{m_N},\quad 
     \vec{v}^{\perp} \equiv \vec{v} + \frac{\vec{q}}{2\mu},\quad 
     \vec{S}_{\chi},\quad 
     \vec{S}_{N}, \label{basis}
\end{equation}
where $\vec{q}$ is the momentum transfer, $m_N$ is the nucleon mass, $\vec{v}$ is the WIMP–nucleon relative velocity, $\mu$ is the reduced mass of the WIMP–nucleon system, and $\vec{S}_{\chi}$ and $\vec{S}_N$ are the WIMP and nucleon spin operators, respectively. 

By forming all allowed Galilean-invariant combinations up to second order in $\vec{q}$, the most general non-relativistic interaction Lagrangian can be written as
\begin{equation}
\mathcal{L}_{\rm int} = \sum_{N=n,p} \sum_i c_i^{(N)} \mathcal{O}_i \,
\chi^+ \chi^- N^+ N^-,
\end{equation}
where $\mathcal{O}_i$ denote the linearly independent dimensionless NREFT operators listed in Table~\ref{tab:NREFT_operators}, and $c_i^{(N)}$ are the corresponding coupling constants to protons and neutrons. The superscripts ($+,-$) label outgoing and incoming states; for elastic scattering, $\chi^+ \equiv \chi^-$ and $N^+ \equiv N^-$.
\begin{table}[t]
\centering
\small
\setlength{\tabcolsep}{6pt}   
\renewcommand{\arraystretch}{0.9} 
\captionsetup{justification=raggedright,singlelinecheck=false}
\begin{tabular}{l l}
\hline\hline
\multicolumn{2}{c}{NREFT operators relevant for elastic WIMP--nucleon scattering} \\
\hline
$\mathcal{O}_1 = 1_\chi 1_N$ 
& $\mathcal{O}_9 = i\, \vec{S}_\chi \cdot ( \vec{S}_N \times \vec{q}/m_N )$ \\

$\mathcal{O}_3 = i\, \vec{S}_N \cdot ( \vec{q}/m_N \times \vec{v}^{\perp} )$ 
& $\mathcal{O}_{10} = i\, \vec{S}_N \cdot \vec{q}/m_N$ \\

$\mathcal{O}_4 = \vec{S}_\chi \cdot \vec{S}_N$ 
& $\mathcal{O}_{11} = i\, \vec{S}_\chi \cdot \vec{q}/m_N$ \\

$\mathcal{O}_5 = i\, \vec{S}_\chi \cdot ( \vec{q}/m_N \times \vec{v}^{\perp} )$ 
& $\mathcal{O}_{12} = \vec{S}_\chi \cdot ( \vec{S}_N \times \vec{v}^{\perp} )$ \\

$\mathcal{O}_6 = ( \vec{S}_\chi \cdot \vec{q}/m_N )
                 ( \vec{S}_N \cdot \vec{q}/m_N )$
& $\mathcal{O}_{13} = i ( \vec{S}_\chi \cdot \vec{v}^{\perp} )
                       ( \vec{S}_N \cdot \vec{q}/m_N )$ \\

$\mathcal{O}_7 = \vec{S}_N \cdot \vec{v}^{\perp}$ 
& $\mathcal{O}_{14} = i ( \vec{S}_\chi \cdot \vec{q}/m_N )
                       ( \vec{S}_N \cdot \vec{v}^{\perp} )$ \\

$\mathcal{O}_8 = \vec{S}_\chi \cdot \vec{v}^{\perp}$ 
& $\mathcal{O}_{15} = - ( \vec{S}_\chi \cdot \vec{q}/m_N )
                       [ ( \vec{S}_N \times \vec{v}^{\perp} )
                       \cdot \vec{q}/m_N ]$ \\
\hline\hline
\end{tabular}
\caption{Hermitian and Galilean invariant operators defining the non-relativistic effective theory of WIMP--nucleon elastic interactions~\cite{Fan:2010gt,Fitzpatrick:2012ix}. The operators $\mathcal{O}_{1}$ and $\mathcal{O}_{4}$ correspond to canonical spin-independent (SI) and spin-dependent (SD) interactions, respectively. Operator $\mathcal{O}_{2}$ is quadratic in $\vec{v}^{\perp}$ and $\mathcal{O}_{16}$ is a linear combination of $\mathcal{O}_{12}$ and $\mathcal{O}_{15}$ and are therefore not considered here~\cite{Anand:2013yka}.}
\label{tab:NREFT_operators}
\end{table}

Operators $\mathcal{O}_1$ and $\mathcal{O}_4$ correspond to the standard spin-independent (SI) and spin-dependent (SD) interactions, respectively. 
Throughout this work, we refer to $\mathcal{O}_1$ and
$\mathcal{O}_4$ as the standard SI and standard SD operators,
respectively. For the remaining, non-standard operators, we write
``spin-independent'' or ``spin-dependent'' explicitly rather than using the abbreviations SI or SD.
In contrast to the standard SI and SD operators, the non-standard operators introduce explicit momentum- and/or velocity-dependence in the scattering amplitude, leading to recoil spectra with shapes that differ qualitatively from the standard SI and SD cases. These differences in spectral shape are central to our analysis of operator discrimination and mass reconstruction.

We exclude from our discussion $\mathcal{O}_2 = (v^{\perp})^2$, as it does not arise at leading order from the non-relativistic reduction of a relativistic operator~\cite{Anand:2013yka}. Operators with quadratic or higher powers of $v^{\perp}$ are considered negligible in the non-relativistic limit. The interaction considered in this work therefore takes the form
\begin{equation}
\sum_{N=n,p} \sum_{i=1}^{15} c_i^{(N)} \mathcal{O}_i^{(N)},
\qquad c_2^{(N)} \equiv 0.
\end{equation}

For simplicity and to reduce the dimensionality of the parameter space, we work in the isoscalar basis, assuming $c_i^{p}=c_i^{n} \equiv c_i$, and denote the operators in this basis as $\mathcal{O}_i^s$. This assumption allows us to focus on the interplay between operator structure and dark matter mass in shaping the recoil spectrum, without introducing additional degeneracies associated with isospin-violating couplings. 

Realistic WIMP dark matter models typically predict nuclear interactions mediated by a combination of NREFT operators, for which interference effects may be significant (see, e.g., Ref.~\cite{Brenner:2022qku}). To maintain consistency with the projected paleo-detector sensitivities derived in TFKS, we restrict our analysis to interactions generated by individual NREFT operators. A more complete study including multiple operators and their interference is deferred to future work.

\subsection{Inelastic WIMP--Nucleon Interactions}

We also consider inelastic scattering processes in which the incoming and outgoing dark matter states differ in mass~\cite{Tucker-Smith:2001myb}. Defining the mass splitting as $\delta_m \equiv m_{\chi,\mathrm{out}} - m_{\chi,\mathrm{in}}$, energy conservation requires~\cite{Barello:2014uda}
\begin{equation}
  \delta_m + \vec{v}\!\cdot\vec{q} + \frac{|\vec{q}|^2}{2\mu_N} = 0,
\end{equation}
where $\mu_N$ is the reduced mass of the WIMP–nucleon system. The presence of a nonzero $\delta_m$ introduces a kinematic threshold, suppressing low-energy recoils and shifting the spectrum toward higher energies.

Within the NREFT formalism, this modification is implemented by replacing $\vec{v}^{\perp}$ with
\begin{equation}
\vec v_{\text{inel}}^{\perp}
= \vec v^{\perp} + \frac{\delta_m}{|\vec q|^2}\vec q ,
\label{u_inelastic}
\end{equation}
as shown in Ref.~\cite{Barello:2014uda}. The inelastic operators are then obtained from the elastic ones by substituting $\vec v^{\perp} \rightarrow \vec v_{\text{inel}}^{\perp}$.

\subsection{Operators Considered in Paleo-Detector Studies}

The dark matter recoil spectra and projected sensitivities of paleo-detectors to elastic and inelastic WIMP--nucleon scattering were studied in TFKS, for a wide range of target minerals and read-out resolution scenarios. In the present work, we move beyond sensitivity projections and instead investigate the ability of paleo-detectors to discriminate between different NREFT operator hypotheses and to reconstruct the dark matter mass for both elastic and inelastic scattering.

Target minerals with dominant isotopes in spin-zero ground states are sensitive to the standard SI interaction
$\mathcal{O}_{1}^{s}$, as well as to the non-standard spin-independent interactions $\mathcal{O}_{5}^{s}$, $\mathcal{O}_{8}^{s}$, and
$\mathcal{O}_{11}^{s}$, while remaining insensitive to most spin-dependent operators. The spin-dependent operators $\mathcal{O}_{3}^{s}$, $\mathcal{O}_{12}^{s}$, and
$\mathcal{O}_{15}^{s}$ constitute exceptions, since these operators couple to nuclear spin-orbit responses that remain nonvanishing for spin-zero nuclei, as demonstrated in TFKS. By contrast, minerals containing abundant isotopes with nonzero-spin ground states provide sensitivity to both spin-independent and spin-dependent interactions.

In the main text, we consider the interactions
$\mathcal{O}_{1}^{s}$, $\mathcal{O}_{3}^{s}$,
$\mathcal{O}_{5}^{s}$, $\mathcal{O}_{8}^{s}$,
$\mathcal{O}_{11}^{s}$, $\mathcal{O}_{12}^{s}$, and
$\mathcal{O}_{15}^{s}$, using gypsum as a representative target mineral with dominant isotopes in spin-zero ground states. 
In the Appendix, we present the corresponding analysis for the standard SD interaction $\mathcal{O}_{4}^{s}$ and the non-standard spin-dependent interactions $\mathcal{O}_{6}^{s}$,
$\mathcal{O}_{7}^{s}$, $\mathcal{O}_{9}^{s}$,
$\mathcal{O}_{10}^{s}$, $\mathcal{O}_{13}^{s}$, and
$\mathcal{O}_{14}^{s}$, using halite as the target mineral. Halite contains abundant isotopes with nonzero-spin ground states and therefore provides sensitivity to spin-dependent interactions. 
We adopt gypsum and halite as representative targets because both are common naturally occurring minerals, can be found in radiopure marine-evaporite environments, and were projected in TFKS to provide comparatively strong sensitivity among the minerals considered. Together, gypsum and halite also provide complementary sensitivity to interactions involving spin-zero and nonzero-spin nuclei.
Both gypsum and halite have been widely considered as target minerals in the paleo-detector literature~\cite{Drukier:2018pdy,Baum:2021jak,
Theodosopoulos:2026ehn,Baum:2023cct}.

\section{Dark Matter Signals in Paleo-Detectors}\label{DMpaleo}

In this section, we make predictions for dark matter signals in paleo-detectors arising from WIMP–nucleon interactions via the NREFT operators introduced in the previous section.
We particularly focus on the features of the dark matter signals relevant for DM mass reconstruction and discrimination between different interaction hypotheses. 

The signatures of WIMP dark matter in paleo-detectors are damage tracks as well as color center vacancies 
produced by nuclei recoiling in a mineral following WIMP--nucleus scattering. Ancient minerals have accumulated such damage signatures over geological timescales. The damage track length is determined by the stopping power of a given nucleus in a given material, and we therefore use the track length as a proxy for the nuclear recoil energy. The spectrum of damage track lengths is hence our observable.  Calorimetric read-out of color center vacancies induced by WIMP scattering in paleo-detectors was considered in Ref.~\cite{Hedges:2026pgf}. For simplicity, we only consider damage-track read-out and leave a study of dark matter parameter reconstruction with calorimetric read-out for future work.
 
In TFKS, we presented detailed predictions of the track-length spectra for a variety of target minerals and read-out resolution scenarios, adopting representative coupling strengths compatible with current limits from conventional direct-detection experiments. 
In the main body of the text in this work, we discuss projections for the DM-induced track-length spectra in paleo-detectors for interactions via  operators $\mathcal{O}^{s}_{1}$, $\mathcal{O}^{s}_{3}$, $\mathcal{O}^{s}_{5}$, $\mathcal{O}^{s}_{8}$, $\mathcal{O}^{s}_{11}$, $\mathcal{O}^{s}_{12}$, and $\mathcal{O}^{s}_{15}$, and focus on those spectral features most relevant for mass reconstruction and operator discrimination. Track-length spectra for the remaining NREFT operators,  are deferred to the Appendix, in order to keep the number of figures in the main text manageable.

For a WIMP of mass $m_{\chi}$ scattering off a target nucleus $(Z,A)$ of mass $m_{T}$, the differential recoil rate in the NREFT framework is~\cite{Fitzpatrick:2012ix,Anand:2013yka}
\begin{equation}
    \left(\frac{dR}{dE_R}\right)_{(Z,A)} = N_T \frac{\rho_\chi m_T}{32 \pi m_\chi^3 m_N^2} 
    \left\langle \frac{1}{v} \sum_{i j} \sum_{N,N'=p,n} c_i^{(N)} c_j^{(N')} 
    F_{ij,\;(Z,A)}^{(N,N')}\!\left(v^2, q^2\right) \right\rangle~,
    \label{recoilrate}
\end{equation}
where $R$ is the number of recoiling nuclei per unit exposure,\footnote{We assume a one-to-one correspondence between recoils and observable tracks; under this assumption, $R$ can be interpreted as the number of tracks per unit exposure.}
$E_{R}$ is the recoil energy, $N_T$ is the number of target nuclei per detector mass, $m_{N}$ is the nucleon mass, $\rho_{\chi}$ is the local DM density, $v$ is the WIMP speed in the laboratory frame, $F_{ij,\;(Z,A)}^{(N, N')}$ are the nuclear response form factors~\cite{Fitzpatrick:2012ix,Anand:2013yka}, and $\langle...\rangle$ denotes the average over the halo velocity distribution. 
Following Ref.~\cite{Baxter:2021pqo}, we take $\rho_{\chi}=0.3~\text{GeV}/\text{cm}^{3}$~\cite{Read:2014qva} and adopt the Standard Halo Model for the WIMP velocity distribution, with $\vec{v}_{\odot}=(11.1, 12.24,7.25)~\text{km}/\text{s}$~\cite{10.1111/j.1365-2966.2010.16253.x}, 
$\vec{v}_{0}=(0,238,0)~\text{km}/\text{s}$~\cite{Sch_nrich_2012,Bland_Hawthorn_2016,2021}, 
and $v_{\text{esc}}=544~\text{km}/\text{s}$~\cite{10.1111/j.1365-2966.2007.11964.x}.
We compute recoil spectra $dR/dE_R$ using \textit{WimPyDD}~\cite{Jeong:2021bpl} and \textit{dmscatter}~\cite{Gorton:2022eed}, with cross-checks against \textit{DMFormFactor}~\cite{Anand:2013yka} (see TFKS for more details about the computation of $dR/dE_R$). 

The range $x_T$ of a recoiling nucleus $(Z,A)$ with recoil energy $E_{R}$ in a target material is
\begin{equation}
    x_T(E_{R})=\int_{0}^{E_{R}}dE~\left| \frac{dE}{dx} \right|_{(Z,A)}^{-1}~, 
    \label{length}
\end{equation}
where $(dE/dx)_{(Z,A)}$ is the stopping power in the target material. We compute stopping powers using \textit{SRIM}~\cite{ZIEGLER20101818}. The observed damage length may differ from the kinematic range if permanent damage forms only along part of the recoiling nucleus trajectory, or if the trajectory deviates significantly from a straight line. Nevertheless, existing studies indicate that Eq.~\eqref{length} provides a suitable phenomenological description for forecasting and inference studies~\cite{Drukier:2018pdy}. We exclude tracks from recoiling nuclei with charge $Z \leq 2$, since these tracks are either not sufficiently long-lived or the recoiling nuclei are not sufficiently ionizing to form observable tracks~\cite{Drukier:2018pdy}. 

In a paleo-detector, the target consists of a mineral crystal composed of several different nuclear species (and isotopes). Each species $(Z,A)$ contributes to the total target mass through its mass fraction $\xi_{(Z,A)}$, defined as the mass of that nuclear species per unit mass of the mineral. The total event rate is therefore obtained by summing the contributions from all constituent nuclei, weighted by $\xi_{(Z,A)}$.
A measured track length $x_T$ does not uniquely identify the recoiling species.
Instead, a given track length may arise from scattering on any of the nuclear species present in the mineral, with the mapping between recoil energy and track length determined separately for each species via Eq.~\eqref{length}.

We compute track-length spectra for a representative target mineral, gypsum. In the Appendix, we present the spectra for the target mineral halite. The chemical compositions and assumed fiducial ${}^{238}$U concentrations of these minerals are summarized in Table~\ref{tab:U238_concentration}. In order to mitigate radiogenic backgrounds (see Sec.~\ref{background}), we restrict our attention to gypsum and halite which are typically found in relatively radiopure geological environments, such as Marine Evaporites (MEs). These environments are known to exhibit substantially lower levels of radioactive contaminants, including uranium, compared to average crustal materials.
Gypsum contains hydrogen, which plays an important role in shaping the neutron-induced background. Both the uranium content and the presence of hydrogen significantly influence the expected background rates in paleo-detector experiments. In particular, for dark matter masses $m_\chi > 10 \, \mathrm{GeV}/c^{2}$, where radiogenic neutrons constitute the dominant background component, hydrogen-bearing minerals are advantageous, as hydrogen efficiently moderates fast neutrons and suppresses neutron-induced recoil events. 

Olivine is another particularly promising paleo-detector target because of its natural abundance, geological stability, and the availability of ancient samples. It is also the subject of substantial ongoing experimental effort aimed at characterizing recoil-induced damage and developing suitable read-out techniques~\cite{Baum:2023cct,Galelli:2025gss,
Hirose:2025jht,Calabrese-Day:2026soq,Hedges:2026pgf}.
Although the absence of hydrogen leads to a larger radiogenic-neutron background than in gypsum, previous projections nevertheless show that olivine can provide competitive sensitivity to a broad range of NREFT interactions. We restrict the present analysis to gypsum and halite in order to study two representative and complementary target minerals while keeping the number of signal, mass-reconstruction, and operator-discrimination projections manageable. Detailed DM and
background spectra for olivine and other paleo-detector targets, together with projected sensitivities to NREFT interactions, are presented in TFKS.

\begin{table}[ht]
\captionsetup{justification=raggedright,singlelinecheck=false}
\centering
\begin{tabular}{l l c}
\hline
\textbf{Mineral} & \textbf{Composition} & \textbf{Fiducial ${}^{238}$U Concentration [per Weight, g/g]} \\
\hline
Gypsum        & Ca(SO$_4$)$\cdot$2(H$_2$O)     & $10^{-11}$ \\
Halite        & NaCl                          & $10^{-11}$ \\
\hline
\end{tabular}
\caption{Minerals considered in this work, together with their chemical compositions and the fiducial ${}^{238}$U concentrations $(C^{238})$ assumed for radiopure samples.}
\label{tab:U238_concentration}
\end{table}

The primary observable in a paleo-detector is the spectrum of track lengths, which serves as the counterpart to the recoil energy spectrum measured in conventional direct-detection experiments. We therefore construct the differential event rate with respect to the track length, $dR/dx_T$, defined per unit time, per unit detector mass, and per unit track length, and obtained by summing over all nuclear species $(Z,A)$ present in the mineral, weighted by $\xi_{(Z,A)}$.

The relation between recoil energy and track length is determined by the stopping power. Expressed in terms of $x_T$, the event rate reads
\begin{equation}
    \frac{dR}{dx_T}
    = \sum_{(Z,A)} \xi_{(Z,A)}
      \left(\frac{dR}{dE_R}\right)_{(Z,A)}
      \left(\frac{dE_R}{dx_T}\right)_{(Z,A)} ,
    \label{dRdx}
\end{equation}
where $(dR/dE_R)_{(Z,A)}$ is given in Eq.~\eqref{recoilrate}, and $(dE_R/dx_T)_{(Z,A)}$ is the stopping-power calculated using \textit{SRIM}~\cite{ZIEGLER20101818}.

Nuclear recoil tracks in a paleo-detector sample may be read out using a range of advanced microscopy techniques~\cite{Drukier:2018pdy}. To capture the interplay between spatial resolution and analyzable target mass, we consider two benchmark read-out configurations that span a representative range of experimental capabilities.

\bigskip

\textbullet $\,\,\,\,\,$ {\it{High-Resolution (HR) Scenario}}: We assume that $10~\mathrm{mg}$ of material can be analyzed with a track-length resolution of $\sigma_x = 1~\mathrm{nm}$. Such performance could potentially be achieved using Helium Ion Beam Microscopy in combination with pulsed-laser and fast-ion-beam ablation techniques~\cite{HILL201265,VANGASTEL20122104,Joens2013,ECHLIN20151,PFEIFENBERGER2017109,10.1116/1.5047806}. The excellent spatial resolution in this configuration enhances sensitivity to fine spectral features at short track lengths, which is particularly relevant for resolving low-energy recoil structures.

\bigskip

\textbullet $\,\,\,\,\,$ {\it{High-Exposure (HE) Scenario}}: We assume that $100~\mathrm{g}$ of material can be analyzed with a resolution of $\sigma_x = 15~\mathrm{nm}$. This configuration may be realized using Small-Angle X-ray Scattering tomography at a synchrotron facility~\cite{RODRIGUEZ2014150,Schaff2015,Holler2014}. The substantially larger analyzable mass increases the total number of recorded events, improving statistical precision in regions of parameter space where recoil spectra extend to longer track lengths, for instance for DM masses larger than $10\ \mathrm{GeV}/c^2$.

\bigskip

These benchmark scenarios are consistent with those adopted in previous paleo-detector studies~\cite{Drukier:2018pdy,Baum:2018tfw,Edwards:2018hcf,Baum:2019fqm,Baum:2021jak}. In the context of the present work, they provide complementary regimes in which to assess how experimental resolution and exposure impact the ability to distinguish between interaction hypotheses and to reconstruct the dark matter mass. Although technically demanding, ongoing experimental efforts continue to explore the feasibility and scalability of these read-out approaches~\cite{Baum:2023cct,Baum:2024eyr,Hirose:2025jht}.

A central observable in our analysis is the binned track-length spectrum, defined as the number of WIMP-induced tracks grouped according to their measured lengths. This quantity represents the experimentally accessible signal and incorporates the finite spatial resolution of the read-out procedure.
For a mineral sample of mass $M$ and age $t_{\text{age}}$, the expected number of tracks in the $k$-th bin, corresponding to $x_T \in [x_{T,k}^{\min}, x_{T,k}^{\max}]$, is given by
\begin{equation}
    \mathcal{N}_k = 
    M \times t_{\mathrm{age}} 
    \int dx_T' \,
    W\!\left(x_T'; x_{T,k}^{\min}, x_{T,k}^{\max}\right)
    \frac{dR}{dx_T}(x_T')~,
    \label{bin}
\end{equation}
where $dR/dx_T$ is defined in Eq.~(\ref{dRdx}) and is assumed to be time independent.

Finite detector resolution is modeled by assuming that the measured track length $x_T$ is Gaussian-distributed about the true length $x_T'$ with variance $\sigma_x^2$, corresponding to the square of the read-out resolution. The associated window function reads
\begin{equation}
    W\!\left(x_T'; x_{T,k}^{\min}, x_{T,k}^{\max}\right)
    = \frac{1}{2}
    \left[
    \operatorname{erf}
    \left(
    \frac{x_T' - x_{T,k}^{\min}}{\sqrt{2}\,\sigma_x}
    \right)
    -
    \operatorname{erf}
    \left(
    \frac{x_T' - x_{T,k}^{\max}}{\sqrt{2}\,\sigma_x}
    \right)
    \right] .
\end{equation}
To prevent artificial information arising from track lengths below the spatial resolution, we exclude contributions from true track lengths $x_T' < \sigma_x/2$ before performing the bin integration. In all numerical analyses presented in this work, we employ 100 logarithmically spaced bins between $\sigma_x/2$ and $1000~\mathrm{nm}$.

\begin{figure}
    \captionsetup{justification=raggedright,singlelinecheck=false}
    \centering
    \includegraphics[width=0.41\textwidth]{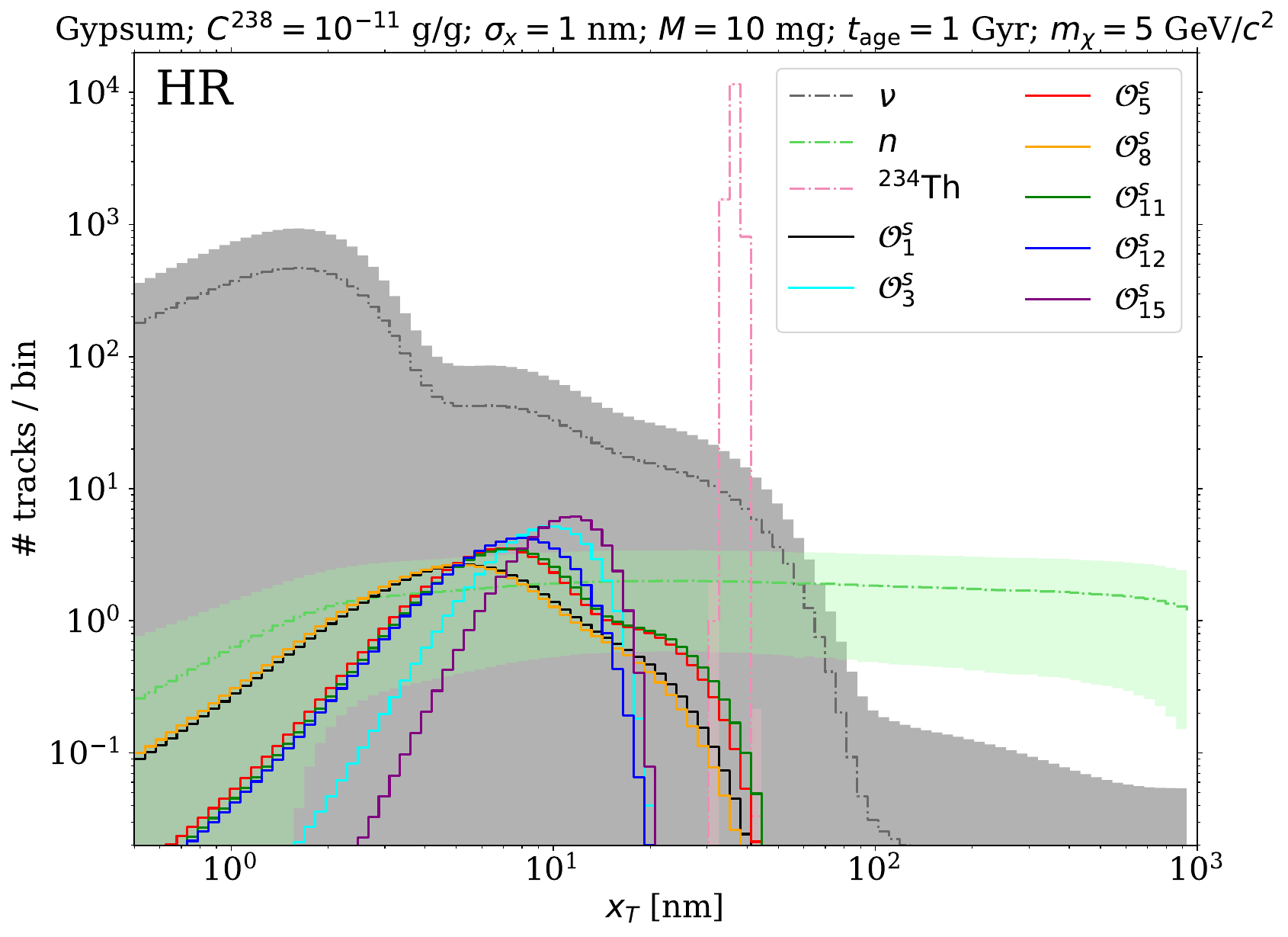}
    \includegraphics[width=0.415\textwidth]{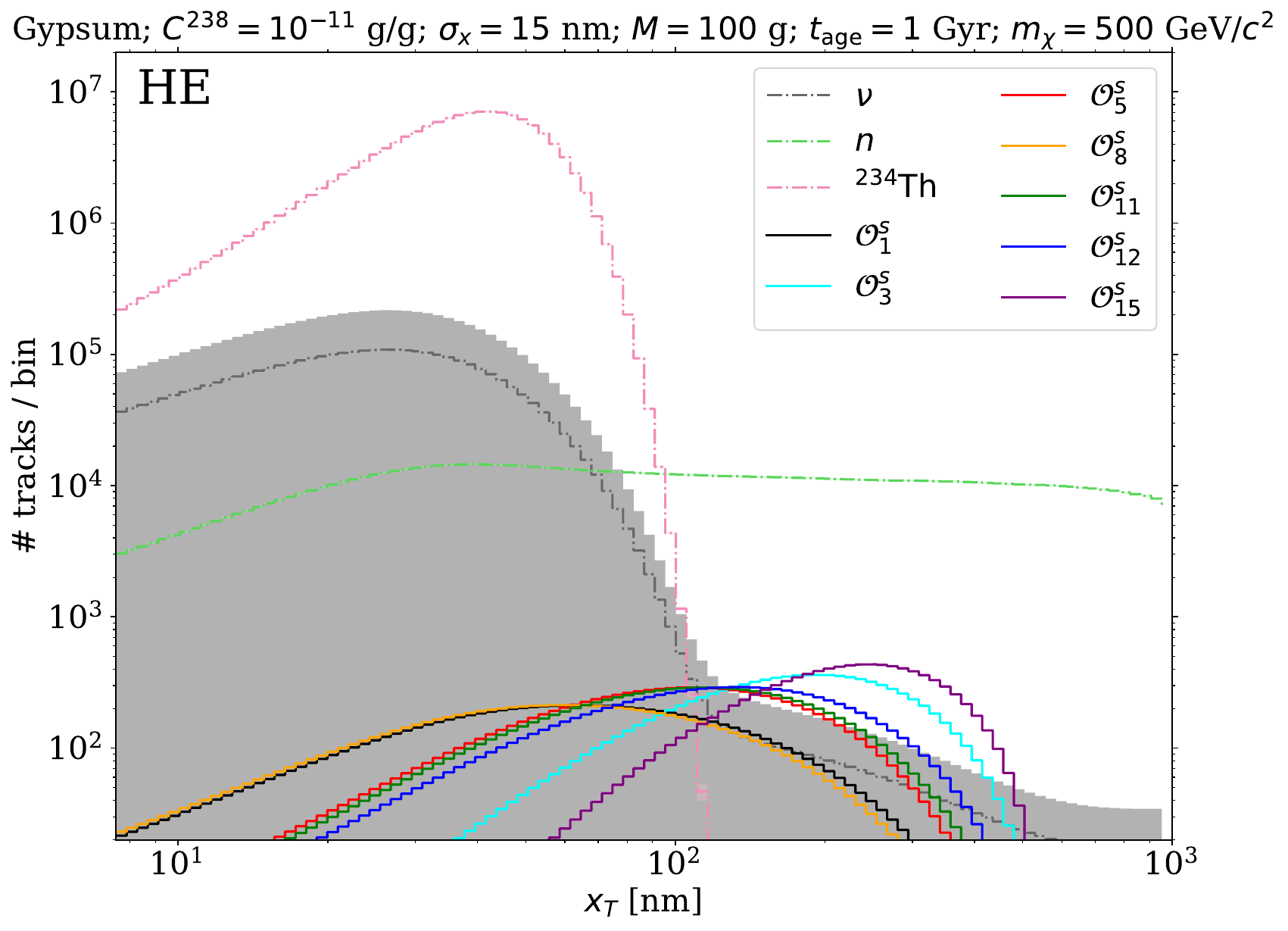}
    \caption{Track-length spectra for the NREFT operators $\mathcal{O}^{s}_{1}$, 
    $\mathcal{O}^{s}_{3}$,
    $\mathcal{O}^{s}_{5}$, $\mathcal{O}^{s}_{8}$, $\mathcal{O}^{s}_{11}$, $\mathcal{O}^{s}_{12}$, and $\mathcal{O}^{s}_{15}$, assuming elastic isoscalar interactions ($c^{p}=c^{n}$) on a gypsum target. Left panel: high-resolution (HR) scenario, with read-out resolution $\sigma_{x}=1\,\mathrm{nm}$, mineral mass $M=10\,\mathrm{mg}$, and DM mass $m_{\chi}=5\,\mathrm{GeV}/c^{2}$. The DM track-length spectra are normalized such that $R = \int dx_T\, dR/dx_T=6\times10^3\ \mathrm{kg}^{-1}\mathrm{Myr}^{-1}$ events per unit exposure, compatible with projected sensitivities for gypsum paleo-detectors~\cite{Theodosopoulos:2026ehn}. Right panel: high-exposure (HE) scenario, with $\sigma_{x}=15\ \mathrm{nm}$, $M=100\ \mathrm{g}$, $m_{\chi}=500\ \mathrm{GeV}/c^{2}$, and $R = 10^2\ \mathrm{kg}^{-1}\mathrm{Myr}^{-1}$ events per unit exposure, compatible with projected paleo-detector sensitivities~\cite{Theodosopoulos:2026ehn}. For comparison, background spectra induced by neutrinos ($\nu$), radiogenic neutrons ($n$), and ${}^{238}\text{U}\to{}^{234}\text{Th}+\alpha$ recoils (${}^{234}\text{Th}$) are also shown; see Sec.~\ref{background}. Here, we include shaded bands around background components, representing the combined statistical (Poisson) and systematic uncertainties in the background predictions. Note that we assume 100\% uncertainty in the neutrino backgrounds which leads to the large, grey-shaded regions. Results are presented for gypsum with a ${}^{238}$U concentration of $10^{-11}\,\mathrm{g/g}$. In both read-out scenarios, the spectra for $\mathcal{O}_{1}^s$ and $\mathcal{O}_{8}^s$ are nearly indistinguishable, while $\mathcal{O}_{3}^s$ and $\mathcal{O}_{15}^s$ generate distinct spectral features relative to the standard SI operator $\mathcal{O}_{1}^s$.}
    \label{fig:Spectrum_binned_SI}
\end{figure}

\begin{figure}
    \captionsetup{justification=raggedright,singlelinecheck=false}
    \centering
    \includegraphics[width=0.41\textwidth]{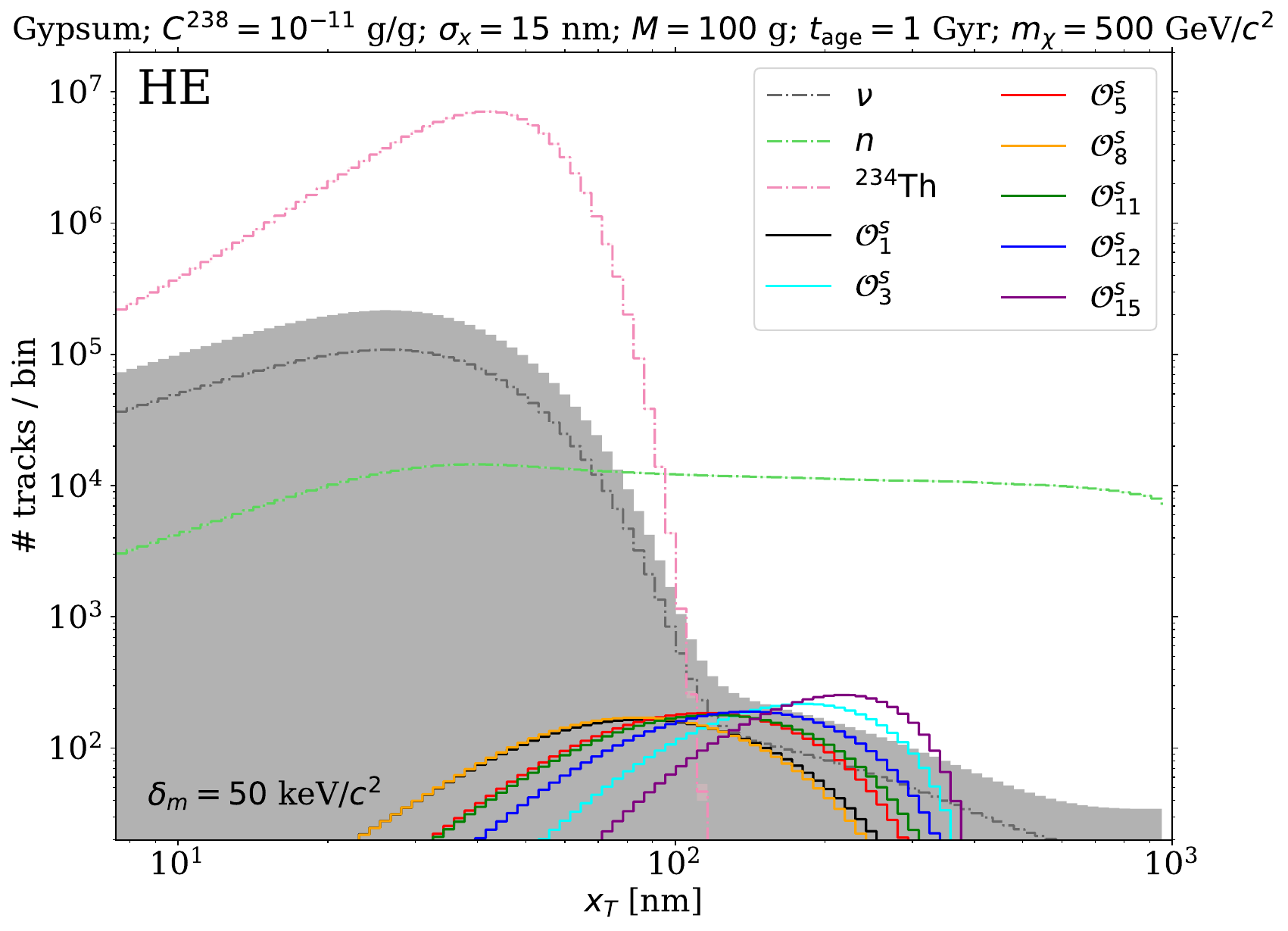}
    \includegraphics[width=0.415\textwidth]{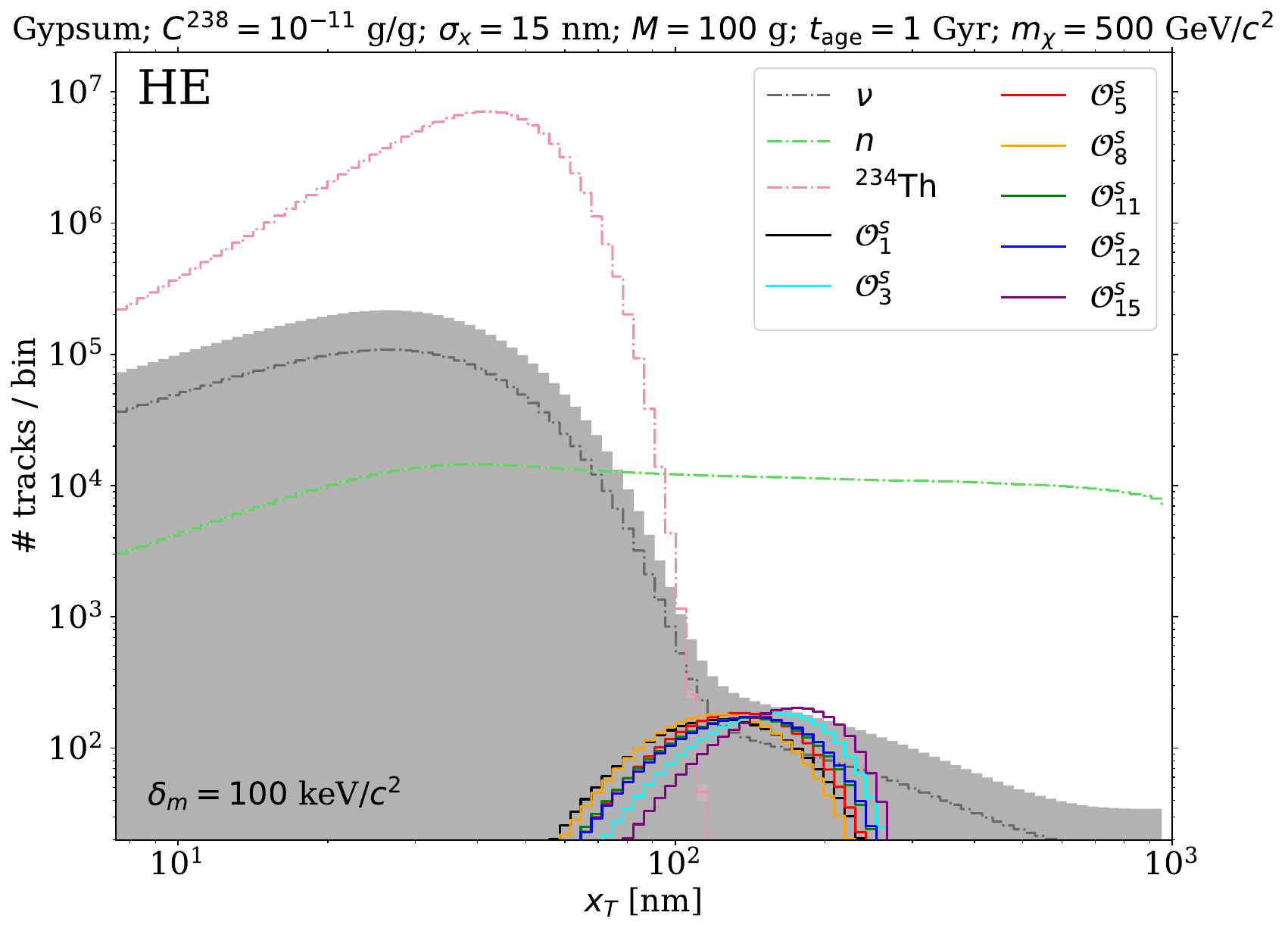}
    \caption{Track length spectra for inelastic scattering via the NREFT operators $\mathcal{O}^{s}_{1}$, 
    $\mathcal{O}^{s}_{3}$,
    $\mathcal{O}^{s}_{5}$, $\mathcal{O}^{s}_{8}$, $\mathcal{O}^{s}_{11}$, $\mathcal{O}^{s}_{12}$, and $\mathcal{O}^{s}_{15}$. The mass-splitting is $\delta_m=50\ \mathrm{keV}/c^2$ (left) and $\delta_m=100\ \mathrm{keV}/c^2$ (right), and the spectra are normalized such that $R = \int dx_T\, dR/dx_T=50\ \mathrm{kg}^{-1}\mathrm{Myr}^{-1}$ (left) and $R =30\ \mathrm{kg}^{-1}\mathrm{Myr}^{-1}$ (right) events per unit exposure, compatible with projected sensitivities for gypsum paleo-detectors~\cite{Theodosopoulos:2026ehn}. The calculations assume isoscalar interactions ($c^{p}=c^{n}$) in the high-exposure (HE) scenario, with read-out resolution $\sigma_{x}=15\ \mathrm{nm}$, mineral mass $M=100\ \mathrm{g}$, and DM mass $m_{\chi}=500\ \mathrm{GeV}/c^{2}$. For comparison, background spectra induced by neutrinos ($\nu$), radiogenic neutrons ($n$), and ${}^{238}\text{U}\to{}^{234}\text{Th}+\alpha$ recoils (${}^{234}\text{Th}$) are also shown; see Sec.~\ref{background}. Here, we include shaded bands around background components, representing the combined statistical (Poisson) and systematic uncertainties in the background predictions. Note that we assume 100\% uncertainty in the neutrino backgrounds which leads to the large, grey-shaded regions. Results are presented for gypsum with a ${}^{238}$U concentration of $10^{-11}\,\mathrm{g/g}$. For $\delta_m=50\ \mathrm{keV}$, the spectra for $\mathcal{O}_{1}^s$ and $\mathcal{O}_{8}^s$ are nearly indistinguishable, as well as the spectra for $\mathcal{O}_{5}^s$ and $\mathcal{O}_{11}^s$. The operators $\mathcal{O}_{3}^s$ and $\mathcal{O}_{15}^s$ generate distinct spectral features relative to $\mathcal{O}_{1}^s$. For $\delta_m=100\ \mathrm{keV}$, which is near the upper limit of $\delta_m$ paleo-detectors can probe (see Ref.~\cite{Theodosopoulos:2026ehn}), the spectra for all the operators look similar.}
    \label{fig:Spectrum_binned_SI_inelastic}
\end{figure}

The dominant isotopes in gypsum have spin-zero ground states, implying that only a subset of NREFT operators contribute to the recoil spectrum. In particular, non-vanishing contributions arise from $\mathcal{O}^{s}_{1}$, $\mathcal{O}^{s}_{3}$, $\mathcal{O}^{s}_{5}$, $\mathcal{O}^{s}_{8}$, $\mathcal{O}^{s}_{11}$, $\mathcal{O}^{s}_{12}$, and $\mathcal{O}^{s}_{15}$, as discussed in TFKS.  Figs.~\ref{fig:Spectrum_binned_SI} and \ref{fig:Spectrum_binned_SI_inelastic} show the non-vanishing binned track-length spectra in gypsum for elastic and inelastic WIMP--nucleon scattering, respectively, together with the background spectra discussed in Section~\ref{background}. 
We also include shaded bands around the background components representing the combined statistical (Poisson) and systematic uncertainties in the background predictions.

As discussed in detail in TFKS, a high read-out resolution is crucial for probing DM particles with masses $m_\chi\lesssim10\ \mathrm{GeV}/c^2$, whereas a high exposure is essential for probing $m_\chi\gtrsim10\ \mathrm{GeV}/c^2$ with sensitivities comparable to or exceeding those of conventional direct-detection experiments.
Accordingly, we focus on the high-resolution (HR) scenario for $m_\chi\lesssim10\ \mathrm{GeV}/c^2$ and the high-exposure (HE) scenario for $m_\chi\gtrsim10\ \mathrm{GeV}/c^2$.
In Fig.~\ref{fig:Spectrum_binned_SI}, the left panel assumes a DM mass of $m_{\chi}=5~\mathrm{GeV}/c^{2}$ in the HR read-out scenario, while the right panel corresponds to $m_{\chi}=500~\mathrm{GeV}/c^{2}$ in the HE scenario. For inelastic scattering, paleo-detectors lose sensitivity to lighter DM particles ($m_\chi<40\ \mathrm{GeV}/c^2$) as shown in TFKS.
Consequently, in Fig.~\ref{fig:Spectrum_binned_SI_inelastic} we consider only the HE scenario and $m_{\chi}=500~\mathrm{GeV}/c^{2}$.

Because heavier DM particles produce larger recoil energies, they also generate longer damage tracks in the target mineral. Accordingly, the track-length spectrum for $m_{\chi}=500\,\mathrm{GeV}/c^{2}$ in Fig.~\ref{fig:Spectrum_binned_SI} is shifted toward larger track lengths compared to the $m_{\chi}=5\,\mathrm{GeV}/c^{2}$
case. 
The dependence of the spectral shape on the DM mass is the key feature enabling mass reconstruction. Given a paleo-detector signal generated by DM with mass $M_\chi$, one can determine the mass interval $\Delta m_\chi$ around $M_\chi$ that cannot be excluded, i.e., for which the corresponding track-length spectra cannot be distinguished by the paleo-detector. This interval, $\Delta m_\chi$, can therefore be interpreted as the uncertainty in the reconstructed DM mass. A detailed mass-reconstruction analysis is presented in Sec.~\ref{mass_reconstruction}.

In addition to mass reconstruction, another major goal of this work is to assess the ability of paleo-detectors to discriminate between different NREFT operators. To this end, we compare the track-length spectra predicted by different operators, focusing exclusively on the shape of the spectra rather than the overall normalizations, i.e., the total number of WIMP-induced tracks, which is assumed to be fixed for a given dataset. 
We therefore normalize the spectra in Fig.~\ref{fig:Spectrum_binned_SI} such that each operator predicts the same total number of WIMP–induced tracks per unit exposure, $R = \int dx_T \, dR/dx_T$.
This procedure isolates differences in spectral shape from differences in overall rate.
In order for the gypsum paleo-detector to be able to discriminate different WIMP-nucleon interactions, the detector should be sensitive to these interactions. Thus, we consider coupling strengths slightly above the projected 90\% C.L. exclusion limits for elastic scattering in gypsum paleo-detectors from TFKS. For this set of coupling strengths, the expected number of events per unit exposure is typically $R=10^2\ \mathrm{kg}^{-1}\mathrm{Myr}^{-1}$ for $m_{\chi} > 10~\mathrm{GeV}/c^{2}$ and $R=6\times10^3\ \mathrm{kg}^{-1}\mathrm{Myr}^{-1}$ for $m_{\chi} < 10~\mathrm{GeV}/c^{2}$, for a gypsum paleo-detector. 

For both a lighter (left panel) and a heavier (right panel) dark matter particle in Fig.~\ref{fig:Spectrum_binned_SI}, the spectra corresponding to $\mathcal{O}^s_{1}$ and $\mathcal{O}^s_{8}$ exhibit very similar shapes, whereas $\mathcal{O}^s_{3}$ and $\mathcal{O}^s_{15}$ produce qualitatively distinct spectral features relative to $\mathcal{O}^s_{1}$. 
This pattern suggests that a gypsum paleo-detector may be capable of distinguishing signals generated by the standard SI operator $\mathcal{O}^s_{1}$ from those induced by $\mathcal{O}^s_{3}$ or $\mathcal{O}^s_{15}$, while discrimination between $\mathcal{O}^s_{1}$ and $\mathcal{O}^s_{8}$ is expected to be more challenging. We quantify these statements in Section~\ref{comparison}.

Fig.~\ref{fig:Spectrum_binned_SI_inelastic} shows the binned track-length spectra for inelastic
scattering in gypsum, considering mass splittings of $\delta_m=50\ \mathrm{keV}/c^2$ (left panel) and $\delta_m=100\ \mathrm{keV}/c^2$ (right panel). All spectra shown in Fig.~\ref{fig:Spectrum_binned_SI_inelastic} are normalized to the same total number of WIMP-induced tracks per unit exposure, $R$. We consider coupling strengths slightly above the projected 90\% C.L. exclusion limits for inelastic scattering in gypsum paleo-detectors from TFKS. For this set of coupling strengths, the expected number of events per unit exposure is typically $R=50\ \mathrm{kg}^{-1}\mathrm{Myr}^{-1}$ for $\delta_m=50\ \mathrm{keV}/c^2$ and $R=30\ \mathrm{kg}^{-1}\mathrm{Myr}^{-1}$ for $\delta_m=100\ \mathrm{keV}/c^2$, for a gypsum paleo-detector. 
As shown in Fig.~\ref{fig:Spectrum_binned_SI_inelastic}, for $\delta_m=50\ \mathrm{keV}$, the spectra predicted by $\mathcal{O}_{1}^s$ and $\mathcal{O}_{8}^s$ are nearly identical, as are those for $\mathcal{O}_{5}^s$ and $\mathcal{O}_{11}^s$. By contrast, $\mathcal{O}_{3}^s$ and $\mathcal{O}_{15}^s$
generate distinctive spectral features relative  $\mathcal{O}_{1}^s$. For $\delta_m=100\ \mathrm{keV}$, which lies near the upper reach of paleo-detector sensitivity, the spectra for all operators become similar.  

In this section, we have introduced the binned track-length spectrum, which constitutes the experimentally accessible observable and forms the basis of the subsequent statistical analysis. Explicit results were presented for the NREFT operators $\mathcal{O}^{s}_{1}$, $\mathcal{O}^{s}_{3}$, $\mathcal{O}^{s}_{5}$, $\mathcal{O}^{s}_{8}$, $\mathcal{O}^{s}_{11}$, $\mathcal{O}^{s}_{12}$, and $\mathcal{O}^{s}_{15}$, for elastic and inelastic scattering in gypsum. Similar analyses for elastic and inelastic scattering in halite mediated by the remaining NREFT operators are presented in the Appendix.

In Section~\ref{background}, we describe the background components relevant for paleo-detector experiments. These background spectra are included in Figs.~\ref{fig:Spectrum_binned_SI} and \ref{fig:Spectrum_binned_SI_inelastic} to enable a direct comparison with the signal-induced track-length spectra and to provide the necessary input for the likelihood analyses performed in the following sections.
In Section~\ref{mass_reconstruction}, we utilize the spectral differences associated with DM mass, to reconstruct the DM mass. In Section~\ref{comparison}, we quantify how the qualitative differences between the spectra predicted by different NREFT operators translate into statistically significant operator discrimination.

\section{Backgrounds in Paleo-Detectors}\label{background}

The types of backgrounds relevant for dark matter searches with paleo-detectors are similar to those encountered in conventional direct-detection experiments, although their relative impact differs significantly. Paleo-detectors utilize comparatively small target masses ($\lesssim 0.1~\mathrm{kg}$) but integrate over geological timescales ($t_{\mathrm{age}}\sim0.1$--$1~\mathrm{Gyr}$), without event-by-event timing information. The observable is a population of nuclear damage tracks, providing essentially perfect rejection of electronic recoils. As a result, only nuclear recoils constitute relevant backgrounds. 
In this section, we summarize the three dominant sources of nuclear-recoil backgrounds: cosmogenic muons, astrophysical neutrinos, and radiogenic processes. Detailed discussions can be found in Refs.~\cite{Drukier:2018pdy,Baum:2019fqm}.

Cosmogenic muon–induced backgrounds are suppressed by selecting minerals shielded by substantial overburden during their geological history, analogous to deep-underground operation in conventional experiments. Since paleo-detectors require only gram-scale samples, suitable minerals may be obtained from boreholes at depths exceeding those of typical underground laboratories. At depths of order $5~\mathrm{km}$, the muon-induced neutron flux is $\mathcal{O}(10^{2})~\mathrm{cm}^{-2}\,\mathrm{Gyr}^{-1}$~\cite{Mei:2005gm}, rendering this contribution subdominant compared to other nuclear-recoil backgrounds. After extraction, storage at modest depth (e.g.\ $\sim50~\mathrm{m}$) results in negligible additional exposure over the timescale of laboratory analysis.

Astrophysical neutrinos produce nuclear recoils through coherent elastic neutrino–nucleus scattering. While neutrino-induced tracks can themselves provide valuable astrophysical information~\cite{Baum:2019fqm,Jordan:2020gxx,Tapia-Arellano:2021cml}, in the present work they are treated as a background to WIMP-induced signals. We include contributions from solar neutrinos, diffuse supernova neutrinos (DSNB), neutrinos from Galactic core-collapse supernovae, and atmospheric neutrinos~\cite{OHare:2020lva}. Over geological integration times, paleo-detectors are sensitive not only to the DSNB but also to neutrinos from local supernovae, given the estimated Galactic rate of $2$--$3$ per century~\cite{Cappellaro:2003eg,Diehl:2006cf,Strumia:2006db,Leaman_2011,Botticella_2012,Adams:2013ana}. 

At short track lengths ($x_T \lesssim 100~\mathrm{nm}$), the neutrino background is dominated by solar components, while supernova neutrinos contribute at intermediate track lengths ($x_T \sim 100~\mathrm{nm}$) and atmospheric neutrinos dominate at the longest track lengths ($x_T \gtrsim 100~\mathrm{nm}$). The overlap between neutrino-induced and WIMP-induced spectral features plays a central role in determining the achievable precision of mass reconstruction and the robustness of operator discrimination.

Radiogenic backgrounds originate from trace radioactive elements embedded in natural minerals, with $^{238}\mathrm{U}$ providing the leading contribution. In typical crustal rocks, uranium concentrations are of order $C^{238}\sim10^{-6}\,\mathrm{g/g}$, which would produce unacceptably large nuclear-recoil backgrounds for dark matter searches. Consequently, paleo-detector studies focus on minerals formed in radiopure environments, such as marine evaporites (MEs) produced by seawater evaporation (see Ref.~\cite{Drukier:2018pdy} and the appendix of Ref.~\cite{Baum:2019fqm}). Following earlier analyses~\cite{Baum:2018tfw,Drukier:2018pdy,Edwards:2018hcf,Baum:2019fqm,Baum:2021jak}, we assume benchmark uranium concentrations of $C^{238}=10^{-11}\,\mathrm{g/g}$ for MEs.

The principal radiogenic processes are $\alpha$ decays and neutron production through spontaneous fission and $(\alpha,n)$ reactions~\cite{Drukier:2018pdy,Baum:2021jak}. The dominant source of isolated recoil events arises from the first decay in
the ${}^{238}\mathrm{U}$ chain, $^{238}\mathrm{U}\to{}^{234}\mathrm{Th}+\alpha$, which produces a $^{234}\mathrm{Th}$ nucleus with a fixed recoil energy of $72~\mathrm{keV}$. These recoils form a monochromatic feature in track-length space with number density
$n(^{234}\mathrm{Th})\simeq10^{6}~\mathrm{g}^{-1}\,(C^{238}/10^{-11}~\mathrm{g/g})$~\cite{Drukier:2018pdy}. 
A $^{234}\mathrm{Th}$ recoil contributes to this background only while the decay chain has undergone the first, but not the second, $\alpha$ decay. Once the second $\alpha$ decay ${}^{234}\mathrm{U}\rightarrow{}^{230}\mathrm{Th}+\alpha$ occurs, the resulting recoil-track pattern can in principle be
identified as a uranium-chain event, and rejected. Since the mineral ages considered here ($\mathcal{O}(1\ \mathrm{Gyr})$) are much longer than the $^{234}\mathrm{U}$ half-life ($\mathcal{O}(0.1\ \mathrm{Myr})$), the population of such single-$\alpha$ events reaches secular equilibrium and is therefore effectively independent of $t_{\mathrm{age}}$ at
fixed present-day $^{238}\mathrm{U}$ concentration.

Neutrons generated in spontaneous fission and $(\alpha,n)$ reactions induce a broad spectrum of nuclear recoils. We model the neutron production using SOURCES-4A~\cite{sources4a1999} and compute the resulting recoil spectra with neutron–nucleus cross sections from the TENDL-2017 library~\cite{KONING20122841}, accessed via JANIS4.0~\cite{SOPPERA2014294}. The neutron background is substantially reduced in hydrogen-containing minerals, since neutrons lose a significant fraction kinetic energy in single elastic scatters with hydrogen nuclei, making such materials particularly advantageous for paleo-detector applications~\cite{Baum:2018tfw,Drukier:2018pdy}.

Radiogenic backgrounds therefore shape the track-length spectrum over a broad range and introduce potential degeneracies with WIMP-induced signals, particularly for operators that enhance higher-energy recoils. Accurate modeling of these components is essential for reliable parameter inference.

\section{Mass reconstruction}\label{mass_reconstruction}

In this section, we investigate the prospects for reconstructing the dark matter mass with paleo-detectors under the assumption that a dark matter signal is present. For a hypothetical signal with true mass $M_{\chi}$, we determine the range of test masses $m_{\chi}$ that cannot be excluded at the $2\sigma$ confidence level. In the low dark mass regime $\mathcal{O}(\mathrm{GeV})$ where paleo-detectors exhibit strong sensitivity and the track-length spectrum significantly depends on DM mass, the allowed mass interval is expected to be narrow compared to $M_{\chi}$. Conversely, at large DM masses of $\mathcal{O}(\mathrm{TeV})$, the track-length spectrum depends only weakly on $M_{\chi}$, causing the reconstructed mass interval to broaden substantially and potentially become unbounded from above.  
A previous study of parameter reconstruction with paleo-detectors was presented in Ref.~\cite{Edwards:2018hcf}, which focused on standard elastic SI scattering and studied WIMP-mass reconstruction using the Euclideanized-signal method~\cite{Edwards:2017kqw,Edwards:2018lsl}. The present work generalizes this analysis to both elastic and inelastic scattering for a broad set of NREFT operators, using a profile-likelihood-ratio approach.

In TFKS, we obtained projected 90\% C.L. upper limits on the isoscalar WIMP--nucleon NREFT coupling constants ($c^p=c^n$) for elastic and inelastic scattering in the DM mass range $1$--$5000~\mathrm{GeV}/c^2$.
This was achieved by testing a signal-plus-background hypothesis against a hypothetical data set containing only background events. For each tested value of the DM mass $m_\chi$, the signal hypothesis assumed $m_\chi$ to be known.

In the present work, we investigate the ability of paleo-detectors to reconstruct the DM mass.
We consider a hypothetical data set
$\mathbf{D}(M_\chi,c_j)$
generated from a DM signal produced by particles of true mass $M_\chi$ interacting through the operator $\mathcal{O}_j^s$ with coupling strength $c_j$, together with the expected background contributions.
To quantify the mass reconstruction capability, we compare two signal-plus-background hypotheses.
The null hypothesis assumes a DM signal coming from $\mathcal{O}_j^s$ interactions with a fixed test mass $m_\chi$ and the expected backgrounds, while the alternative hypothesis allows the DM mass to vary freely.
In both hypotheses, the normalization of the DM spectrum is fixed such that the total DM event rate per unit exposure,
$R = \int dx_T\, dR/dx_T$,
is equal to the rate inferred from the data set $\mathbf{D}(M_\chi,c_j)$. Consequently, our analysis probes the ability of paleo-detectors to reconstruct the DM mass based solely on differences in the spectral shape of the DM-induced track-length distribution as a function of the DM mass, rather than on differences in the overall interaction strength, since both hypotheses predict the same total rate $R$.

Paleo-detectors operate as counting experiments, so the number of observed events in each bin is assumed to follow Poisson statistics. Under this assumption, the log-likelihood for observing a data set $\mathbf{D}(M_{\chi},c_j)$ given the parameter set $\{\vec{\theta}, m_{\chi},\tilde{c}_j\},$ reads
\begin{equation}
    \ln \mathcal{L}_{\text{Poisson}}\bigl(\mathbf{D}(M_{\chi},c_j)\,|\,\vec{\theta}; m_{\chi},\tilde{c}_j\bigr)
    = \sum_{i} \left[ D_{i}(M_{\chi},c_j) \ln \mathcal{N}_{i}\bigl(\vec{\theta}; m_{\chi},\tilde{c}_j\bigr)
    - \mathcal{N}_{i}\bigl(\vec{\theta}; m_{\chi},\tilde{c}_j\bigr) \right]~,\label{loglikelihood}
\end{equation}
where $\mathcal{N}_i(\vec{\theta}; m_{\chi},\tilde{c}_j)$ denotes the predicted number of tracks in the $i^{\text{th}}$ bin, including both background and DM contributions. Here, $m_{\chi}$ is the test DM mass, and $\vec{\theta}$ represents the set of nuisance parameters,
\begin{equation}
    \vec{\theta}=\left\{M, t_{\text{age}},\Phi_{\nu}^{\mathrm{sol}},\Phi_{\nu}^{\mathrm{DSNB}},\Phi_{\nu}^{\mathrm{GSNB}},\Phi_{\nu}^{\mathrm{atm}},C^{238}\right\}~,
\end{equation}
with $M$ and $t_{\text{age}}$ the mass and age of the mineral sample, $\Phi_{\nu}$ the various neutrino fluxes, and $C^{238}$ the uranium concentration. We choose the coupling constant $\tilde{c}_j$ such that the total DM event rate per unit exposure, $R=\int dx_T\,\frac{dR}{dx_T}$,
matches the rate inferred from the data set $\mathbf{D}(M_\chi,c_j)$. Constant terms are omitted, as they cancel in the likelihood ratios considered below.

The total binned track-length spectrum, defined as the number of background- and DM-induced tracks in the $i^{th}$~bin with track length $x_T \in [x_{T,i}^{\min}, x_{T,i}^{\max}]$, is given by
\begin{equation}
\begin{aligned}
    \mathcal{N}_i(\vec{\theta}; m_{\chi},\tilde{c}_j) &=
    \mathcal{N}_{\nu,i}^{\mathrm{sol}}\!\left(\Phi_{\nu}^{\mathrm{sol}}\right)
    + \mathcal{N}_{\nu,i}^{\mathrm{DSNB}}\!\left(\Phi_{\nu}^{\mathrm{DSNB}}\right)
    +\mathcal{N}_{\nu,i}^{\mathrm{GSNB}}\!\left(\Phi_{\nu}^{\mathrm{GSNB}}\right)
    + \mathcal{N}_{\nu,i}^{\mathrm{atm}}\!\left(\Phi_{\nu}^{\mathrm{atm}}\right) \\
    &\quad + \mathcal{N}_{\mathrm{rad},i}^{^{234}\mathrm{Th}}\!\left(C^{238}\right)
    + \mathcal{N}_{\mathrm{rad},i}^{n}\!\left(C^{238}\right)
    + \mathcal{N}_{\mathrm{DM},i}\!\left(m_{\chi},\tilde{c}_j\right) ~,
\end{aligned}
\end{equation}
where each $\mathcal{N}_{s,i}$ is obtained from Eq.~(\ref{bin}) for the corresponding signal ($s=\text{DM}$) or background ($s=\nu~\text{or}~\text{rad}$) component. The neutrino terms account for solar ($\mathcal{N}_\nu^{\mathrm{sol}}$), DSNB ($\mathcal{N}_\nu^{\mathrm{DSNB}}$), Galactic Supernova Neutrino Background ($N_\nu^{\mathrm{GSNB}}$), and atmospheric neutrinos ($\mathcal{N}_\nu^{\mathrm{atm}}$). Radiogenic backgrounds include isolated $^{234}$Th recoils ($\mathcal{N}_{\mathrm{rad}}^{^{234}\mathrm{Th}}$) from the first $\alpha$ decay in the $^{238}$U chain, as well as neutron-induced recoils ($N_{\mathrm{rad}}^{n}$) from spontaneous fission and $(\alpha,n)$ reactions.
The DM contribution $\mathcal{N}_{\mathrm{DM},k}(m_{\chi},\tilde{c}_j)$ depends on the DM mass $m_{\chi}$ and coupling constant $\tilde{c}_j$; for $\vec{S}_{\chi}$-dependent operators, we assume a DM spin of $1/2$.  
All components scale linearly with the sample mass $M$. 
With the exception of $\mathcal{N}_{\mathrm{rad},k}^{^{234}\mathrm{Th}}$, they also scale linearly with the mineral age $t_{\mathrm{age}}$, whereas
$\mathcal{N}_{\mathrm{rad},k}^{^{234}\mathrm{Th}}$ is effectively independent of $t_{\mathrm{age}}$ for Gyr-old minerals (see Sec.~\ref{background}).

External information constrains the nuisance parameters. Neutrino fluxes are informed by experimental measurements and theoretical modeling, while $M$, $t_{\text{age}}$, and $C^{238}$ can be independently determined for a given sample. We incorporate these constraints through Gaussian penalty terms,
\begin{equation}
\ln \mathcal{L}_{\text{ext. const.}}\left(\vec{\theta}\right)
= -\frac{1}{2}\sum_{j}  
\left( \frac{\theta_{j} - \bar{\theta}_{j}}{\ell_{j}\,\bar{\theta}_{j}} \right)^{2} ~,
\end{equation}
where $\bar{\theta}_{j}$ denotes the central value and $\ell_{j}$ the relative uncertainty of the $j$-th nuisance parameter. The full likelihood is then
\begin{equation}
\ln \mathcal{L}\bigl(\mathbf{D}(M_{\chi},c_j)\,|\,\vec{\theta}; m_{\chi},\tilde{c}_j\bigr)
= \ln \mathcal{L}_{\text{Poisson}}\bigl(\mathbf{D}(M_{\chi},c_j)\,|\,\vec{\theta};  m_{\chi},\tilde{c}_j\bigr)
+ \ln \mathcal{L}_{\text{ext. const.}}(\vec{\theta})~.
\end{equation}

To estimate the allowed mass range, we use the profile log-likelihood ratio~\cite{Cowan:2010js,Billard_2012,Conrad:2014nna}
\begin{equation}
q(m_{\chi},c_j)
= -2 \ln \left[
\frac{ \mathcal{L}\left( \mathbf{D}(M_{\chi},c_j) \big| \hat{\hat{\vec{\theta}}}; m_{\chi},\tilde{c}_j \right)}
{ \mathcal{L}\left( \mathbf{D}(M_{\chi},c_j) \big| \hat{\vec{\theta}}; \hat{m}_{\chi}, \tilde{c}_j'\right)}
\right],
\end{equation}
where the nuisance parameters $\hat{\hat{\vec{\theta}}}$ maximize the likelihood in the numerator for fixed test  $m_{\chi}$, while in the denominator the likelihood is maximized over both $\hat{\vec{\theta}}$ and $\hat{m}_{\chi}$.
In both the null and alternative hypotheses, the coupling constants $\tilde{c}_j$ and $\tilde{c}_j'$ are chosen such that the total DM event rate per unit exposure, $R=\int dx_T\,\frac{dR}{dx_T}$,
matches the rate inferred from the data set $\mathbf{D}(M_\chi,c_j)$.
To determine the mass range which cannot be rejected at the $2\sigma$ confidence level, we construct an \emph{Asimov} data set~\cite{Cowan:2010js} assuming a signal with mass $M_{\chi}$ and coupling constant $c_j$. The Asimov data are defined as 
\begin{equation}
D_i = \mathcal{N}_{i}\left(\bar{\vec{\theta}}; M_\chi,c_j\right),
\end{equation}
where $\bar{\vec{\theta}}$ represents a fiducial choice of nuisance parameters.

In our analysis, we adopt sample masses of $\bar{M}=10\,\mathrm{mg}$ and $\bar{M}=100\,\mathrm{g}$ for the high-resolution (HR) and high-exposure (HE) scenarios, respectively. In both cases, we assume a sample age of $\bar{t}_{\mathrm{age}}=1\,\mathrm{Gyr}$, corresponding to the duration over which the sample has recorded damage tracks.
These quantities are treated as externally constrained nuisance parameters with relative uncertainties $\ell_M=0.01\%$ and $\ell_{t_{\mathrm{age}}}=5\%$.
While the target mass can be measured with high precision, the sample age can typically be determined to within a few percent using standard geological dating techniques~\cite{Gradstein2012,Gallagher,vandenHaute1998}.
The fiducial solar and atmospheric neutrino fluxes are taken from Ref.~\cite{OHare:2020lva}, while the diffuse supernova neutrino background (DSNB) spectrum and the contribution from Galactic supernovae are calculated following Ref.~\cite{Baum:2019fqm}. The assumed ${}^{238}\mathrm{U}$
concentrations in the target minerals are summarized in Table~\ref{tab:U238_concentration}. As benchmark values, we adopt relative uncertainties $\ell_{\Phi^i_\nu}=100\%$ and $\ell_{C^{238}}=1\%$ for the corresponding nuisance parameters, consistent with the systematic uncertainties assumed in previous studies~\cite{Baum:2018tfw,Drukier:2018pdy,Edwards:2018hcf,Baum:2019fqm,Jordan:2020gxx}. Nonetheless, as discussed in Ref.~\cite{Baum:2021jak}, the projected sensitivity of paleo-detectors to DM signals is largely insensitive to these external nuisance-parameter constraints.

Assuming Wilks’ theorem~\cite{Wilks:1938dza} applies, the $2\sigma$ compatibility region is defined by $p\text{-value} > 4.55 \times 10^{-2}$. We quantify the mass reconstruction accuracy through the fractional interval $\Delta m_{\chi}/M_{\chi}$, where $\Delta m_{\chi}$ is the difference between the maximum and minimum masses $m_\chi$ satisfying the $2\sigma$ criterion.

\begin{figure}
    \captionsetup{justification=raggedright,singlelinecheck=false}
    \centering
    \begin{subfigure}{0.35\textwidth}
    \includegraphics[width=\linewidth]{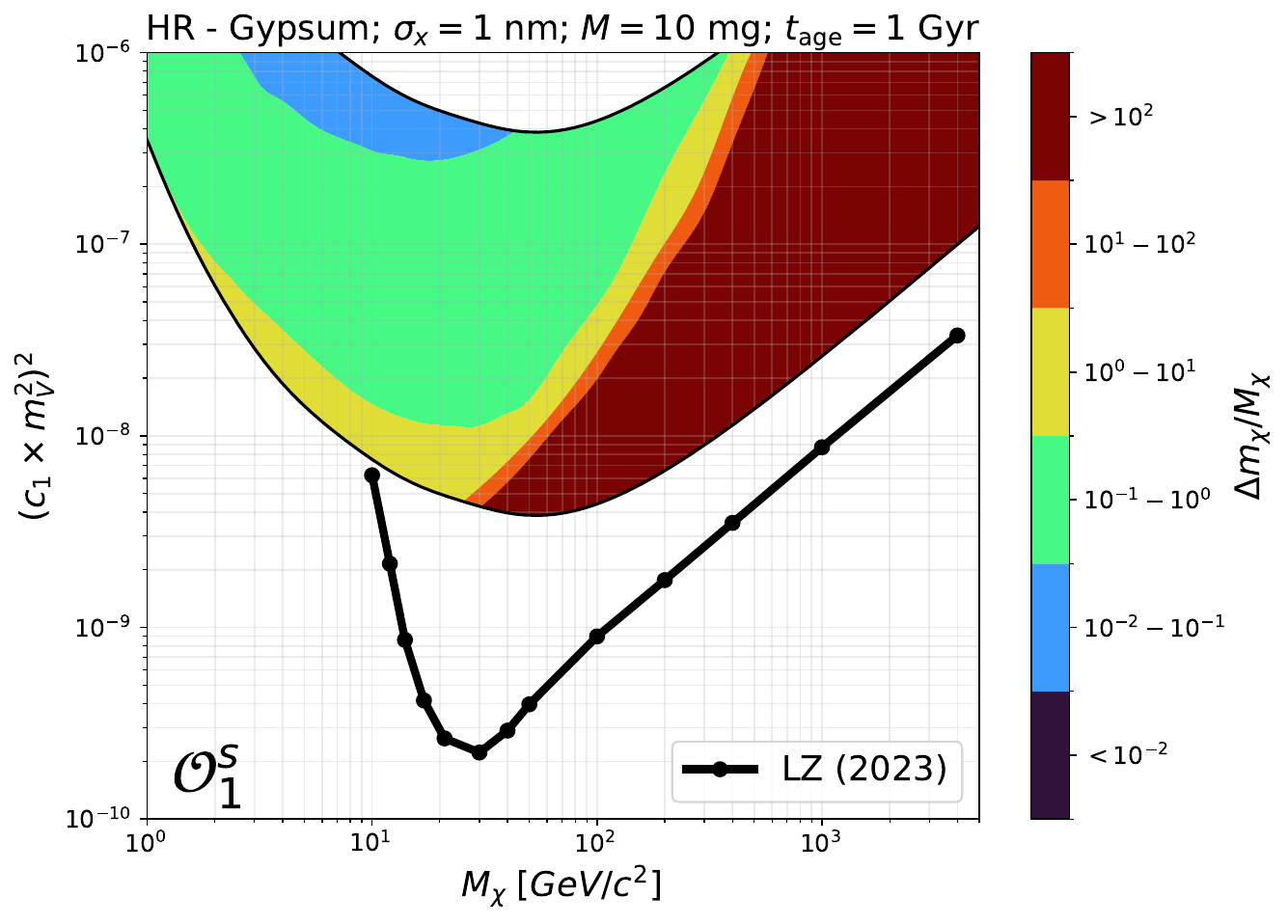}
    \end{subfigure}
    \begin{subfigure}{0.35\textwidth}
    \includegraphics[width=\linewidth]{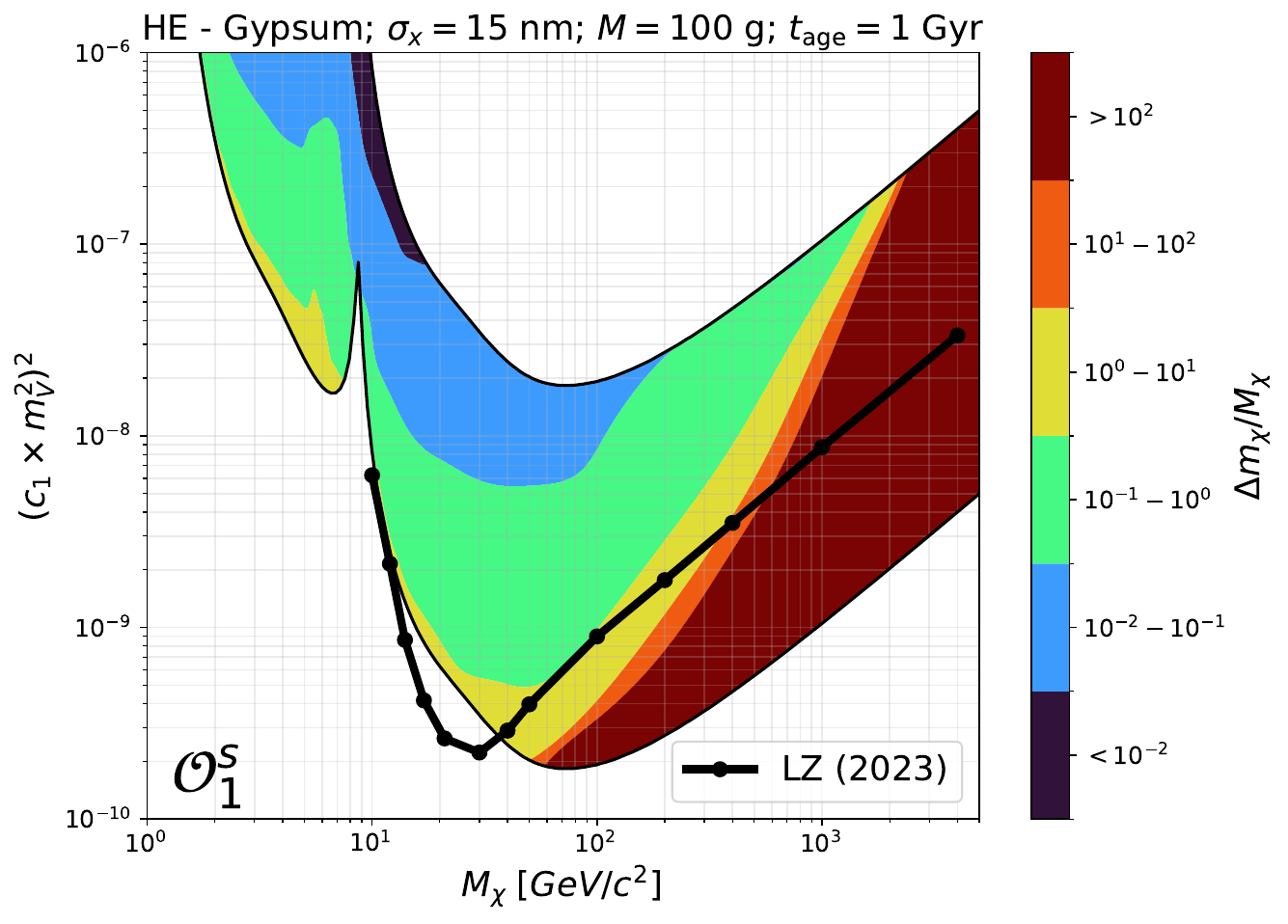}
    \end{subfigure}

    \vspace{0.1ex}

    \begin{subfigure}{0.35\textwidth}
    \includegraphics[width=\linewidth]{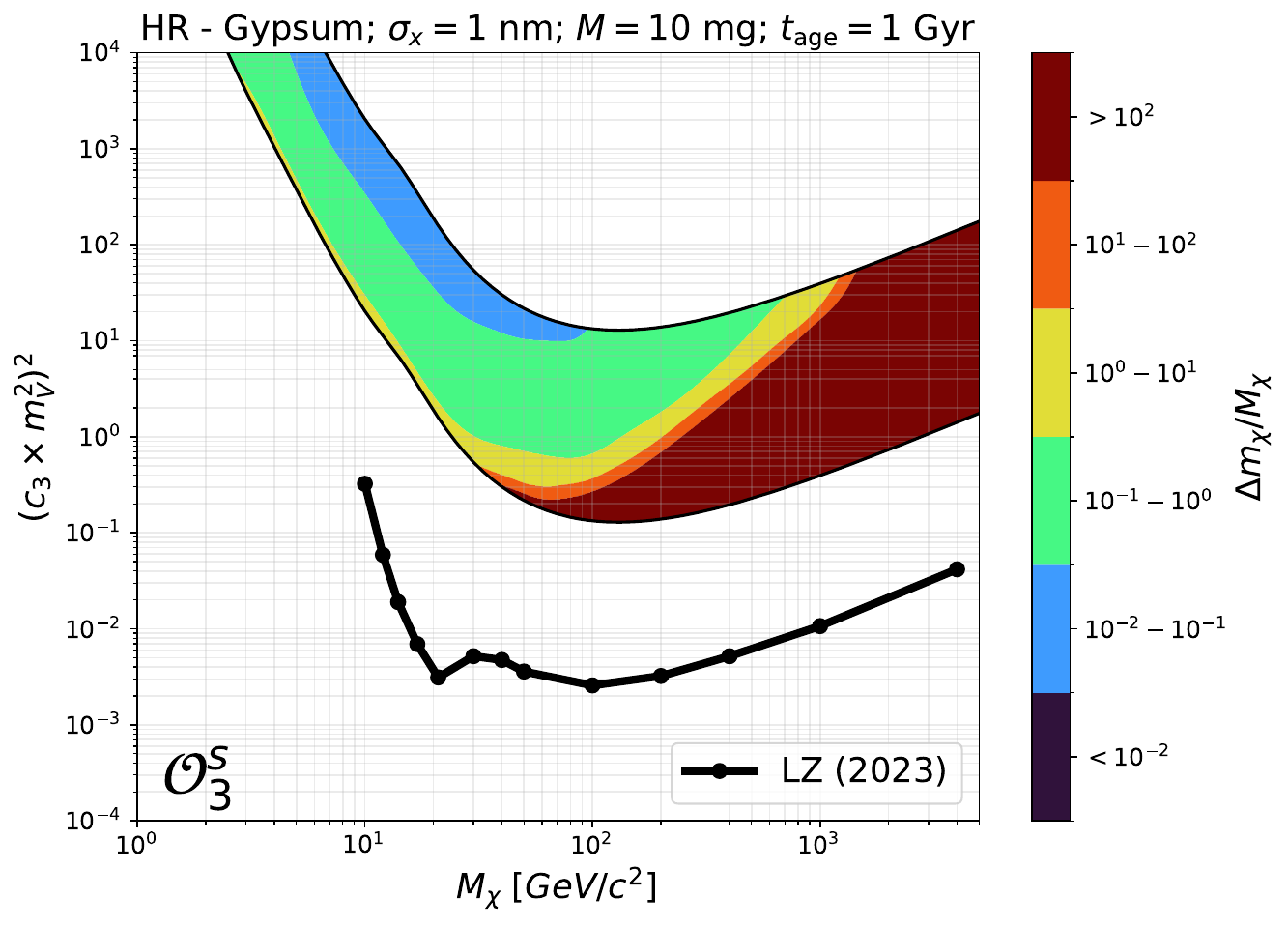}
    \end{subfigure}
    \begin{subfigure}{0.35\textwidth}
    \includegraphics[width=\linewidth]{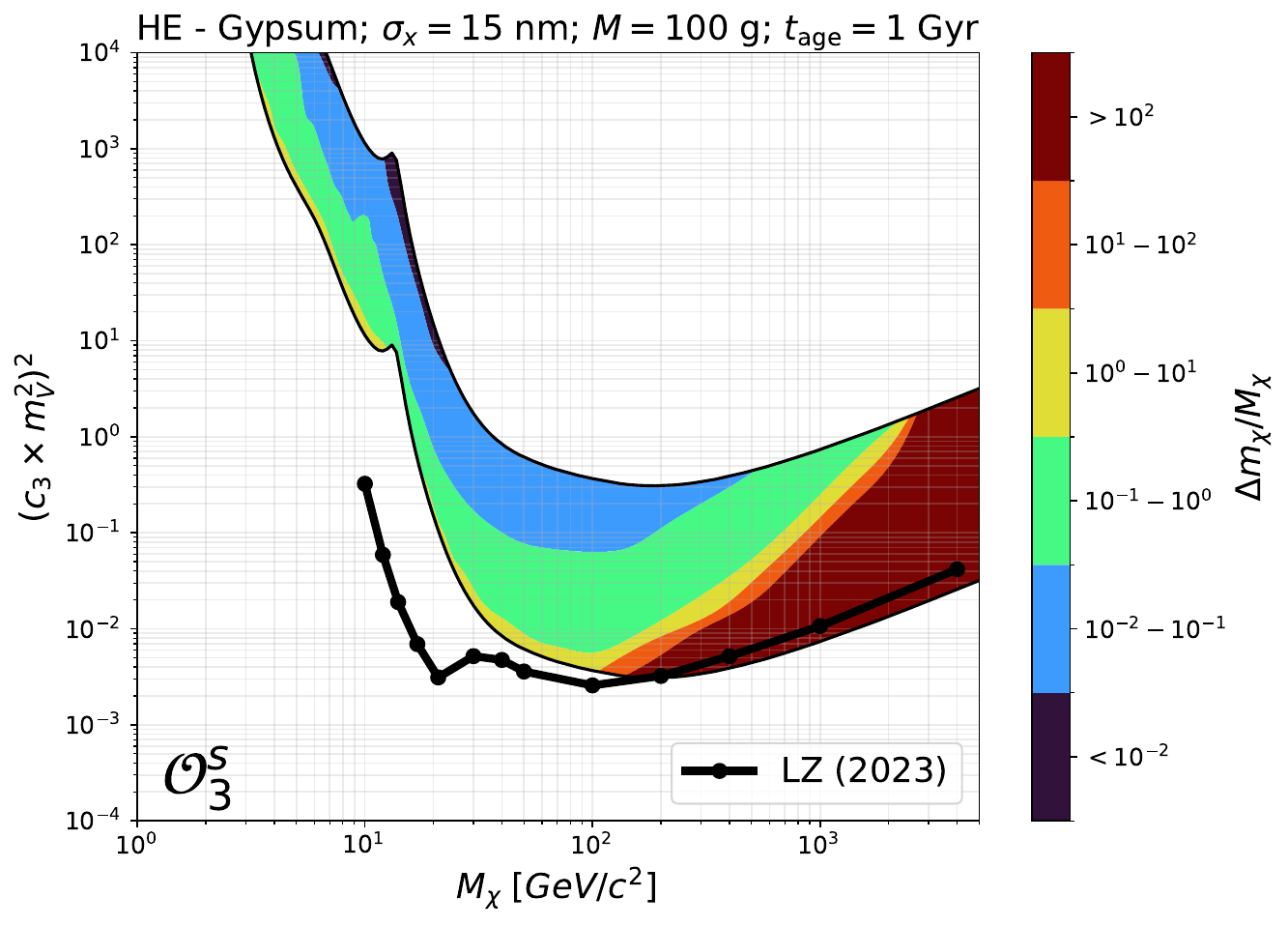}
    \end{subfigure}

    \vspace{0.1ex}
    
    \begin{subfigure}{0.35\textwidth}
    \includegraphics[width=\linewidth]{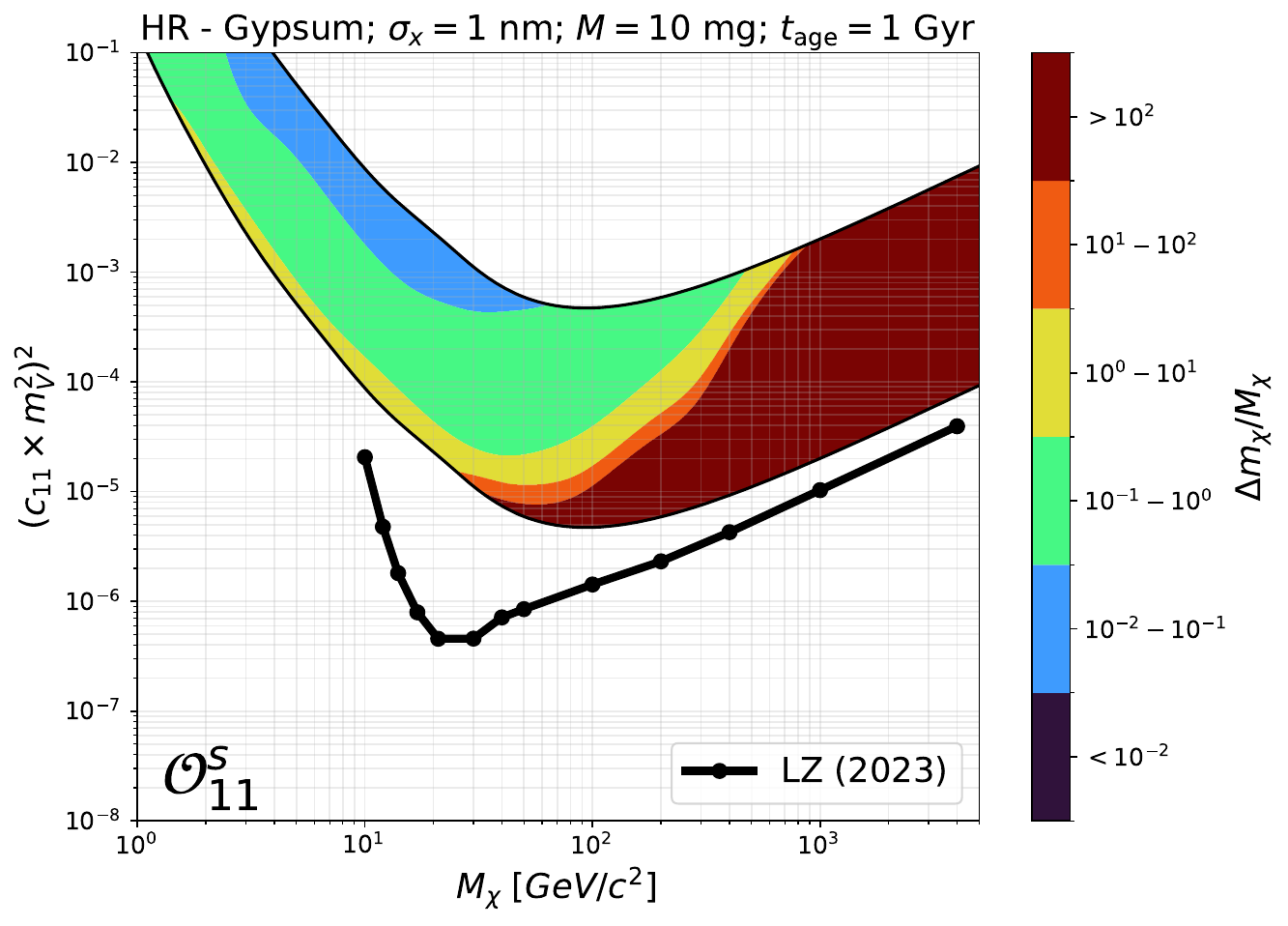}
    \end{subfigure}
    \begin{subfigure}{0.35\textwidth}
    \includegraphics[width=\linewidth]{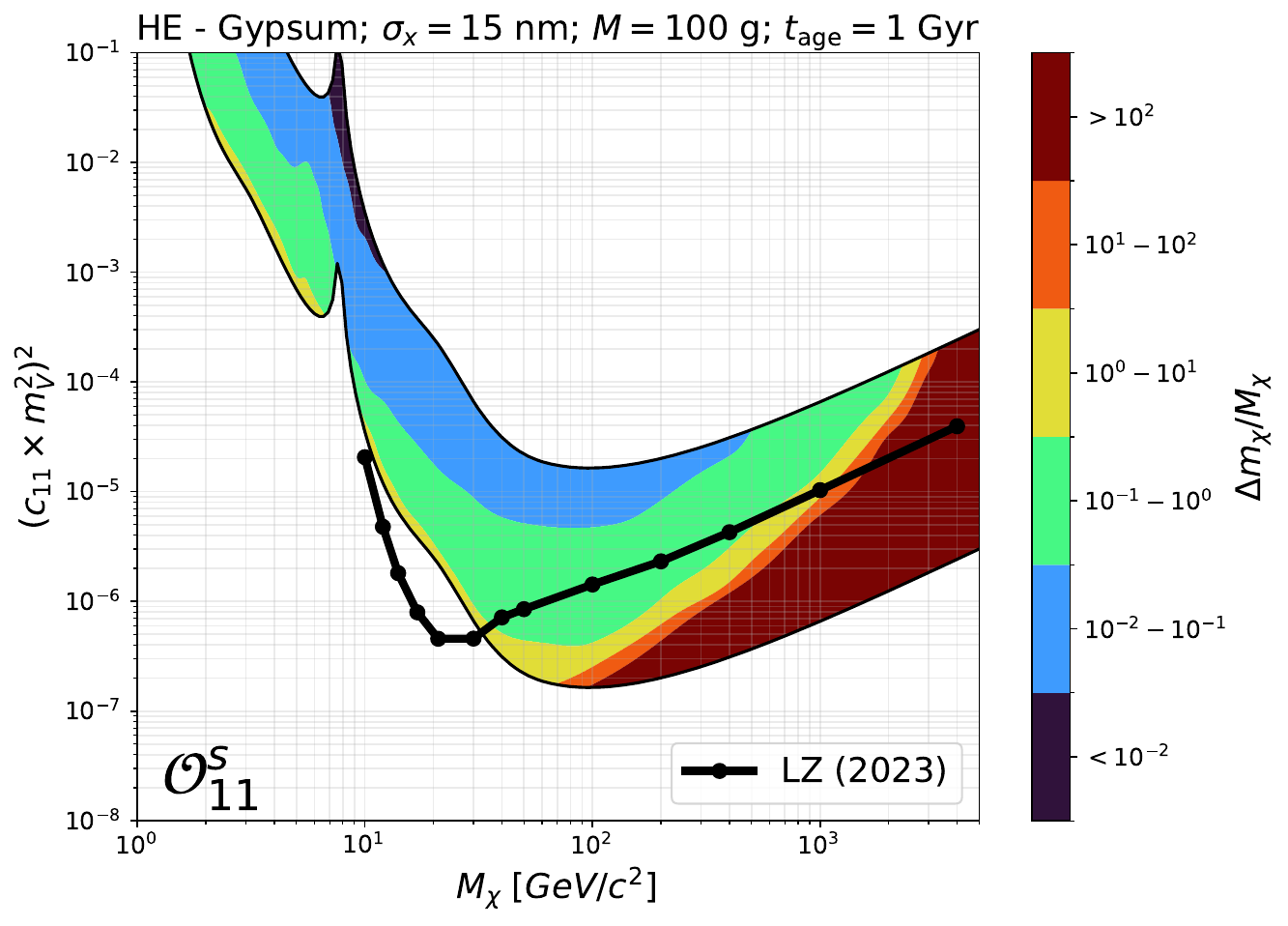}
    \end{subfigure}

    \vspace{0.1ex}

    \begin{subfigure}{0.35\textwidth}
    \includegraphics[width=\linewidth]{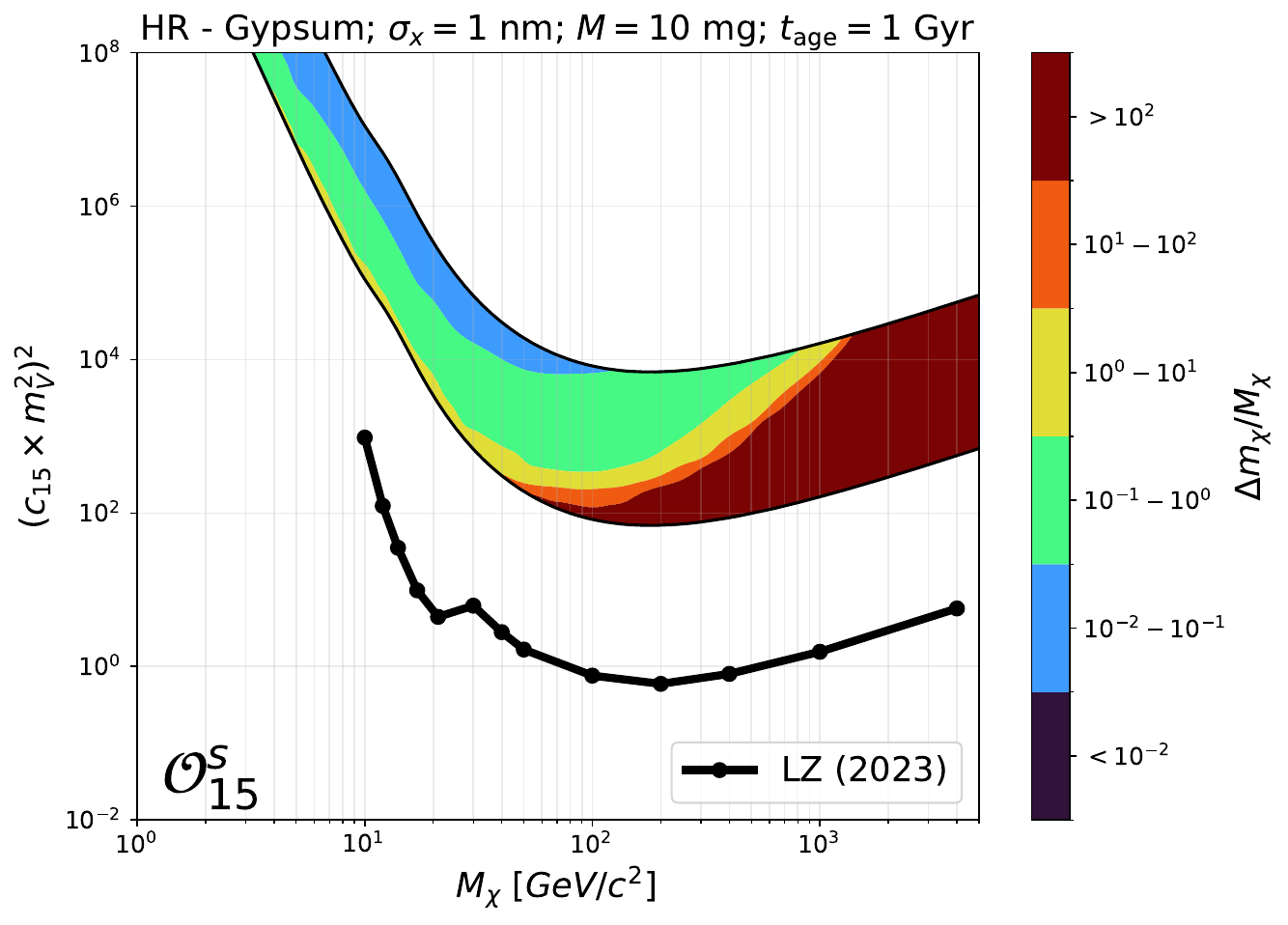}
    \end{subfigure}
    \begin{subfigure}{0.35\textwidth}
    \includegraphics[width=\linewidth]{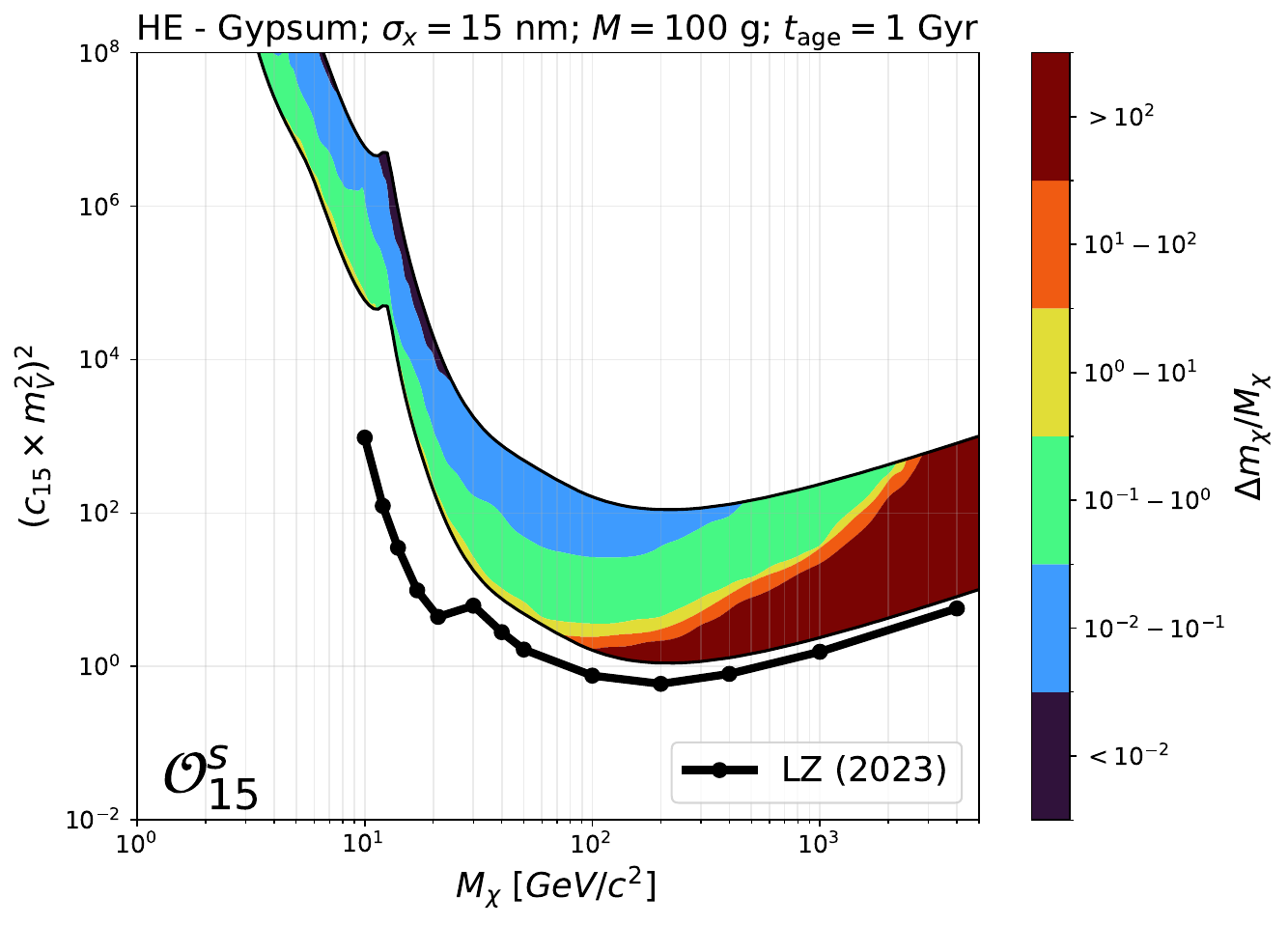}
    \end{subfigure}

    \caption{Projected constraints on the DM mass $M_{\chi}$ from hypothetical signals generated by elastic interactions via the NREFT operators $\mathcal{O}_1^s$, $\mathcal{O}_3^s$, $\mathcal{O}_{11}^s$, and $\mathcal{O}_{15}^s$ in gypsum, with $C^{238}=10^{-11}\ \mathrm{g/g}$ and $t_{\rm age}=1\ \mathrm{Gyr}$. 
    Results are shown for the high-resolution (HR: $\sigma_x=1\ \mathrm{nm}$, $M=10\ \mathrm{mg}$, left) and high-exposure (HE: $\sigma_x=15\ \mathrm{nm}$, $M=100\ \mathrm{g}$, right) scenarios. Colored regions display contours of the relative DM mass uncertainty $\Delta m_{\chi}/M_{\chi}$, ranging from $10^{-2}$ to $10^{2}$, for DM masses $M_{\chi}$ between $1$ and $5000\ \mathrm{GeV}/c^2$ and coupling strengths $(c_j m_V^2)^2$ where $m_V = 246.2\ \mathrm{GeV}$; the contours are bounded between the projected 90\% C.L. exclusion limits for gypsum paleo-detectors from TFKS and a factor of 100 above them. 
    Portions of the parameter space are already excluded by existing experiments, such as LUX–ZEPLIN~\cite{LZ:2023lvz} (upper bounds shown as thick black line with circular markers). Gypsum paleo-detectors are projected to reconstruct the DM mass with a relative uncertainty as small as $\Delta m_\chi/M_\chi \sim 10^{-2}$ (light blue regions) for $M_\chi \simeq 1$--$10\ \mathrm{GeV}/c^2$ , a mass range that is difficult to probe with conventional direct-detection experiments. In the HE scenario, mass reconstruction remains possible over the range $M_\chi \simeq 40$--$1000\ \mathrm{GeV}/c^2$ with a relative uncertainty $\Delta m_\chi/M_\chi \gtrsim 1$ for the operators $\mathcal{O}_{1}^s$ and $\mathcal{O}_{11}^s$.}

    \label{fig:mass_reconstruction_gypsum_elastic}
\end{figure}

The reconstructed mass intervals, assuming DM signals generated by the NREFT operators $\mathcal{O}_1^s$, $\mathcal{O}_3^s$, $\mathcal{O}_{11}^s$, and $\mathcal{O}_{15}^s$, are shown in Fig.~\ref{fig:mass_reconstruction_gypsum_elastic} for gypsum as the target mineral. The left and right panels correspond to the high-resolution (HR) and high-exposure (HE) scenario, respectively. The HR (HE) scenario assumes a read-out resolution of $\sigma_x=1\ \mathrm{nm}$ ($\sigma_x=15\ \mathrm{nm}$) and a target mass of $M=10\ \mathrm{mg}$ ($M=100\ \mathrm{g}$).
Fig.~\ref{fig:mass_reconstruction_gypsum_elastic} displays contours of the relative DM mass uncertainty $\Delta m_{\chi}/M_{\chi}$, ranging from $10^{-2}$ to $10^{2}$, for DM masses $M_{\chi}$ between $1$ and $5000\ \mathrm{GeV}/c^2$ and coupling strengths $(c_j m_V^2)^2$ between the projected 90\% C.L. exclusion limits for gypsum paleo-detectors from TFKS and a factor of 100 larger. Here, $m_V = 246.2\ \mathrm{GeV}$ denotes the electroweak scale. 

The reconstructed mass intervals broadly follow the same trends as the projected $90\%$ confidence-level upper limits on the coupling constant obtained in TFKS. We therefore omit mass-reconstruction projections for the remaining operators to which gypsum paleo-detectors are sensitive, namely $\mathcal{O}_{5}^s$, $\mathcal{O}_{8}^s$, and $\mathcal{O}_{12}^s$. For the operators $\mathcal{O}_{4}^s$, $\mathcal{O}_{6}^s$, $\mathcal{O}_{10}^s$, and $\mathcal{O}_{13}^s$, we present mass-reconstruction projections for halite paleo-detectors in the Appendix.

For large DM masses ($m_{\chi} \gtrsim 1\ \mathrm{TeV}/c^2$), the reconstructed mass intervals become significantly broadened ($\Delta m_{\chi}/M_{\chi}>10^2$), reflecting the weak dependence of the recoil spectrum on $m_{\chi}$ in this regime. This behavior arises from the dependence of the spectral shape on the DM mass through the reduced WIMP-nucleus mass $\mu_T$, which becomes nearly independent of DM mass when it is significantly larger than the target nucleus mass $m_{T}$. 
For the candidate minerals considered in the paleo-detector literature, the target nuclei have masses $m_T \lesssim 50\ \mathrm{GeV}/c^2$.
In the regime $m_\chi\gg m_T$, the reduced mass approaches $\mu_T\simeq m_T$, causing the spectral shape to become effectively independent of $m_\chi$. As a consequence, the reconstructed DM mass becomes unbounded from above.

In Fig.~\ref{fig:mass_reconstruction_gypsum_elastic}, we also 
depict the 90\% C.L. exclusion limits from LUX–ZEPLIN~\cite{LZ:2023lvz} to illustrate that portions of the parameter space are already excluded by existing experiments. Nevertheless, in the high-exposure (HE) scenario and for operators $\mathcal{O}_1^s$ and $\mathcal{O}_{11}^s$ in parameter space not yet ruled out by conventional experiments, a gypsum paleo-detector can reconstruct $M_\chi$ roughly in the range $40-1000\ \mathrm{GeV}/c^2$ with relative uncertainty up to $\Delta m_{\chi}/M_{\chi}=10^2$. In the mass range $1-10\ \mathrm{GeV}/c^2$, which lies beyond the reach of the exclusion limits from LUX–ZEPLIN, paleo-detectors are projected to reconstruct DM mass, especially in the high-resolution scenario (HR), with relative uncertainty up to $\Delta m_{\chi}/M_{\chi}\sim10^{-1}$.   

\begin{figure}
    \captionsetup{justification=raggedright,singlelinecheck=false}
    \centering
    \begin{subfigure}{0.46\textwidth}
    \includegraphics[width=\linewidth]{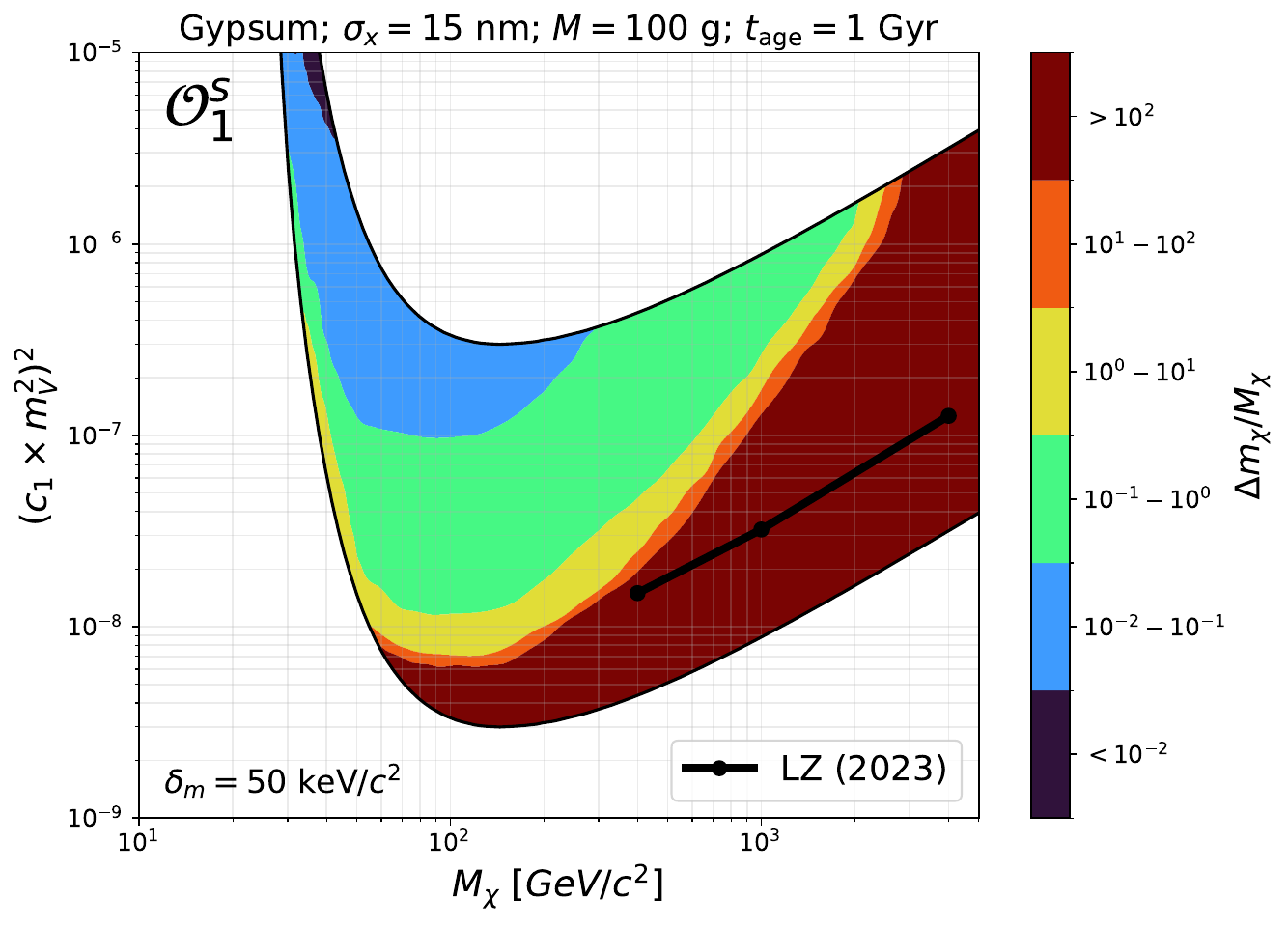}
    \end{subfigure}
    \begin{subfigure}{0.46\textwidth}
    \includegraphics[width=\linewidth]{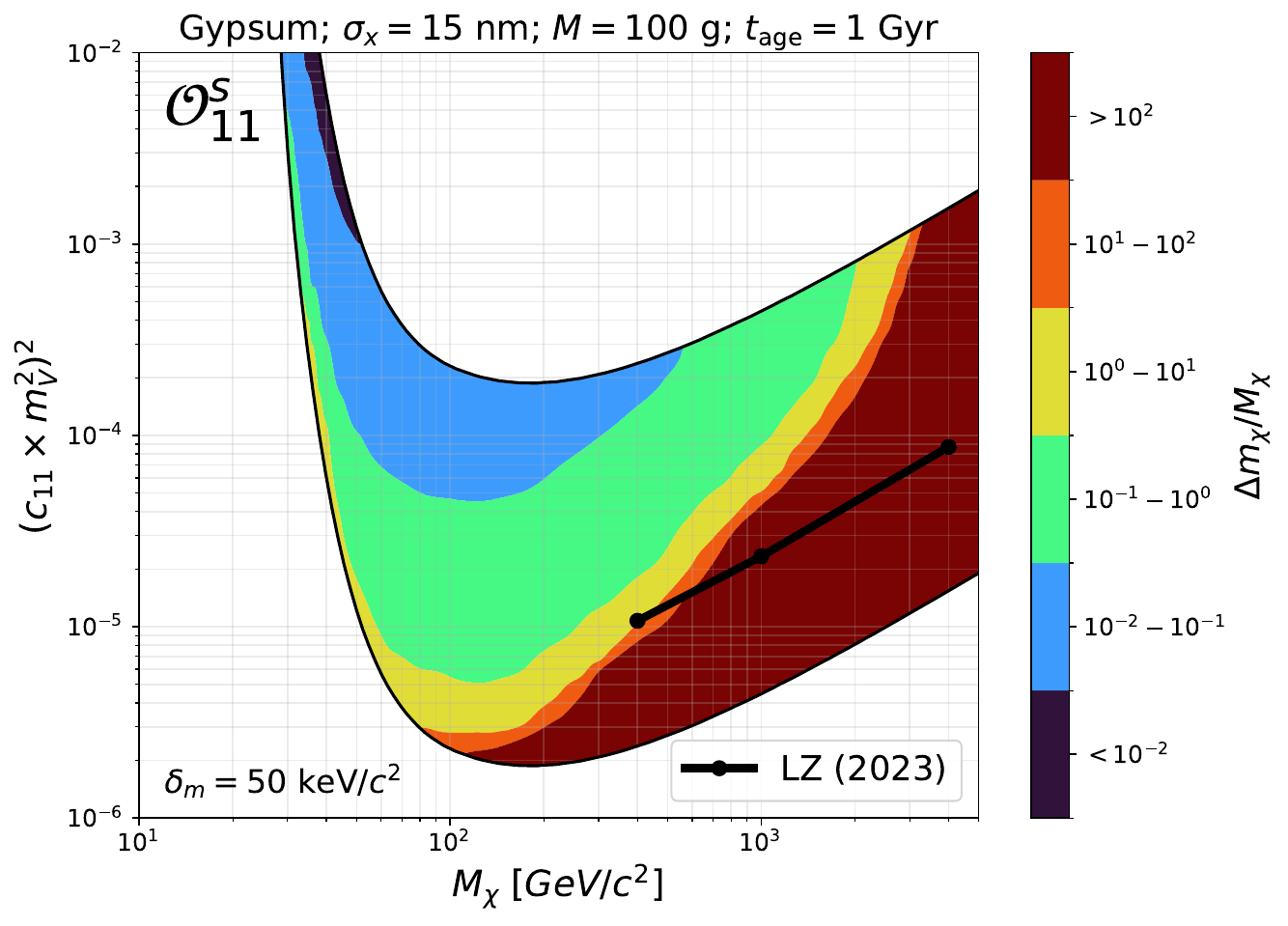}
    \end{subfigure}

    \caption{Projected constraints on the DM mass $M_{\chi}$ from a hypothetical signal generated by inelastic interactions via the $\mathcal{O}_{1}^s$ operator (left) and $\mathcal{O}_{11}^s$ (right) in gypsum, with $C^{238}=10^{-11}\ \mathrm{g/g}$ and $t_{\rm age}=1\ \mathrm{Gyr}$. Axes and contours are as in Figure 3, but for the case of inelastic scattering.
    Results are shown for a mass-splitting $\delta_m=50\ \mathrm{keV}/c^2$ in the high-exposure (HE: $\sigma_x=15\ \mathrm{nm}$, $M=100\ \mathrm{g}$) scenario. 
    Gypsum paleo-detectors are projected to reconstruct $M_\chi$ with relative uncertainties down to  $\Delta m_{\chi}/m_{\chi}\sim10^{-2}$ (light blue region) for DM masses between $30$ and $400\,\mathrm{GeV}/c^2$, a regime that is difficult to probe with conventional direct-detection experiments.}

    \label{fig:mass_reconstruction_Gypsum_inelastic}
\end{figure}

In the case of inelastic scattering, Fig.~\ref{fig:mass_reconstruction_Gypsum_inelastic} shows the reconstructed mass intervals assuming DM signals generated by the NREFT operators $\mathcal{O}_1^s$ and $\mathcal{O}_{11}^s$ in gypsum, with a mass splitting $\delta_m=50\ \mathrm{keV}/c^2$.
As discussed in detail in TFKS,
paleo-detectors lose sensitivity to DM masses below $\sim30\,\mathrm{GeV}/c^2$ for $\delta_m\sim50\,\mathrm{keV}/c^2$. We therefore focus on the high-exposure (HE) scenario when projecting the mass reconstruction with paleo-detectors in the inelastic case, since it is better suited to larger DM masses. 
We adopt $\delta_m\sim50\,\mathrm{keV}/c^2$ as a representative mass splitting for inelastic scattering and do not consider larger values, since most paleo-detectors lose sensitivity near $\delta_m\sim100\,\mathrm{keV}/c^2$ (see TFKS). 
We omit mass-reconstruction projections for the operators $\mathcal{O}_3^s$ and $\mathcal{O}_{15}^s$,
since gypsum paleo-detectors are insensitive to these operators for inelastic scattering at $M_\chi \gtrsim 400\ \mathrm{GeV}/c^2$ (see TFKS), while at lower masses the reconstructed mass intervals for these operators follow trends similar to those obtained for $\mathcal{O}_1^s$ and $\mathcal{O}_{11}^s$.

In Fig.~\ref{fig:mass_reconstruction_Gypsum_inelastic}, we additionally display the 90\% C.L. exclusion limits from LUX--ZEPLIN~\cite{LZ:2023lvz}, demonstrating that part of the parameter space is already excluded by existing experiments.
Even so, gypsum paleo-detectors are projected to reconstruct $M_\chi$ with relative uncertainties down to $\Delta m_{\chi}/M_{\chi}\sim10^{-2}$ for DM masses between $30$ and $400\,\mathrm{GeV}/c^2$, where conventional direct-detection experiments lose sensitivity.  

In this section, we have made projections for the ability of gypsum paleo-detectors (a representative target mineral) to reconstruct DM mass in both elastic and inelastic scenarios. In particular, gypsum paleo-detectors are projected to reconstruct DM mass $M_{\chi}$ with relative uncertainty up to $\Delta m_{\chi}/M_{\chi}\sim10^{-1}$ in the mass range $1-10\ \mathrm{GeV}/c^2$, which is out of the reach of the 90\% C.L. exclusion limits from LUX--ZEPLIN. Additionally, for some NREFT operators, such as $\mathcal{O}_{1}^s$ and $\mathcal{O}_{11}^s$, gypsum paleo-detectors are projected to reconstruct $M_\chi$ in the range $40-1000\ \mathrm{GeV}/c^2$ with relative uncertainty as small as $\Delta m_{\chi}/M_{\chi}=1$. In the inelastic scattering scenario, for $\delta_m=50\ \mathrm{keV}/c^2$ and DM masses between $30$ and $400\,\mathrm{GeV}/c^2$, gypsum paleo-detectors are projected to reconstruct $M_\chi$ with relative uncertainties down to  $\Delta m_{\chi}/m_{\chi}\sim10^{-2}$, a regime in which conventional direct-detection experiments lose sensitivity. In the Appendix, we present projections for DM mass reconstruction with halite paleo-detectors for elastic and inelastic scattering mediated by selected spin-dependent NREFT operators.

The mass-reconstruction analysis presented here should be interpreted as conditional on the interaction operator, as the hypothetical data set and the tested signal hypothesis are assumed to arise from the same NREFT operator. Consequently, our projections quantify the reconstructed mass intervals once the interaction hypothesis is fixed, or has been identified with sufficient confidence. A fully simultaneous
reconstruction of the DM mass, coupling strengths, and operator content would require a higher-dimensional inference framework, and is left for future work.

\section{Discriminating Between standard SI and non-standard WIMP–nucleus interactions} \label{comparison}

Assuming that a DM signal has been detected, in this section, we investigate the ability of paleo-detectors to discriminate between non-standard NREFT operators and the standard spin-independent (SI) $\mathcal{O}_{1}^{s}$ WIMP--nucleon interactions. 
In particular, for a hypothetical detected signal arising from dark matter interactions via a non-standard operator, we determine the confidence level for rejecting the standard SI interactions, using a profile-likelihood-ratio approach. 
We focus on projections for rejecting the standard SI elastic and
inelastic interaction hypotheses in gypsum paleo-detectors
in the presence of hypothetical signals generated by the non-standard operators $\mathcal{O}^{s}_{3}$,
$\mathcal{O}^{s}_{5}$, $\mathcal{O}^{s}_{8}$,
$\mathcal{O}^{s}_{11}$, $\mathcal{O}^{s}_{12}$, and
$\mathcal{O}^{s}_{15}$. 
The corresponding analysis for rejecting the standard spin-dependent (SD) $\mathcal{O}_{4}^{s}$ elastic and inelastic interaction
hypotheses in halite paleo-detectors,
in the presence of hypothetical signals generated by the non-standard operators $\mathcal{O}^{s}_{6}$,
$\mathcal{O}^{s}_{7}$, $\mathcal{O}^{s}_{9}$,
$\mathcal{O}^{s}_{10}$, $\mathcal{O}^{s}_{13}$, and
$\mathcal{O}^{s}_{14}$,
is presented in the Appendix.

To discriminate between two interactions, we compare two signal-plus-background hypotheses. The null hypothesis assumes background plus DM signals arising solely from the standard SI operator. The alternative hypothesis includes, in addition, a contribution from a non-standard NREFT operator. We denote by $F$ the fraction of the DM signal arising from the non-standard operator, while the remaining fraction $(1-F)$ originates from the standard SI interaction. 
Neither hypothesis assumes a fixed value of the DM mass $m_\chi$. 
For a given hypothetical data set, the normalization of the track-length spectrum is fixed to reproduce the same total number of WIMP-induced tracks per unit exposure, $R = \int dx_T\, dR/dx_T$ in both hypotheses. Consequently, the discrimination power obtained in this analysis arises exclusively from differences in the shape of the predicted track-length spectra, rather than from differences in the overall interaction strength. 

To assess the ability of gypsum paleo-detectors to discriminate between different WIMP--nucleon interaction hypotheses, we consider coupling strengths slightly above the projected 90\% C.L. exclusion limits derived in TFKS, for which gypsum paleo-detectors are expected to be sensitive. For elastic scattering, these coupling strengths correspond to typical event rates per unit exposure of $R=10^2\ \mathrm{kg}^{-1}\mathrm{Myr}^{-1}$ for $m_{\chi}>10\,\mathrm{GeV}/c^{2}$ and $R=6\times10^3\ \mathrm{kg}^{-1}\mathrm{Myr}^{-1}$ for $m_{\chi}<10\,\mathrm{GeV}/c^{2}$. For inelastic scattering, these coupling strengths imply an expected number of events per unit exposure which typically is $R=50\ \mathrm{kg}^{-1}\mathrm{Myr}^{-1}$ for $\delta_m=50\ \mathrm{keV}/c^2$ and $R=30\ \mathrm{kg}^{-1}\mathrm{Myr}^{-1}$ for $\delta_m=100\ \mathrm{keV}/c^2$. 

A similar analysis has been performed for conventional direct-detection experiments in Ref.~\cite{Kavanagh:2015jma}, which explores how both the nuclear recoil energy and recoil direction can be used to distinguish between different NREFT operators.  In that study,
directional sensitivity plays an important role in performing this differentiation. For example, Fig. 7 in Ref.~\cite{Kavanagh:2015jma} shows that, using directional detection, it is possible to distinguish the non-standard NREFT interactions from the standard SI/SD interactions at the 2$\sigma$ level with $\mathcal{O}(100-500)$
events.  
In that work, the $p$-value for rejecting the null hypothesis was shown as a function of the expected number of DM events and assumes a background-free experiment. 

In the case of paleo-detectors, we will instead show the $p$-value for rejecting the null hypothesis as a function of the exposure.
With the much larger exposure (up to $\sim 10^3-10^4$ times larger) of paleo-detectors compared to conventional direct-detection experiments, not only the number of signal events but also the number of background events will be much higher.  The discrimination power is therefore determined by the full signal-plus-background spectrum at a given exposure, rather than by the
number of DM events alone.

The log-likelihood analysis presented in this section closely follows the framework introduced in Section~\ref{mass_reconstruction} for the DM mass-reconstruction study. The primary differences are the introduction of an additional parameter, $F$, which denotes the fraction of the DM signal generated by a non-standard operator, and the fact that neither hypothesis assumes a fixed DM mass in the interaction-discrimination analysis.

To quantify the ability to reject the standard SI hypothesis, we employ the profile log-likelihood ratio
\begin{equation}
q(c_j) = -2 \ln \left[
\frac{ \mathcal{L}\left( \mathbf{D}(M_{\chi},c_j,F) \big| \hat{\hat{\vec{\theta}}};  \hat{\hat{m}}_{\chi},\tilde{c}_j,F=0 \right)}
{ \mathcal{L}\left( \mathbf{D}(M_{\chi},c_j,F) \big| \hat{\vec{\theta}}; \hat{m}_{\chi},\tilde{c}_j',\hat{F} \right)}
\right]~,\label{test_statistic}
\end{equation}
where the numerator corresponds to the best fit under the standard-interaction hypothesis ($F=0$), and the denominator to the global best fit for a given $\tilde{c}_j'$. 
In both the null and alternative hypotheses, the coupling constants $\tilde{c}_j$ and $\tilde{c}_j'$ are chosen such that the total DM event rate per unit exposure, $R=\int dx_T\,\frac{dR}{dx_T}$,
matches the rate inferred from the data set $\mathbf{D}(M_\chi,c_j,F)$.
To determine the confidence levels for rejecting the standard-interaction hypothesis, we construct an \emph{Asimov} data set $\mathbf{D}$ assuming that the true signal arises solely from a non-standard NREFT operator ($F=1$) with DM mass $M_\chi$. The Asimov data are defined as 
\begin{equation}
D_i = \mathcal{N}_{i}\left(\bar{\vec{\theta}}; M_\chi,c_j,F=1 \right),
\end{equation}
where $\bar{\vec{\theta}}$ represents a fiducial choice of nuisance parameters and $\mathcal{N}_i\left(\bar{\vec{\theta}}; M_\chi,c_j,F=1 \right)$ denotes the predicted number of tracks in the $i^{\text{th}}$ bin, including both background and DM contributions.
By Wilks’ theorem~\cite{Wilks:1938dza}, $q$ asymptotically follows a $\chi^2_1$ distribution, from which $p$-values are given by 
\begin{equation}
    p\text{-value}=\int_{q}^{+\infty}P(\chi^2_1)d\chi^2_1~,
\end{equation}
where $P(\chi^2_1)$ is the probability density function for the one-dimensional $\chi^2_1$ distribution.

In our analysis, we assume a fixed sample age of $\bar{t}_{\text{age}} = 1\,\mathrm{Gyr}$, corresponding to the time over which the sample has been recording damage tracks. The sample mass is then related to the exposure via
$\bar{M} = \text{exposure}/\bar{t}_{\text{age}}$.
In practice, for a fixed target mineral age, increasing the experimental exposure corresponds to increasing the mass of the target mineral. Consequently, the test statistic defined in Eq.~\eqref{test_statistic} depends on the exposure, or, equivalently, on the target mineral mass.  

\begin{figure}
    \captionsetup{justification=raggedright,singlelinecheck=false}
    \centering
    \begin{subfigure}{0.46\textwidth}
    \includegraphics[width=\linewidth]{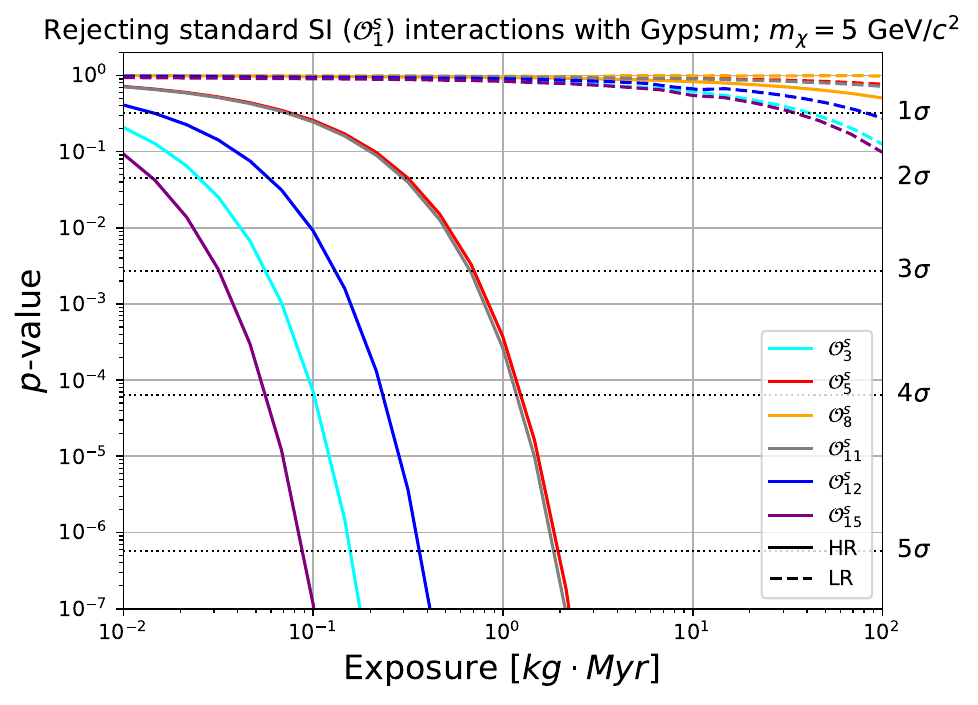}
    \end{subfigure}
    \begin{subfigure}{0.465\textwidth}
    \includegraphics[width=\linewidth]{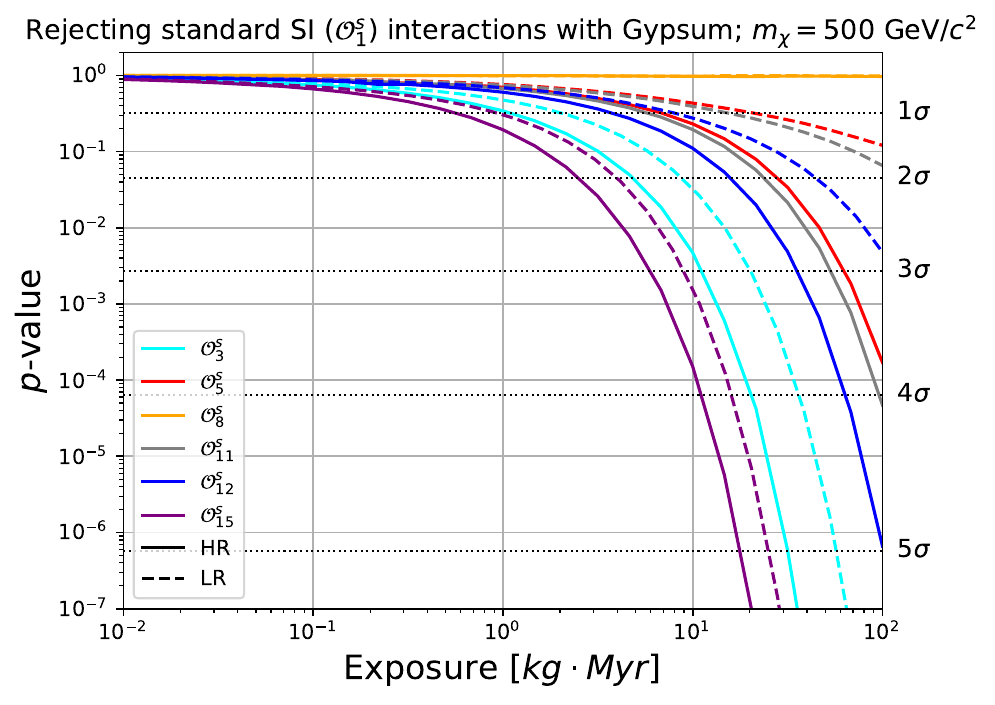}
    \end{subfigure}

    \caption{Projected significance for rejecting the elastic standard SI ($\mathcal{O}_1^s$) interaction hypothesis as a function of exposure. The $p$-value and corresponding significance (in units of $\sigma$) are shown assuming a DM signal generated by non-standard NREFT operators ($\mathcal{O}^{s}_{3}$, $\mathcal{O}^{s}_{5}$, $\mathcal{O}^{s}_{8}$, $\mathcal{O}^{s}_{11}$, $\mathcal{O}^{s}_{12}$, and $\mathcal{O}^{s}_{15}$) in gypsum. The left (right) panel corresponds to $m_{\chi}=5\ \mathrm{GeV}/c^2$ ($500\ \mathrm{GeV}/c^2$). Solid and dashed curves denote the high-resolution (HR, $\sigma_x = 1\ \mathrm{nm}$) and low-resolution (LR, $\sigma_x = 15\ \mathrm{nm}$) scenarios, respectively. The track-length spectra used in the calculations are normalized such that $R = \int dx_T\, dR/dx_T=6\times10^3\ \mathrm{kg}^{-1}\mathrm{Myr}^{-1}$ (left) and $10^2\ \mathrm{kg}^{-1}\mathrm{Myr}^{-1}$ (right) events per unit exposure, compatible with projected sensitivities for gypsum paleo-detectors~\cite{Theodosopoulos:2026ehn}. The standard SI hypothesis can be rejected with the highest significance for scattering via $\mathcal{O}_{15}^s$ and $\mathcal{O}_3^s$, whereas it cannot be rejected in the LR scenario for light DM ($m_{\chi}=5\,\mathrm{GeV}/c^2$) for any interaction operator. For signals generated by $\mathcal{O}_8^s$, the SI hypothesis cannot be rejected at either DM mass. The curves for $\mathcal{O}_5^s$ and $\mathcal{O}_{11}^s$ nearly overlap and may be difficult to distinguish.}

    \label{fig:gypsum}
\end{figure}

\begin{figure}
    \captionsetup{justification=raggedright,singlelinecheck=false}
    \centering
    \begin{subfigure}{0.46\textwidth}
    \includegraphics[width=\linewidth]{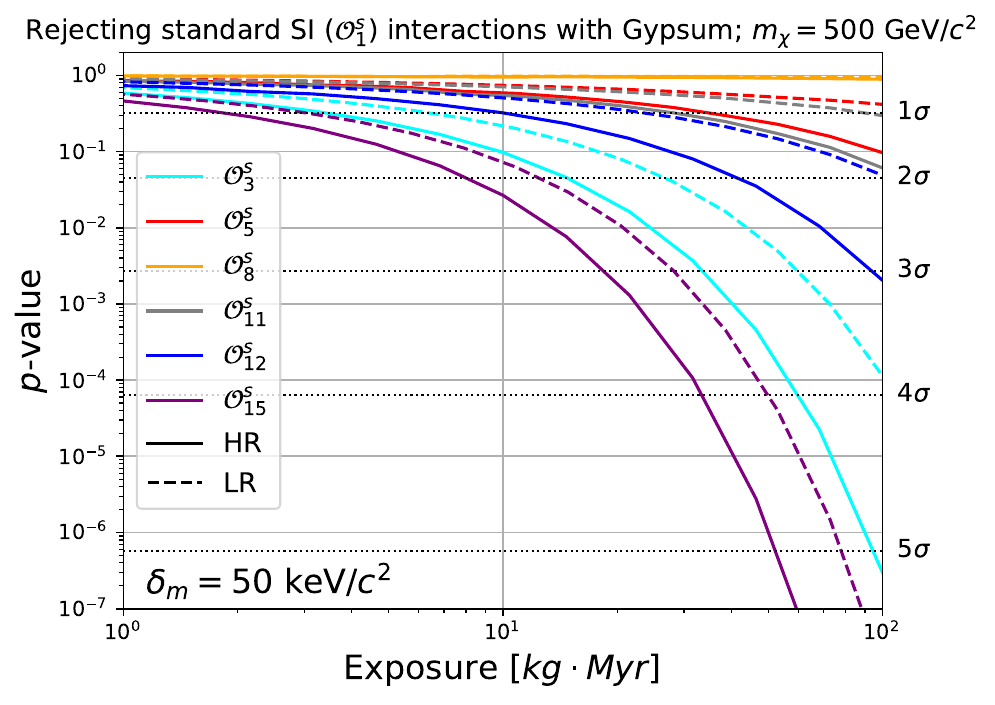}
    \end{subfigure}
    \begin{subfigure}{0.465\textwidth}
    \includegraphics[width=\linewidth]{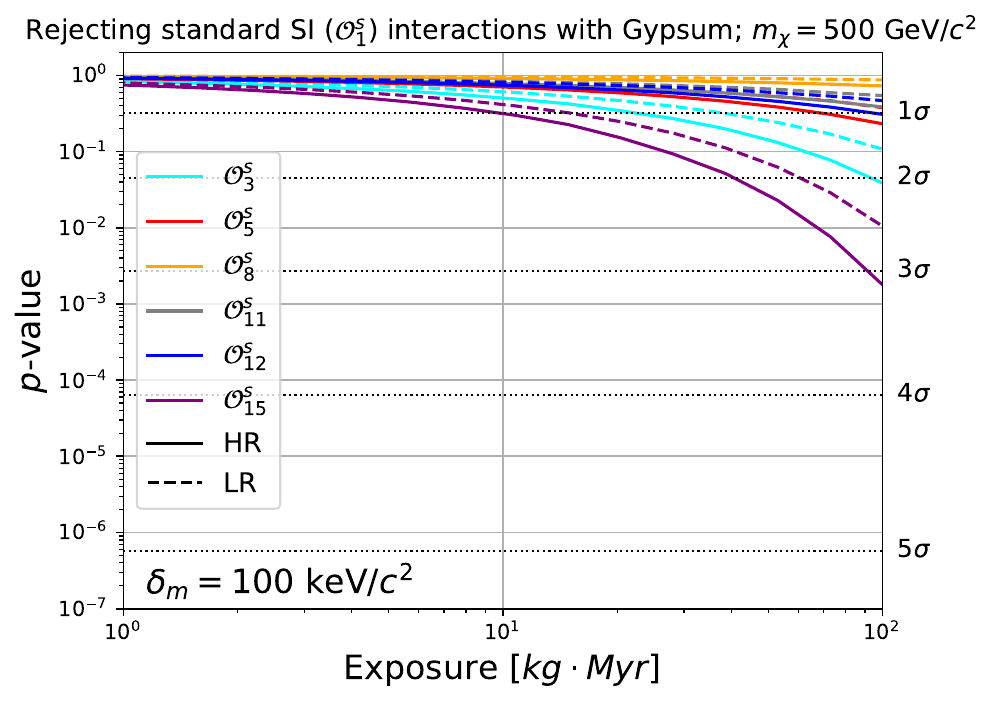}
    \end{subfigure}

    \caption{Projected significance for rejecting the inelastic standard SI ($\mathcal{O}_1^s$) interaction hypothesis as a function of exposure. The $p$-value and corresponding significance (in units of $\sigma$) are shown assuming a DM signal generated by non-standard NREFT operators ($\mathcal{O}^{s}_{3}$, $\mathcal{O}^{s}_{5}$, $\mathcal{O}^{s}_{8}$, $\mathcal{O}^{s}_{11}$, $\mathcal{O}^{s}_{12}$, and $\mathcal{O}^{s}_{15}$) in gypsum. The left (right) panel corresponds to $\delta_{m}=50\ \mathrm{keV}/c^2$ ($100\ \mathrm{keV}/c^2$). Solid and dashed curves denote high-resolution (HR, $\sigma_x = 1\ \mathrm{nm}$) and low-resolution (LR, $\sigma_x = 15\ \mathrm{nm}$) scenarios, respectively. The DM mass is set to $m_{\chi} = 500\ \mathrm{GeV}/c^2$. The track-length spectra used in the calculations are normalized such that $R = \int dx_T\, dR/dx_T=50\ \mathrm{kg}^{-1}\mathrm{Myr}^{-1}$ (left) and $30\ \mathrm{kg}^{-1}\mathrm{Myr}^{-1}$ (right) events per unit exposure, compatible with projected sensitivities for gypsum paleo-detectors~\cite{Theodosopoulos:2026ehn}. The standard SI operator $\mathcal{O}_1^s$ can be rejected with lower confidence levels for inelastic scattering, especially as the mass-splitting $\delta_m$ approaches the maximum values paleo-detectors can probe: $\delta_{m}=100\ \mathrm{keV}$. The curves for $\mathcal{O}_5^s$ and $\mathcal{O}_{11}^s$ nearly overlap and may be difficult to distinguish.}

    \label{fig:gypsum_inelastic}
\end{figure}

Gypsum paleo-detectors are sensitive to the standard SI interaction $\mathcal{O}^{s}_{1}$, as well as to the non-standard interactions $\mathcal{O}^{s}_{3}$, $\mathcal{O}^{s}_{5}$, $\mathcal{O}^{s}_{8}$, $\mathcal{O}^{s}_{11}$, $\mathcal{O}^{s}_{12}$, and $\mathcal{O}^{s}_{15}$, as demonstrated in TFKS. 
Figs.~\ref{fig:gypsum} and \ref{fig:gypsum_inelastic} show the confidence levels for rejecting the standard SI interaction hypothesis as a function of exposure, assuming hypothetical signals generated by the non-standard operators in the elastic and inelastic scattering scenarios, respectively. 
Solid (dashed) curves correspond to the high-resolution $\sigma_x=1\ \mathrm{nm}$ (low-resolution $\sigma_x=15\ \mathrm{nm}$) scenarios. 
In Fig.~\ref{fig:gypsum}, the left (right) panel assumes a DM mass of $5\ \mathrm{GeV}/c^2$ ($500\ \mathrm{GeV}/c^2$), with event-rate normalizations of $R=6\times10^3\ \mathrm{kg}^{-1}\mathrm{Myr}^{-1}$ ($R=10^2\ \mathrm{kg}^{-1}\mathrm{Myr}^{-1}$) DM events per unit exposure. 
In Fig.~\ref{fig:gypsum_inelastic}, the left (right) panel assumes inelastic scattering with a mass-splitting of $50\ \mathrm{keV}/c^2$ ($100\ \mathrm{keV}/c^2$), event-rate normalizations of $R=50\ \mathrm{kg}^{-1}\mathrm{Myr}^{-1}$ ($R=30\ \mathrm{kg}^{-1}\mathrm{Myr}^{-1}$) DM events per unit exposure, and a DM mass of $500\ \mathrm{GeV}/c^2$.
In both scattering scenarios, the highest confidence levels are obtained for signals mediated by $\mathcal{O}^{s}_{15}$, followed by $\mathcal{O}^{s}_{3}$, $\mathcal{O}^{s}_{12}$, $\mathcal{O}^{s}_{11}$, and $\mathcal{O}^{s}_{5}$. In contrast, for signals generated by $\mathcal{O}^{s}_{8}$, the standard SI operator cannot be rejected. 
This behavior is consistent with the qualitative features observed in Figs.~\ref{fig:Spectrum_binned_SI} and \ref{fig:Spectrum_binned_SI_inelastic}: the track-length spectra for $\mathcal{O}^{s}_{3}$ and $\mathcal{O}^{s}_{15}$ differ significantly from that of $\mathcal{O}^{s}_{1}$, whereas the spectrum generated by $\mathcal{O}^{s}_{8}$ is nearly indistinguishable from the standard SI case. 

For elastic scattering and a DM mass of $5\ \mathrm{GeV}/c^2$, the standard SI hypothesis cannot be rejected above the $1\sigma$ level for most operators in the low-resolution (LR) scenario, as shown in the left panel of Fig.~\ref{fig:gypsum}. This is because lighter DM particles ($m_\chi\lesssim10\ \mathrm{GeV}/c^2$) produce relatively short damage tracks ($x_T\lesssim100\ \mathrm{nm}$), requiring higher read-out resolution to accurately reconstruct the track-length spectrum. Additionally, discriminating most operators in the high-resolution $\sigma_x=1\ \mathrm{nm}$ (HR) scenario requires exposures of at least $10^{-1}\ \mathrm{kg}\cdot\mathrm{Myr}$. As discussed in Section~\ref{DMpaleo}, such high read-out resolution is likely achievable only by scanning relatively small mineral samples with total masses of order $M\sim10\ \mathrm{mg}$. Consequently, simultaneously achieving the HR scenario and exposures significantly larger than $10^{-2}\ \mathrm{kg}\cdot\mathrm{Myr}$ is currently expected to be challenging. For this reason, we focus on the LR scenario and on heavier DM particles ($m_\chi>10\ \mathrm{GeV}/c^2$) throughout the remainder of our analysis.
The same conclusion also applies in the inelastic scattering scenario. 

For inelastic scattering, the confidence levels for rejecting the standard SI interaction hypothesis decrease with increasing mass-splitting $\delta_m$, as shown in Fig.~\ref{fig:gypsum_inelastic}. This trend reflects the reduction in paleo-detector sensitivity with increasing $\delta_m$, discussed in TFKS, with most paleo-detectors losing sensitivity near $\delta_m\sim100\,\mathrm{keV}$.
Consequently, for $\delta_m\sim100\,\mathrm{keV}$ (right panel of Fig.~\ref{fig:gypsum_inelastic}), signals generated by most operators require exposures exceeding $10^{2}\,\mathrm{kg}\cdot\mathrm{Myr}$ to reject the standard SI hypothesis. We therefore adopt $\delta_m\sim50\,\mathrm{keV}$ as a representative benchmark for the remainder of the inelastic-scattering analysis. 

\begin{figure}
    \captionsetup{justification=raggedright,singlelinecheck=false}
    \centering
    \begin{subfigure}{0.36\textwidth}
    \includegraphics[width=\linewidth]{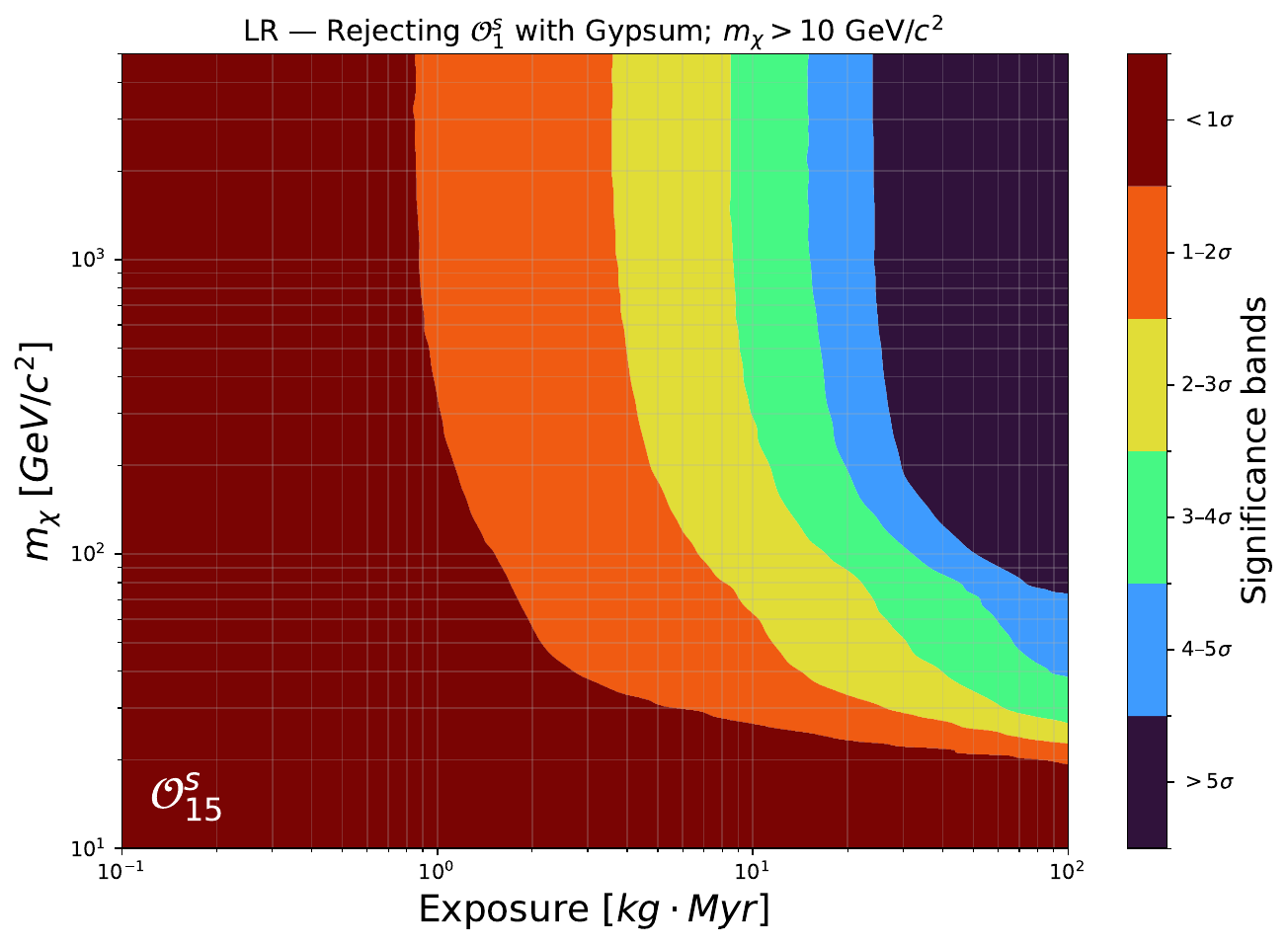}
    \end{subfigure}
    \begin{subfigure}{0.36\textwidth}
    \includegraphics[width=\linewidth]{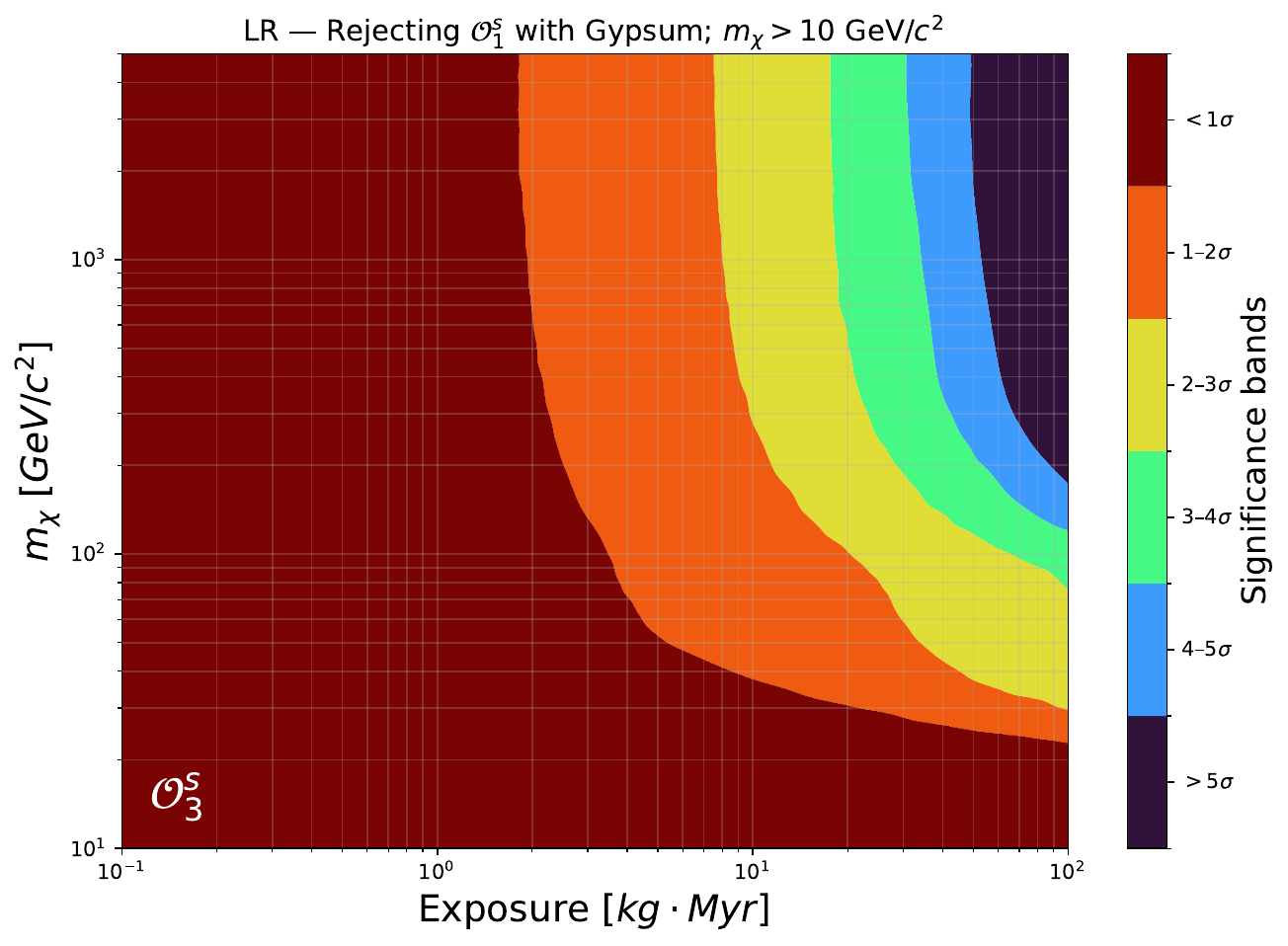}
    \end{subfigure}
    
    \vspace{0.1ex}

    \begin{subfigure}{0.36\textwidth}
    \includegraphics[width=\linewidth]{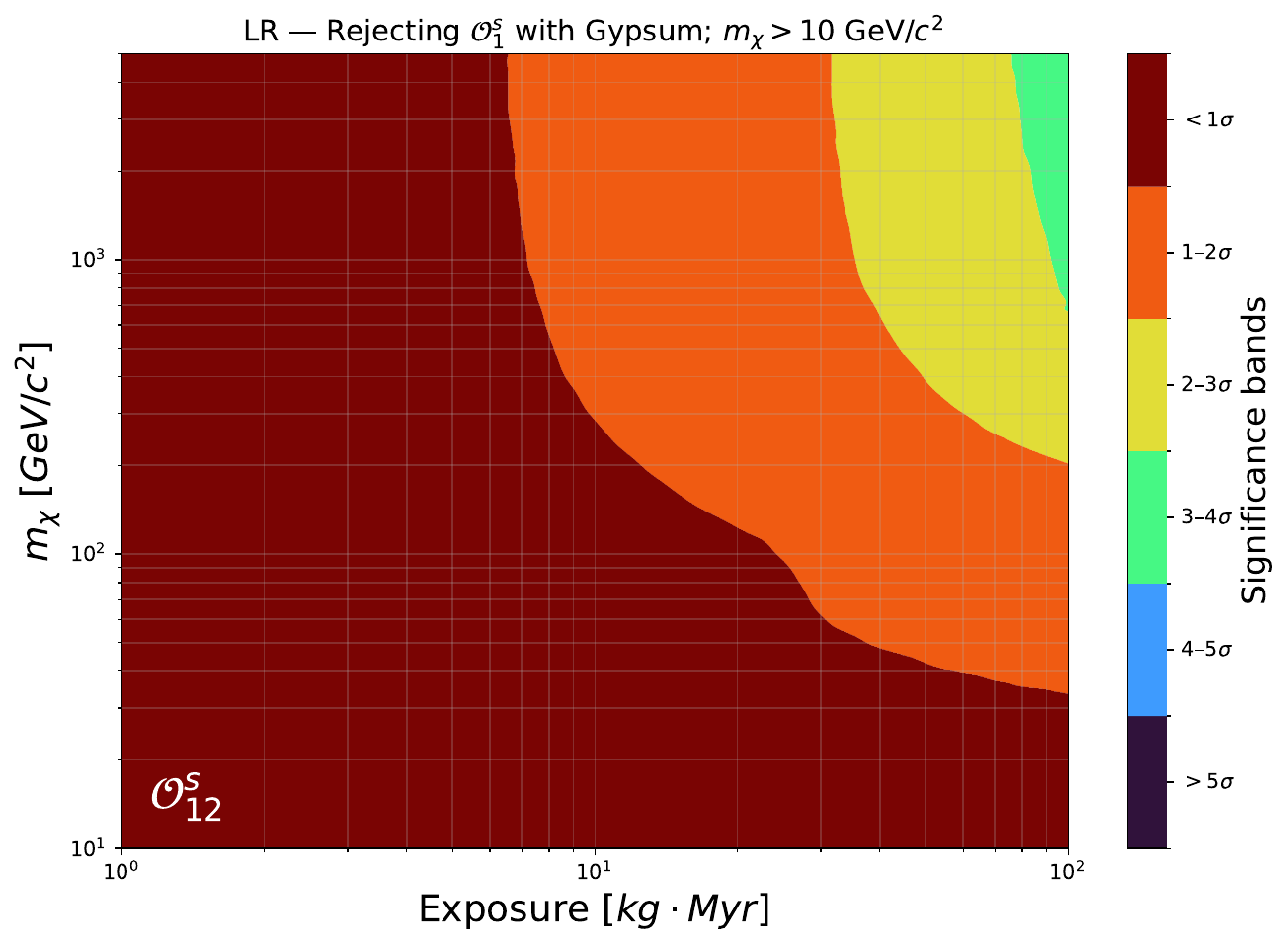}
    \end{subfigure}
    \begin{subfigure}{0.36\textwidth}
    \includegraphics[width=\linewidth]{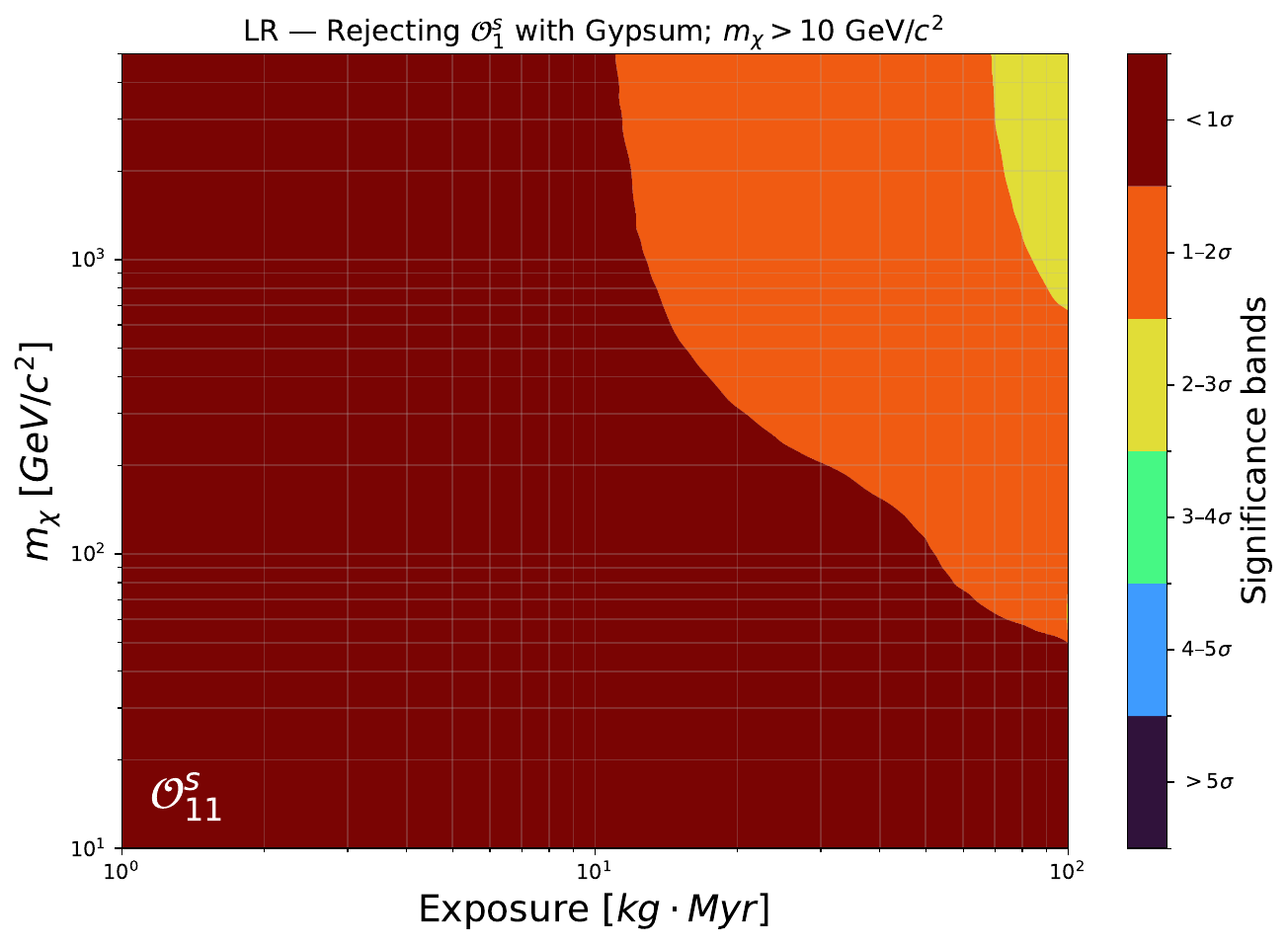}
    \end{subfigure}
    \caption{Projected confidence levels for rejecting the standard SI-only ($\mathcal{O}_1^s$) elastic interaction hypothesis, assuming a DM signal generated entirely by $\mathcal{O}_{15}^{s}$ and $\mathcal{O}_{3}^{s}$ (top panels), and by $\mathcal{O}_{12}^{s}$ and $\mathcal{O}_{11}^{s}$ (bottom panels), for gypsum as the target mineral. Contours corresponding to $1\sigma$–$5\sigma$ significance are shown as a function of DM mass and exposure for the low-resolution (LR, $\sigma_x=15\ \mathrm{nm}$) scenario. The track-length spectra used in the calculations are normalized such that $R = \int dx_T\, dR/dx_T=10^2\ \mathrm{kg}^{-1}\mathrm{Myr}^{-1}$ events per unit exposure, compatible with projected sensitivities for gypsum paleo-detectors~\cite{Theodosopoulos:2026ehn}. The standard SI operator ($\mathcal{O}_1^s$) cannot be rejected for $m_{\chi}\lesssim20\ \mathrm{GeV}/c^2$ for signals generated by $\mathcal{O}_{15}^s$ or $\mathcal{O}_{3}^s$, and for $m_{\chi}\lesssim30\ \mathrm{GeV}/c^2$ for signals generated by $\mathcal{O}_{12}^s$ or $\mathcal{O}_{11}^s$.}

    \label{fig:gypsum_contour}
\end{figure}

\begin{figure}
    \captionsetup{justification=raggedright,singlelinecheck=false}
    \centering
    \begin{subfigure}{0.36\textwidth}
    \includegraphics[width=\linewidth]{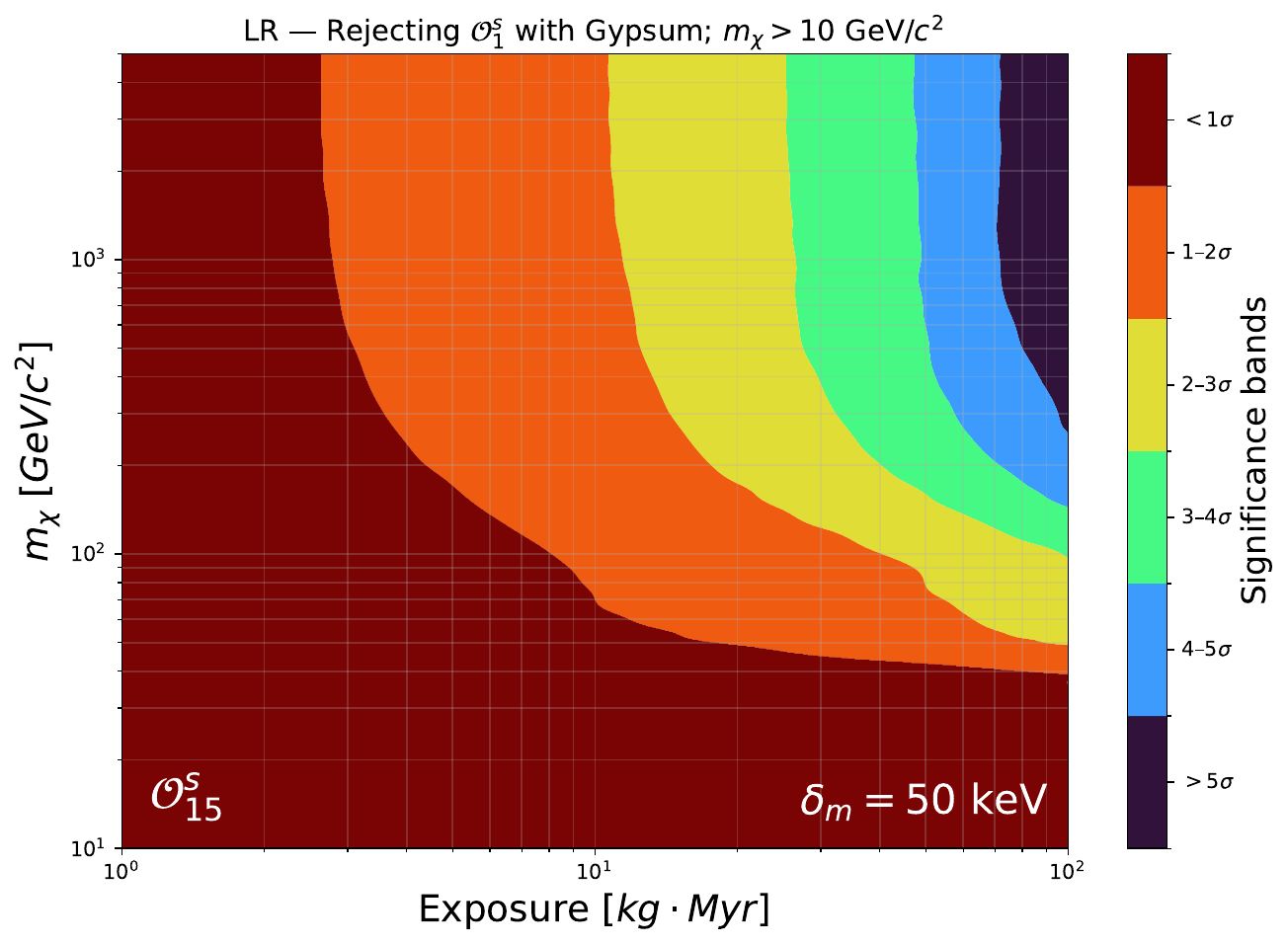}
    \end{subfigure}
    \begin{subfigure}{0.36\textwidth}
    \includegraphics[width=\linewidth]{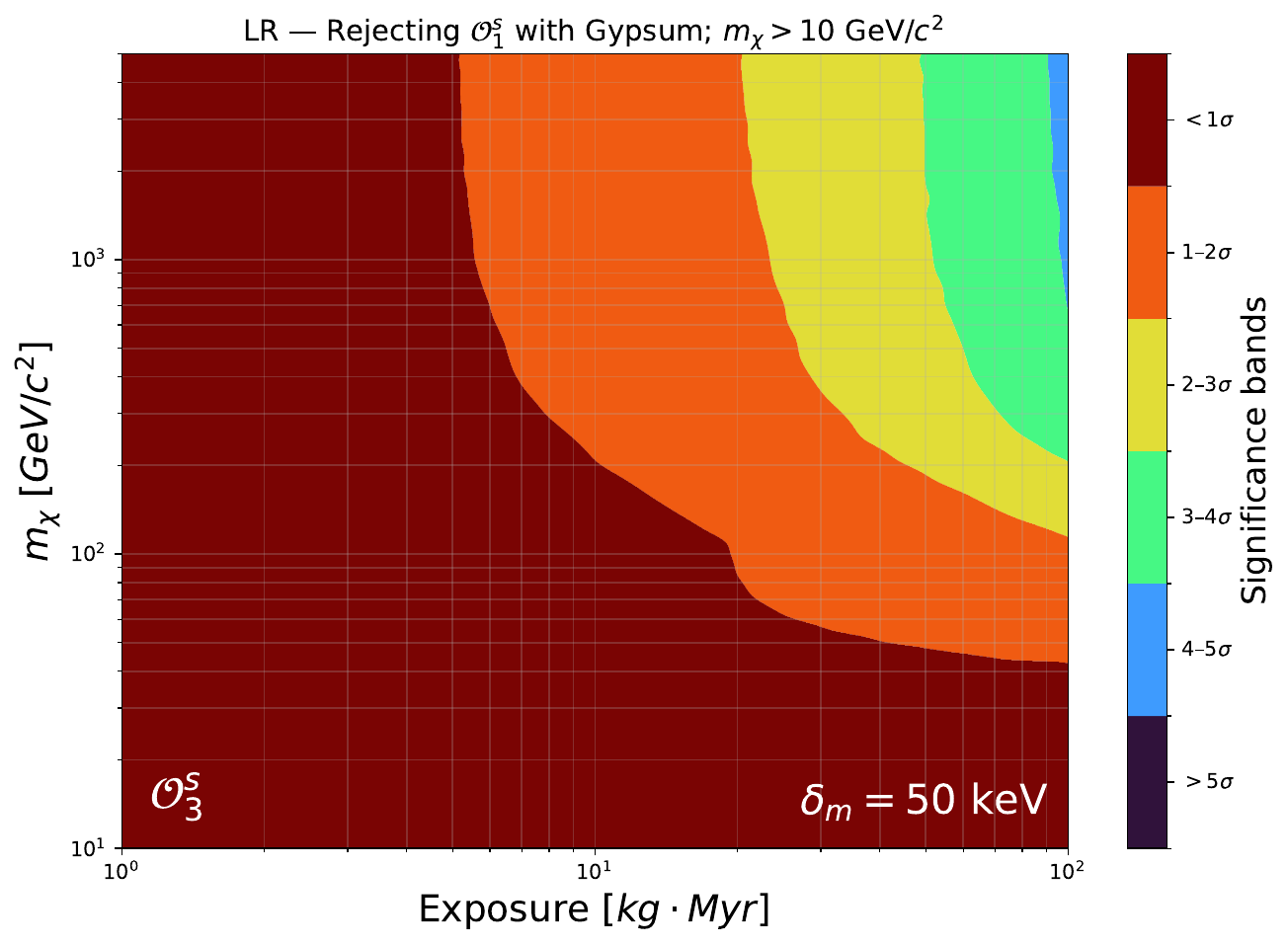}
    \end{subfigure}
    \caption{Projected confidence levels for rejecting the standard SI-only ($\mathcal{O}_1^s$) inelastic interaction hypothesis, assuming a DM signal generated entirely by $\mathcal{O}_{15}^{s}$ (left) and $\mathcal{O}_{3}^{s}$ (right), for gypsum as the target mineral and a mass splitting $\delta_m=50\,\mathrm{keV}/c^2$. Contours corresponding to $1\sigma$--$5\sigma$ significance are shown as a function of DM mass and exposure for the low-resolution (LR, $\sigma_x=15\,\mathrm{nm}$) scenario. The track-length spectra used in the calculations are normalized such that $R = \int dx_T\, dR/dx_T=50\ \mathrm{kg}^{-1}\mathrm{Myr}^{-1}$ events per unit exposure, compatible with projected sensitivities for gypsum paleo-detectors~\cite{Theodosopoulos:2026ehn}. For inelastic scattering, the standard SI operator cannot be rejected for $m_{\chi}\lesssim40\ \mathrm{GeV}/c^2$.}

    \label{fig:gypsum_contour_inelastic}
\end{figure}

Figs.~\ref{fig:gypsum} and \ref{fig:gypsum_inelastic} show the confidence levels for rejecting the standard SI interaction hypothesis only for selected DM masses.
To illustrate how these confidence levels vary continuously with $m_\chi$,
we present contour plots of the projected confidence levels for rejecting the standard SI operator $\mathcal{O}_{1}^{s}$, assuming hypothetical DM signals generated by non-standard operators with $m_\chi>10\,\mathrm{GeV}/c^2$.
The corresponding results for the low-resolution $\sigma_x=15\ \mathrm{nm}$ (LR) scenario are shown in Figs.~\ref{fig:gypsum_contour} and \ref{fig:gypsum_contour_inelastic}, assuming gypsum as the target mineral and exposures between $10^{-2}$ and $10^{2}\,\mathrm{kg}\cdot\mathrm{Myr}$, for the elastic and inelastic scattering, respectively.

Fig.~\ref{fig:gypsum_contour} displays the $1\sigma$--$5\sigma$ significance contours for rejecting the standard SI elastic interaction hypothesis, assuming DM signals generated by the non-standard operators $\mathcal{O}^{s}_{3}$, $\mathcal{O}^{s}_{11}$, $\mathcal{O}^{s}_{12}$, and $\mathcal{O}^{s}_{15}$, with an event-rate normalization of $R=100\ \mathrm{kg}^{-1}\mathrm{Myr}^{-1}$ DM events per unit exposure.
Since $\mathcal{O}_{15}^s$ produces a track-length spectrum that differs significantly from that of the standard SI interaction, the standard SI hypothesis can be rejected above the $5\sigma$ level for $m_\chi>100\,\mathrm{GeV}/c^2$ with exposures of order $40\,\mathrm{kg}\cdot\mathrm{Myr}$.
As also evident from the right panel of Fig.~\ref{fig:gypsum}, the confidence levels for rejecting the standard SI hypothesis decrease successively for signals generated by $\mathcal{O}^{s}_{3}$, $\mathcal{O}^{s}_{12}$, and $\mathcal{O}^{s}_{11}$. This trend is likewise visible in Fig.~\ref{fig:gypsum_contour}, where the contours shift progressively toward larger exposures for these operators.
We omit the results for $\mathcal{O}^{s}_{5}$ and $\mathcal{O}^{s}_{8}$ in Fig.~\ref{fig:gypsum_contour}. The spectra generated by $\mathcal{O}^{s}_{5}$ are similar to those of $\mathcal{O}^{s}_{11}$,
and are therefore expected to exhibit comparable contour features. By contrast, $\mathcal{O}^{s}_{8}$ produces spectra nearly indistinguishable from those of $\mathcal{O}^{s}_{1}$, such that the standard SI hypothesis cannot be rejected in this case.
For low read-out resolution, paleo-detectors have limited sensitivity to DM masses near $m_{\chi}\sim10\,\mathrm{GeV}/c^2$.
Consequently, the discrimination power of gypsum paleo-detectors between standard and non-standard operators is significantly reduced for $m_{\chi}\lesssim30\,\mathrm{GeV}/c^2$. 

The general trend in Fig.~\ref{fig:gypsum_contour} is that the confidence levels for rejecting the standard SI interaction hypothesis decrease with decreasing DM mass. For $m_\chi\gtrsim500\,\mathrm{GeV}/c^2$,
the confidence levels remain nearly unchanged because the track-length spectra depend only weakly on $m_\chi$, as discussed in Sec.~\ref{DMpaleo}.
As the DM mass decreases below $m_\chi\sim500\,\mathrm{GeV}/c^2$, the confidence levels are reduced due to the limited read-out resolution of the LR scenario.
However, this trend temporarily flattens near $m_\chi\sim100\,\mathrm{GeV}/c^2$ before the discrimination power of gypsum paleo-detectors is largely lost for $m_{\chi}\lesssim30\,\mathrm{GeV}/c^2$.
This flattening occurs because gypsum paleo-detectors achieve their best projected sensitivity in this mass range, according to TFKS.

Fig.~\ref{fig:gypsum_contour_inelastic} displays the $1\sigma$--$5\sigma$ significance contours for rejecting the standard SI inelastic interaction hypothesis, assuming DM signals generated by the non-standard operators
$\mathcal{O}^{s}_{3}$ and $\mathcal{O}^{s}_{15}$, for a mass splitting
$\delta_m=50\,\mathrm{keV}$, with an event-rate normalization of
$R=50\ \mathrm{kg}^{-1}\mathrm{Myr}^{-1}$ DM events per unit exposure.
The reduced sensitivity of paleo-detectors to inelastic WIMP--nucleon interactions, relative to the elastic-scattering case, is reflected in the shift of the contours toward larger exposures.
Otherwise, the qualitative features of the contour plots remain similar to those observed for elastic scattering.
Results for the remaining operators to which gypsum paleo-detectors are projected to be sensitive are omitted, as they exhibit the same general behavior.

In this section, we have presented projections for the ability of gypsum paleo-detectors (taken here as a representative target mineral) to discriminate between standard and non-standard WIMP--nucleon interaction hypotheses in both elastic and inelastic scattering scenarios. 
In particular, for elastic scattering, signals generated by the operators
$\mathcal{O}_{15}^{s}$ and $\mathcal{O}_{3}^{s}$ can lead to rejection of the standard SI interaction hypothesis at confidence levels exceeding
$5\sigma$ for achievable exposures and DM masses
$m_\chi\gtrsim100\,\mathrm{GeV}/c^2$.
By contrast, signals generated by $\mathcal{O}_{8}^{s}$
produce track-length spectra nearly indistinguishable from those of the standard SI interaction, preventing meaningful discrimination between the corresponding hypotheses.
For low read-out resolution, the discrimination power of gypsum paleo-detectors decreases significantly for
$m_\chi\lesssim30\,\mathrm{GeV}/c^2$,
reflecting the limited sensitivity of the LR scenario to short damage tracks produced by lighter DM particles.
In the inelastic scattering scenario, the reduced sensitivity of paleo-detectors to interactions with nonzero mass splitting shifts the projected confidence-level contours toward larger exposures, although the qualitative behavior remains similar to that observed for elastic scattering.
In the Appendix, we present corresponding projections for the discrimination of standard spin-dependent (SD) interactions in halite paleo-detectors in the case of detected hypothetical signals from elastic and inelastic scattering mediated by selected spin-dependent NREFT operators. 
Overall, in the low-resolution read-out scenario and for exposures of $1$--$100\ \mathrm{kg\cdot Myr}$, paleo-detectors are projected to distinguish the standard SI or SD interaction hypothesis from hypothetical detected signals generated by a broad range of non-standard NREFT operators for WIMP masses above a few tens of $\mathrm{GeV}/c^2$.

\section{Conclusion} \label{conclusion}

In this work, we investigated the prospects for reconstructing the dark matter (DM) mass and discriminating between different WIMP--nucleon interaction hypotheses with paleo-detectors within the framework of non-relativistic effective field theory (NREFT). Building upon the sensitivity projections derived in Ref.~\cite{Theodosopoulos:2026ehn}, we studied both elastic and inelastic scattering scenarios and explored the extent to which the spectral information encoded in DM-induced damage tracks can be used to infer the underlying particle-physics properties of DM.

We first examined the reconstruction of the DM mass from the track-length spectra predicted by different NREFT operators. For elastic scattering, we showed that gypsum paleo-detectors operating in the high-resolution (HR) scenario can reconstruct DM masses in the range $1$--$10\,\mathrm{GeV}/c^2$ with relative uncertainties as small as $\Delta m_\chi/M_\chi\sim10^{-1}$, probing a mass regime that remains challenging for conventional direct-detection experiments. In the high-exposure (HE) scenario, and for representative operators such as $\mathcal{O}_1^s$ and $\mathcal{O}_{11}^s$, paleo-detectors are projected to reconstruct DM masses in the range $40$--$1000\,\mathrm{GeV}/c^2$ with relative uncertainties of order unity. We also demonstrated that, at very large DM masses, the reconstructed mass intervals broaden substantially because the recoil spectrum becomes only weakly dependent on the DM mass through the reduced WIMP--nucleus mass. In the main text, we focused on interactions mediated by operators to which gypsum paleo-detectors are projected to be sensitive, while the corresponding analysis for selected spin-dependent operators in halite paleo-detectors is deferred to the Appendix.

For inelastic scattering, we focused on a representative mass splitting $\delta_m\sim50\,\mathrm{keV}/c^2$, motivated by the loss of sensitivity of paleo-detectors at larger mass-splittings. In this scenario, gypsum paleo-detectors were shown to reconstruct DM masses in the range $30$--$400\,\mathrm{GeV}/c^2$ with relative uncertainties down to $\Delta m_\chi/M_\chi\sim10^{-2}$, a region of parameter space where conventional direct-detection experiments rapidly lose sensitivity. These results highlight the complementarity between paleo-detectors and conventional direct-detection searches, particularly for low-mass DM and inelastic interactions. Corresponding results for inelastic scattering mediated by selected spin-dependent operators in halite paleo-detectors are presented in the Appendix.

A second goal of this work was to assess the ability of paleo-detectors to discriminate between standard and non-standard WIMP--nucleon interactions. To this end, we performed a profile-likelihood-ratio analysis comparing the standard spin-independent (SI) and spin-dependent (SD) interaction hypotheses against signals generated by non-standard NREFT operators. The SI analysis is presented in the main text, while the corresponding SD analysis is deferred to the Appendix. We demonstrated that the discrimination power is strongly correlated with the qualitative differences in the corresponding track-length spectra. In particular, operators such as $\mathcal{O}_{15}^s$ and $\mathcal{O}_{3}^s$, which generate spectra that differ significantly from those of the standard SI interaction, can lead to rejection of the standard SI hypothesis at confidence levels exceeding $5\sigma$ for achievable exposures and DM masses $m_\chi\gtrsim100\,\mathrm{GeV}/c^2$. By contrast, operators such as $\mathcal{O}_{8}^s$, with spectra nearly indistinguishable from the standard SI case, cannot be discriminated even with large exposures.

We further showed that the discrimination power of paleo-detectors depends strongly on the detector's read-out resolution and the DM mass. In the low-resolution (LR) scenario, the ability to discriminate between interaction hypotheses is significantly reduced for DM masses $m_\chi\lesssim30\,\mathrm{GeV}/c^2$, owing to the short damage tracks produced by light DM particles. Similarly, for inelastic scattering, increasing the mass splitting shifts the projected confidence-level contours toward larger exposures, reflecting the reduced sensitivity of paleo-detectors to inelastic interactions. Nevertheless, for representative mass splittings of $\delta_m\sim50\,\mathrm{keV}/c^2$, paleo-detectors retain substantial discrimination power for DM masses above tens of GeV.

In our analysis, we assumed a one-to-one mapping between the observed track length $x_T$ and recoil energy $E_R$ for DM masses between $1\,\mathrm{GeV}/c^2$ and $5\,\mathrm{TeV}/c^2$, an approximation shown to be accurate in Ref.~\cite{Fung:2025cub}.
A natural extension of this work is to explore lighter DM masses, where this approximation is no longer valid. In that regime, a recoil with energy
$E_R$ can produce a relatively wide distribution of possible track lengths
$x_T$, such that the probability of track formation must be modeled explicitly following for instance the approach of Ref.~\cite{Fung:2025cub}.

Throughout this work, we restricted our discussion to scenarios in which the DM signal is generated by a single NREFT operator at a time. Realistic WIMP models may involve combinations of operators and potentially sizable interference effects (see \cite{Brenner:2022qku}). Extending the present analysis to scenarios involving multiple operators simultaneously, including interference effects, as well as to more general ultraviolet-complete models of dark matter, constitutes an important direction for future work.

Another important direction is to develop a fully Bayesian parameter reconstruction framework in which the DM mass, coupling strengths, operator content, and relevant astrophysical and background parameters are inferred simultaneously from the measured track-length spectrum.
The mass-reconstruction analysis presented here assumes that the interaction operator responsible for the signal is fixed, while the operator-discrimination analysis treats the mass and interaction hypothesis separately. 
A complete analysis of a future paleo-detector signal would instead 
reconstruct the DM parameters jointly in the full NREFT parameter space. 
Since this parameter space is high dimensional, involving the DM mass, the set of NREFT operators, as well as astrophysical and nuisance parameters, standard likelihood scans or Markov-chain Monte Carlo methods may become computationally expensive. 
Machine-learning accelerated simulation-based inference methods, such as truncated marginal neural ratio estimation~\cite{Miller:2021hys}, provide a promising approach for this task by training neural networks on simulated paleo-detector spectra to estimate likelihood ratios or marginal likelihood-to-evidence ratios. 
As demonstrated for conventional direct-detection experiments in Ref.~\cite{Cerdeno:2024uqt}, such methods could enable a global reconstruction of the particle and astrophysics properties of a detected DM signal in paleo-detectors.

Overall, our results demonstrate that paleo-detectors provide a powerful and complementary probe of DM interactions beyond conventional direct-detection techniques. In addition to the projected sensitivity to low event rates and inelastic scattering scenarios, the spectral information encoded in ancient damage tracks may enable both precise DM mass reconstruction and statistically significant discrimination between competing interaction hypotheses. These capabilities highlight the potential of paleo-detectors as a next-generation approach for probing the particle nature of dark matter.

\acknowledgments

This work was supported by the U.S. National Science Foundation Growing Convergence Research award 2428507.
KF is grateful for
support from the Jeff \& Gail Kodosky Endowed Chair in
Physics at the University of Texas. KF acknowledges
support from the Swedish Research Council (Contract
No. 638-2013-8993).  KF and DT acknowledge support by the U.S. Department of Energy, Office
of Science, Office of High Energy Physics program under Award Number DE-SC-0022021.  The work of CK is supported in part by the U.S. Department of Energy, Office of Science, Office of High Energy Physics under Award Number DESC0024693.
The work of PS is co-funded by the European Union's Horizon Europe research and innovation program under the Marie Sklodowska-Curie COFUND Postdoctoral Programme grant agreement No. 101081355-SMASH and by the Republic of Slovenia and the European Union from the European Regional Development Fund. Views and opinions expressed are however those of the authors only and do not necessarily reflect those of the European Union or European Research Executive Agency. Neither the European Union nor the granting authority can be held responsible for them.

\appendix* \label{appendix}
\section{}

In this Appendix, we present the corresponding DM mass-reconstruction and operator-discrimination analyses for elastic and inelastic scattering mediated by selected spin-dependent NREFT operators in halite paleo-detectors. Halite is particularly relevant for probing spin-dependent interactions because of the presence of nuclei with nonzero nuclear spin, enabling sensitivity to operators to which gypsum paleo-detectors are comparatively insensitive. The analysis closely follows the methodology adopted in the main text for gypsum.

\begin{figure}
    \captionsetup{justification=raggedright,singlelinecheck=false}
    \centering
    \includegraphics[width=0.41\textwidth]{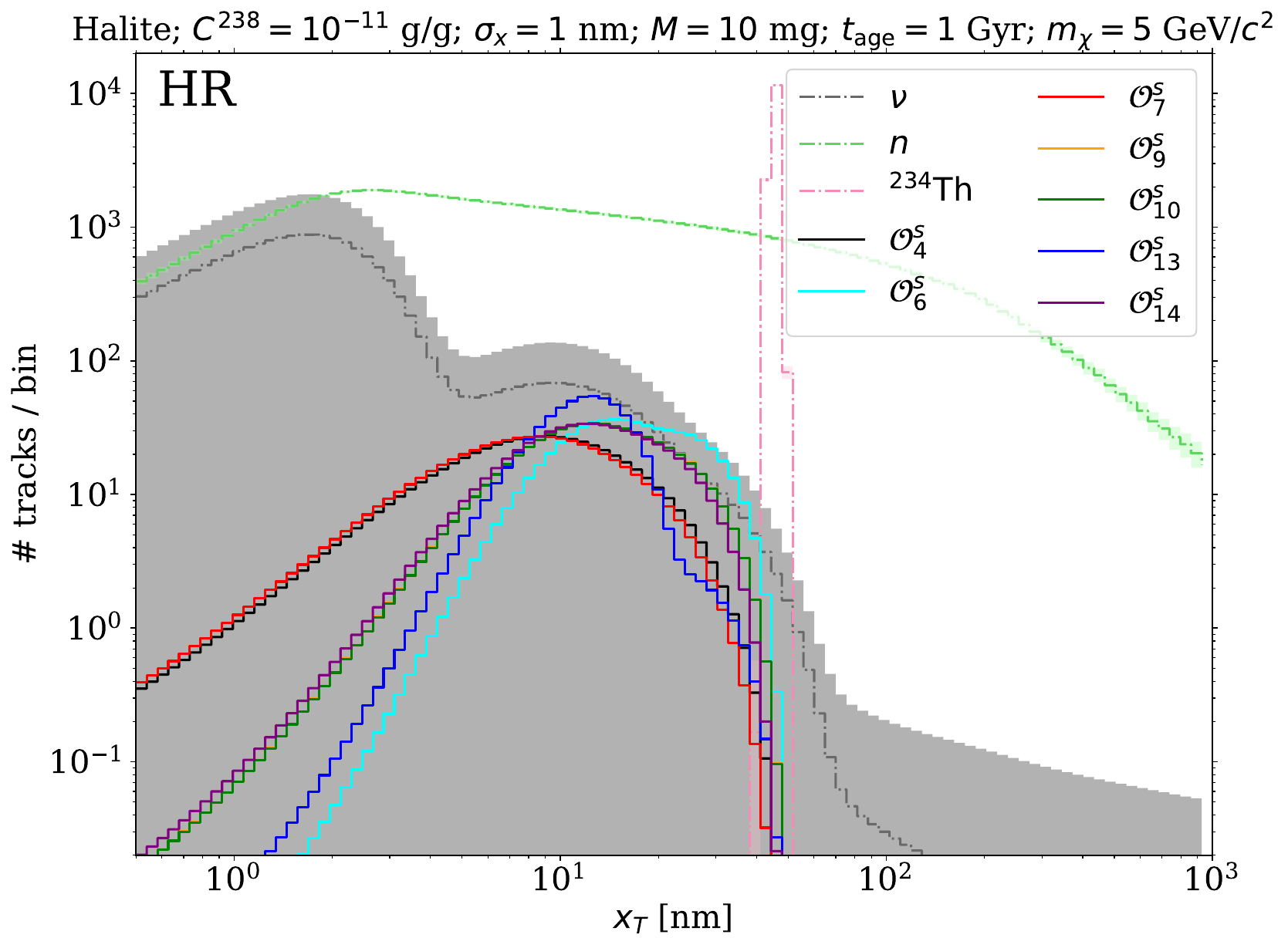}
    \includegraphics[width=0.412\textwidth]{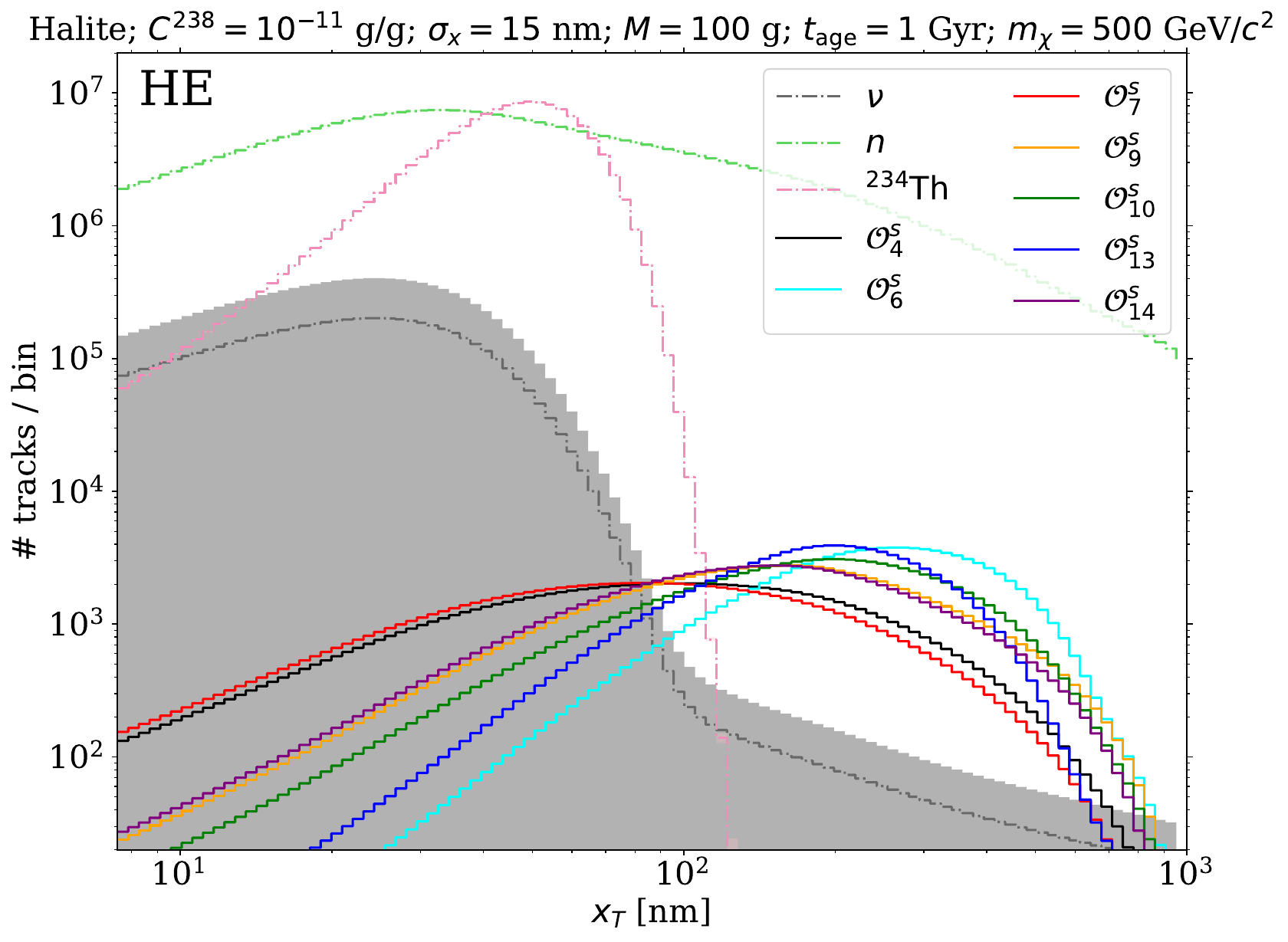}
    \caption{Track length spectra for the NREFT operators $\mathcal{O}^{s}_{4}$, 
    $\mathcal{O}^{s}_{6}$,
    $\mathcal{O}^{s}_{7}$, $\mathcal{O}^{s}_{9}$, $\mathcal{O}^{s}_{10}$, $\mathcal{O}^{s}_{13}$, and $\mathcal{O}^{s}_{14}$, assuming elastic isoscalar interactions ($c^{p}=c^{n}$). Left panel: high-resolution (HR) scenario, with read-out resolution $\sigma_{x}=1\,\mathrm{nm}$, mineral mass $M=10\,\mathrm{mg}$, and DM mass $m_{\chi}=5\,\mathrm{GeV}/c^{2}$. The DM track-length spectra are normalized such that $R = \int dx_T\, dR/dx_T=6\times10^4\ \mathrm{kg}^{-1}\mathrm{Myr}^{-1}$ events per unit exposure, compatible with projected sensitivities for halite paleo-detectors~\cite{Theodosopoulos:2026ehn}. Right panel: high-exposure (HE) scenario, with $\sigma_{x}=15\ \mathrm{nm}$, $M=100\ \mathrm{g}$, $m_{\chi}=500\ \mathrm{GeV}/c^{2}$, and $R = 10^3\ \mathrm{kg}^{-1}\mathrm{Myr}^{-1}$ events per unit exposure, compatible with projected paleo-detector sensitivities~\cite{Theodosopoulos:2026ehn}. For comparison, background spectra induced by neutrinos ($\nu$), radiogenic neutrons ($n$), and ${}^{238}\text{U}\to{}^{234}\text{Th}+\alpha$ recoils (${}^{234}\text{Th}$) are also shown; see Sec.~\ref{background}. Here, we include shaded bands around background components, representing the combined statistical (Poisson) and systematic uncertainties in the background predictions. Note that we assume 100\% uncertainty in the neutrino backgrounds which leads to the large, grey-shaded regions. Results are presented for halite with a ${}^{238}$U concentration of $10^{-11}\,\mathrm{g/g}$. In both read-out scenarios, the spectra for $\mathcal{O}_{4}^s$ and $\mathcal{O}_{7}^s$ are nearly indistinguishable, while $\mathcal{O}_{6}^s$ and $\mathcal{O}_{13}^s$ generate distinct spectral features relative to $\mathcal{O}_{4}^s$.}
    \label{fig:Spectrum_binned_SI_halite}
\end{figure}

\begin{figure}
    \captionsetup{justification=raggedright,singlelinecheck=false}
    \centering
    \includegraphics[width=0.41\textwidth]{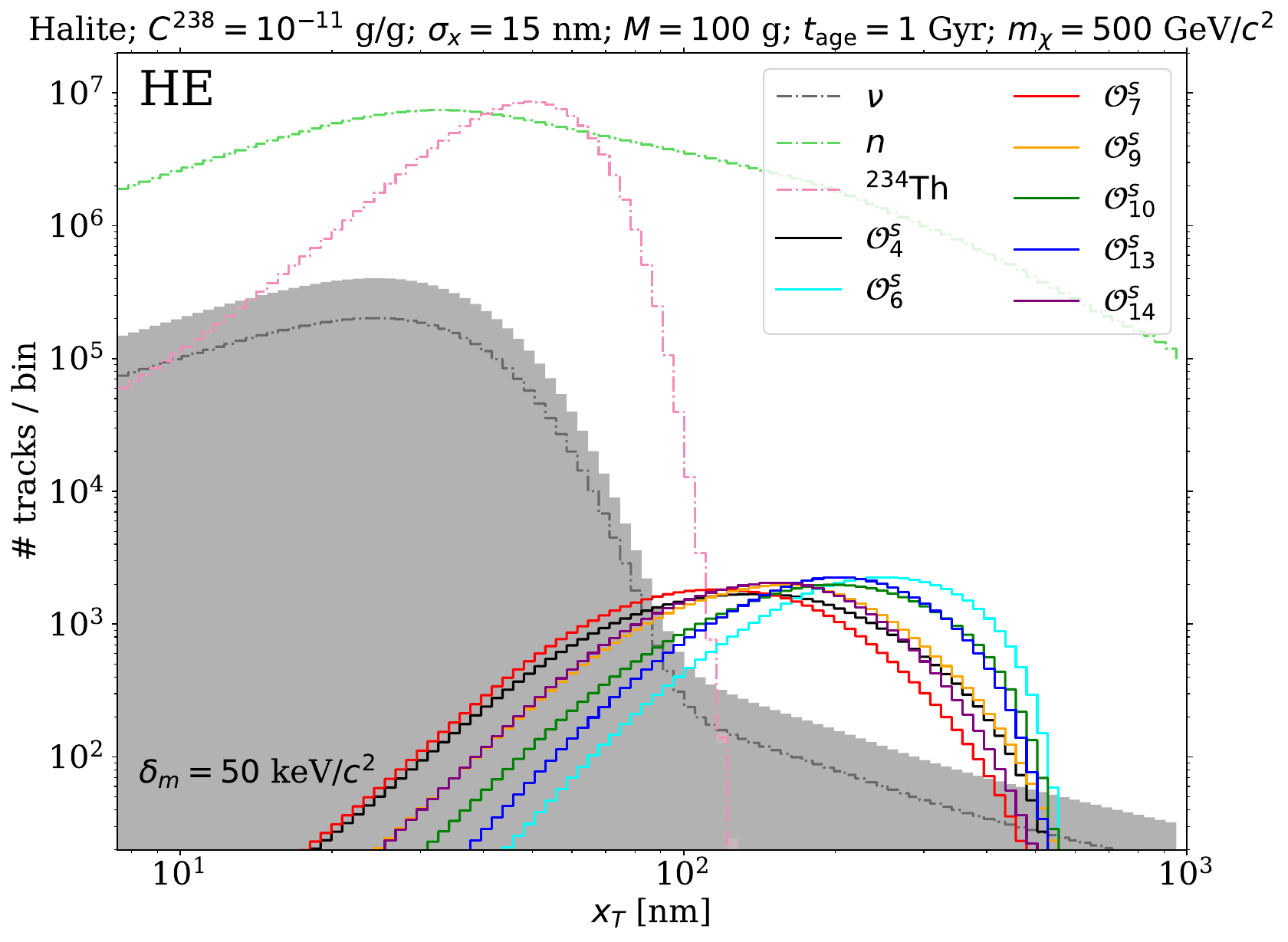}
    \includegraphics[width=0.412\textwidth]{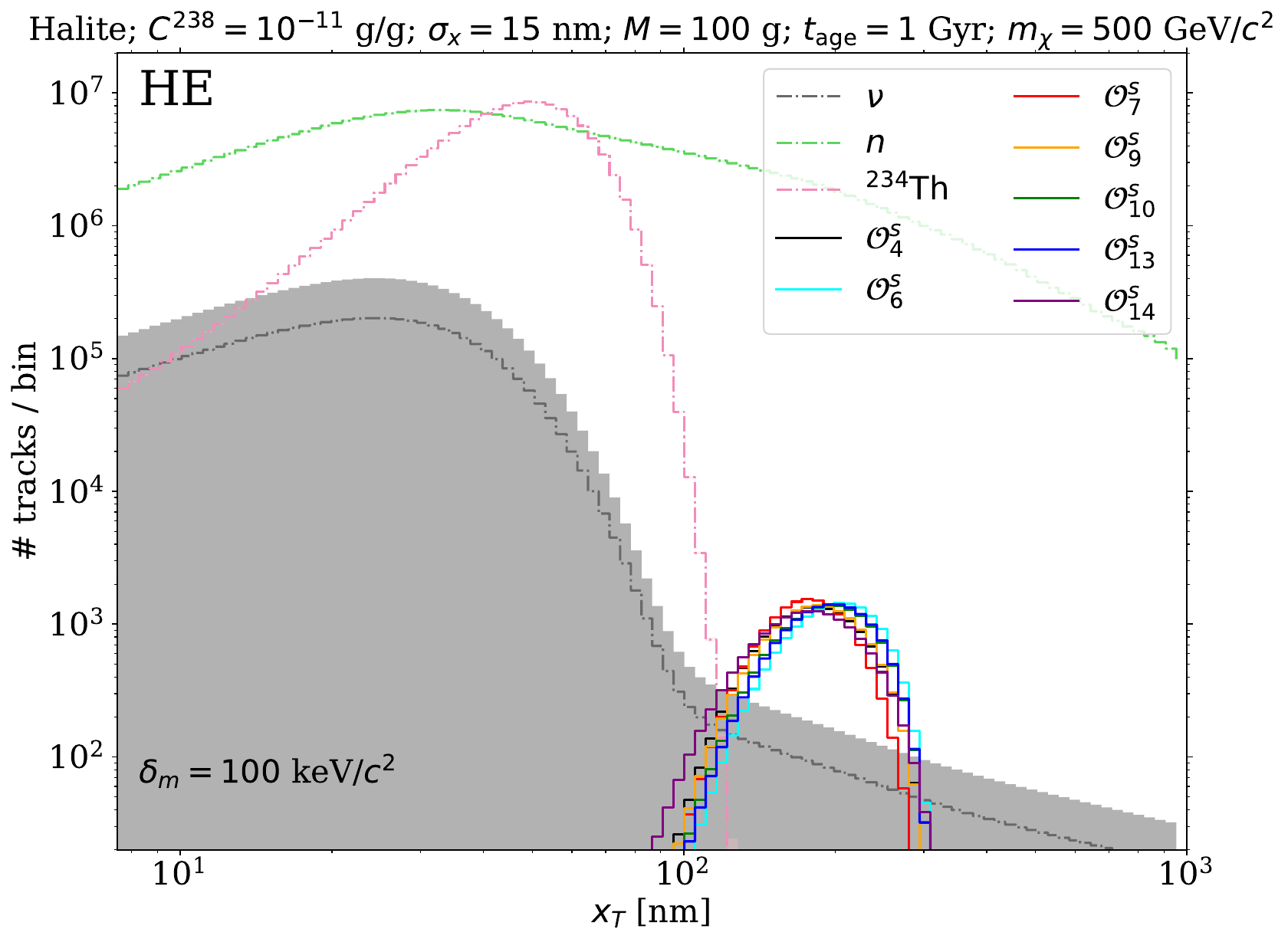}
    \caption{Track length spectra for inelastic scattering via the NREFT operators $\mathcal{O}^{s}_{4}$, 
    $\mathcal{O}^{s}_{6}$,
    $\mathcal{O}^{s}_{7}$, $\mathcal{O}^{s}_{9}$, $\mathcal{O}^{s}_{10}$, $\mathcal{O}^{s}_{13}$, and $\mathcal{O}^{s}_{14}$ on a halite target. The mass-splitting is $\delta_m=50\ \mathrm{keV}/c^2$ (left) and $\delta_m=100\ \mathrm{keV}/c^2$ (right), and the spectra are normalized such that $R = \int dx_T\, dR/dx_T=500\ \mathrm{kg}^{-1}\mathrm{Myr}^{-1}$ (left) and $R =150\ \mathrm{kg}^{-1}\mathrm{Myr}^{-1}$ (right) events per unit exposure, compatible with projected sensitivities for halite paleo-detectors~\cite{Theodosopoulos:2026ehn}. The calculations assume isoscalar interactions ($c^{p}=c^{n}$) in the high-exposure (HE) scenario, with read-out resolution $\sigma_{x}=15\ \mathrm{nm}$, mineral mass $M=100\ \mathrm{g}$, and DM mass $m_{\chi}=500\ \mathrm{GeV}/c^{2}$. For comparison, background spectra induced by neutrinos ($\nu$), radiogenic neutrons ($n$), and ${}^{238}\text{U}\to{}^{234}\text{Th}+\alpha$ recoils (${}^{234}\text{Th}$) are also shown; see Sec.~\ref{background}. Here, we include shaded bands around background components, representing the combined statistical (Poisson) and systematic uncertainties in the background predictions. Note that we assume 100\% uncertainty in the neutrino backgrounds which leads to the large, grey-shaded regions. Results are presented for halite with a ${}^{238}$U concentration of $10^{-11}\,\mathrm{g/g}$. For $\delta_m=50\ \mathrm{keV}$, the spectra for $\mathcal{O}_{4}^s$ and $\mathcal{O}_{7}^s$ are nearly indistinguishable, as well as the spectra for $\mathcal{O}_{9}^s$ and $\mathcal{O}_{14}^s$. The operator $\mathcal{O}_{6}^s$ generates distinct spectral features relative to $\mathcal{O}_{4}^s$. For $\delta_m=100\ \mathrm{keV}$, which is near the upper limit of $\delta_m$ paleo-detectors can probe, the spectra for all the operators look similar.}
    \label{fig:Spectrum_binned_SI_inelastic_halite}
\end{figure}

We first examine the track-length spectra predicted by the standard spin-dependent (SD) operator
$\mathcal{O}_4^s$ and the non-standard operators $\mathcal{O}_6^s$, $\mathcal{O}_7^s$, $\mathcal{O}_9^s$, $\mathcal{O}_{10}^s$, $\mathcal{O}_{13}^s$, and $\mathcal{O}_{14}^s$ for elastic and inelastic scattering in halite. 
The corresponding spectra are shown in Figs.~\ref{fig:Spectrum_binned_SI_halite} and \ref{fig:Spectrum_binned_SI_inelastic_halite} for elastic and inelastic scattering, respectively. 
As in the analysis in the main text, the qualitative differences in the spectral shapes largely determine both the DM mass-reconstruction capabilities and the discrimination power between interaction hypotheses. 
In particular, the predicted track lengths strongly depend on the DM mass, which is crucial for the mass-reconstruction analysis, whereas different operators imply different spectral shapes, enabling the discrimination between different WIMP-nucleon interactions. For instance, the spectra generated by $\mathcal{O}_4^s$ and $\mathcal{O}_7^s$ are nearly indistinguishable, whereas $\mathcal{O}_6^s$ and $\mathcal{O}_{13}^s$ produce track-length spectra with significantly different shapes relative to the standard SD interaction. 
In the inelastic case, the spectra predicted by all operators become increasingly similar as the mass splitting approaches
$\delta_m\sim100\,\mathrm{keV}/c^2$, close to the upper limit of paleo-detector sensitivity.

\begin{figure}
    \captionsetup{justification=raggedright,singlelinecheck=false}
    \centering
    \begin{subfigure}{0.35\textwidth}
    \includegraphics[width=\linewidth]{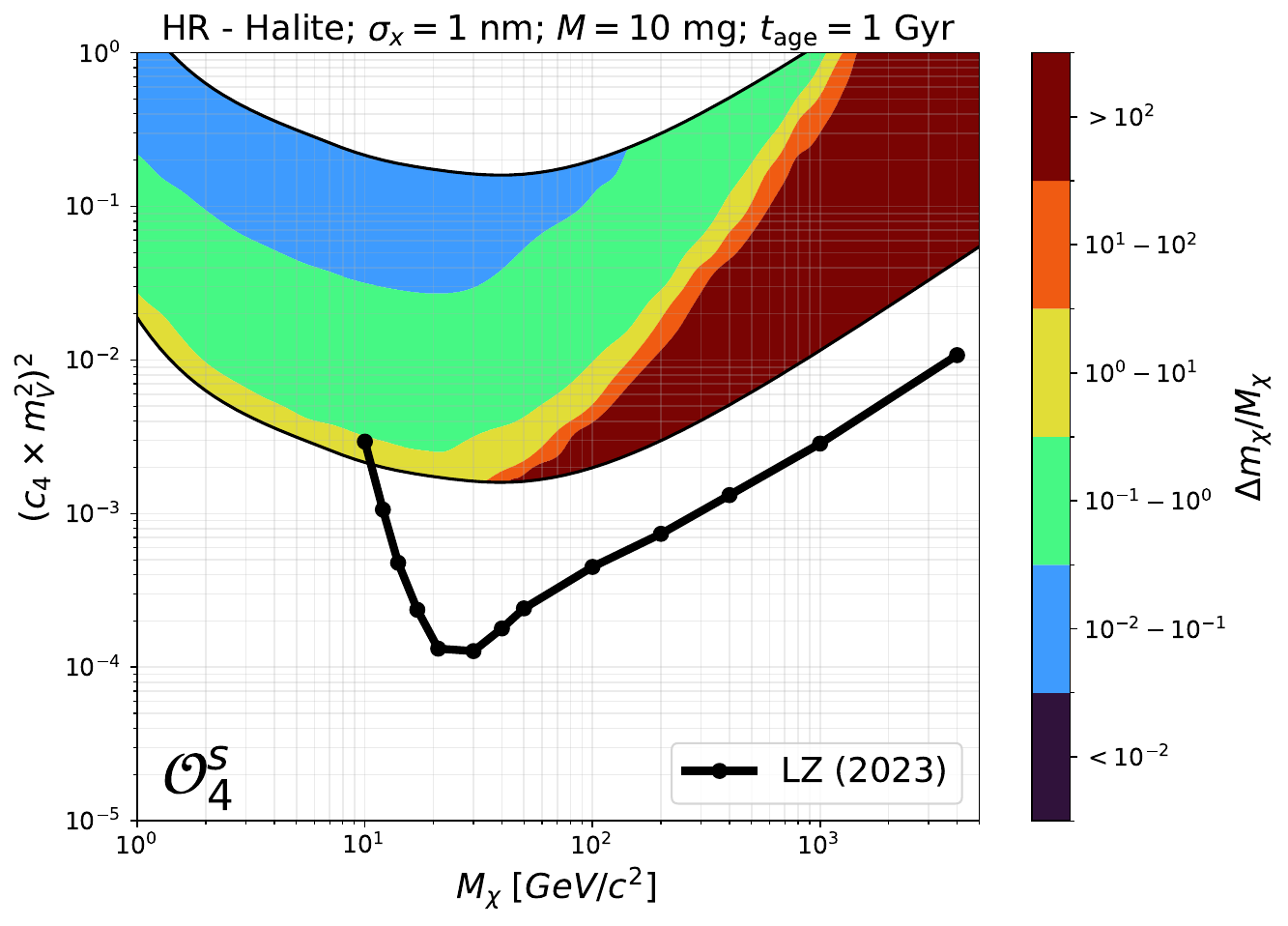}
    \end{subfigure}
    \begin{subfigure}{0.35\textwidth}
    \includegraphics[width=\linewidth]{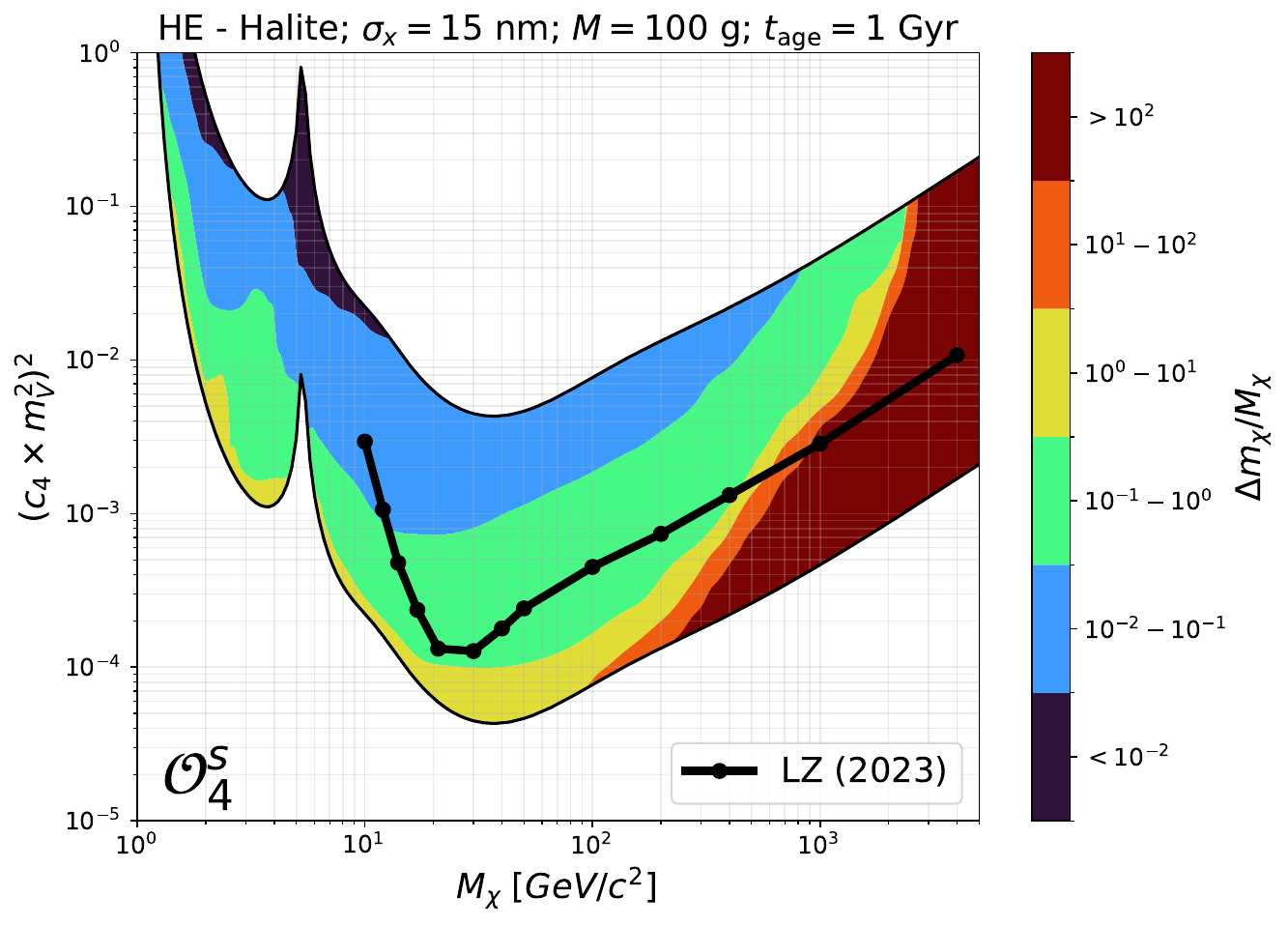}
    \end{subfigure}

    \vspace{0.1ex}

    \begin{subfigure}{0.35\textwidth}
    \includegraphics[width=\linewidth]{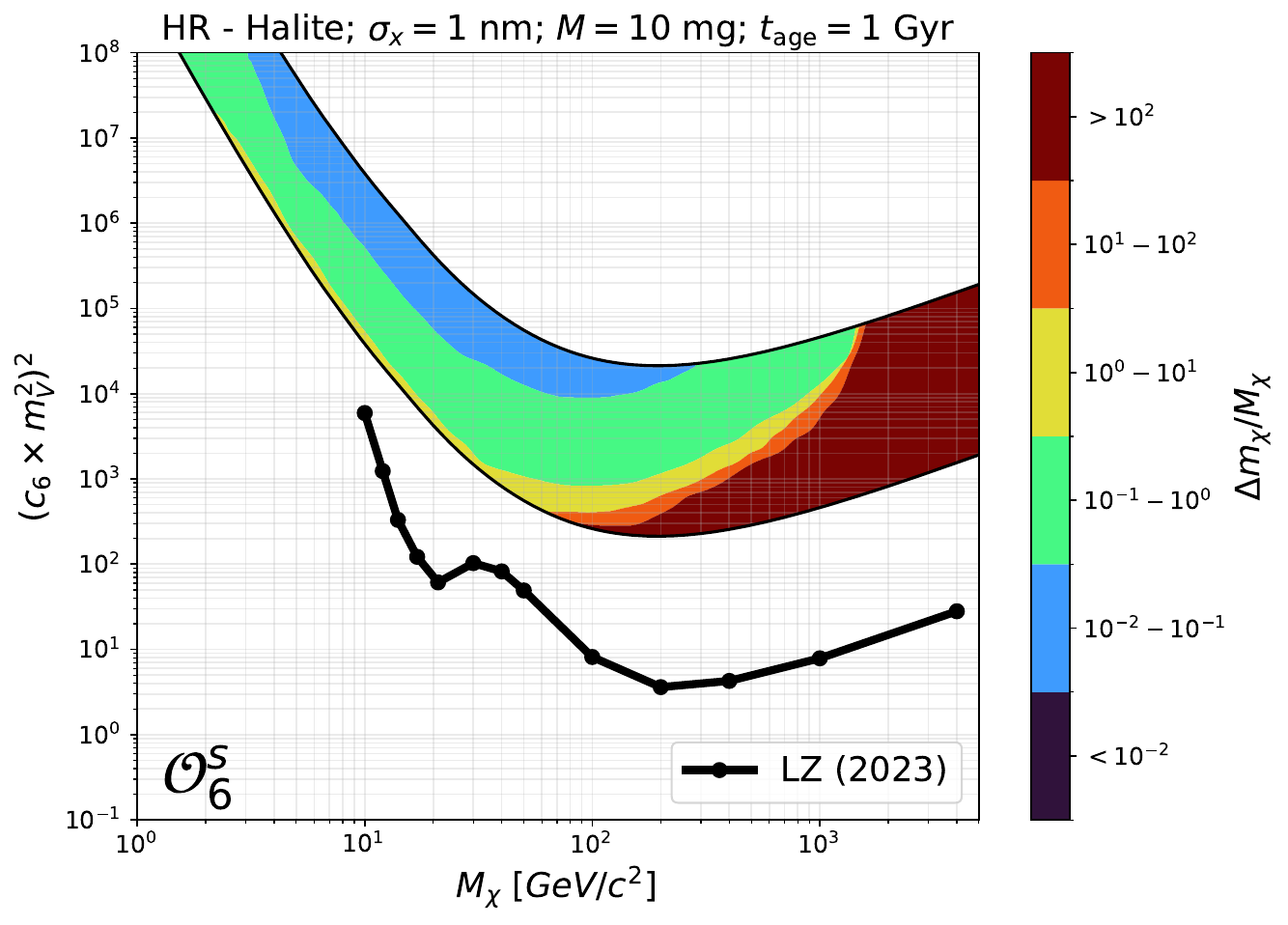}
    \end{subfigure}
    \begin{subfigure}{0.35\textwidth}
    \includegraphics[width=\linewidth]{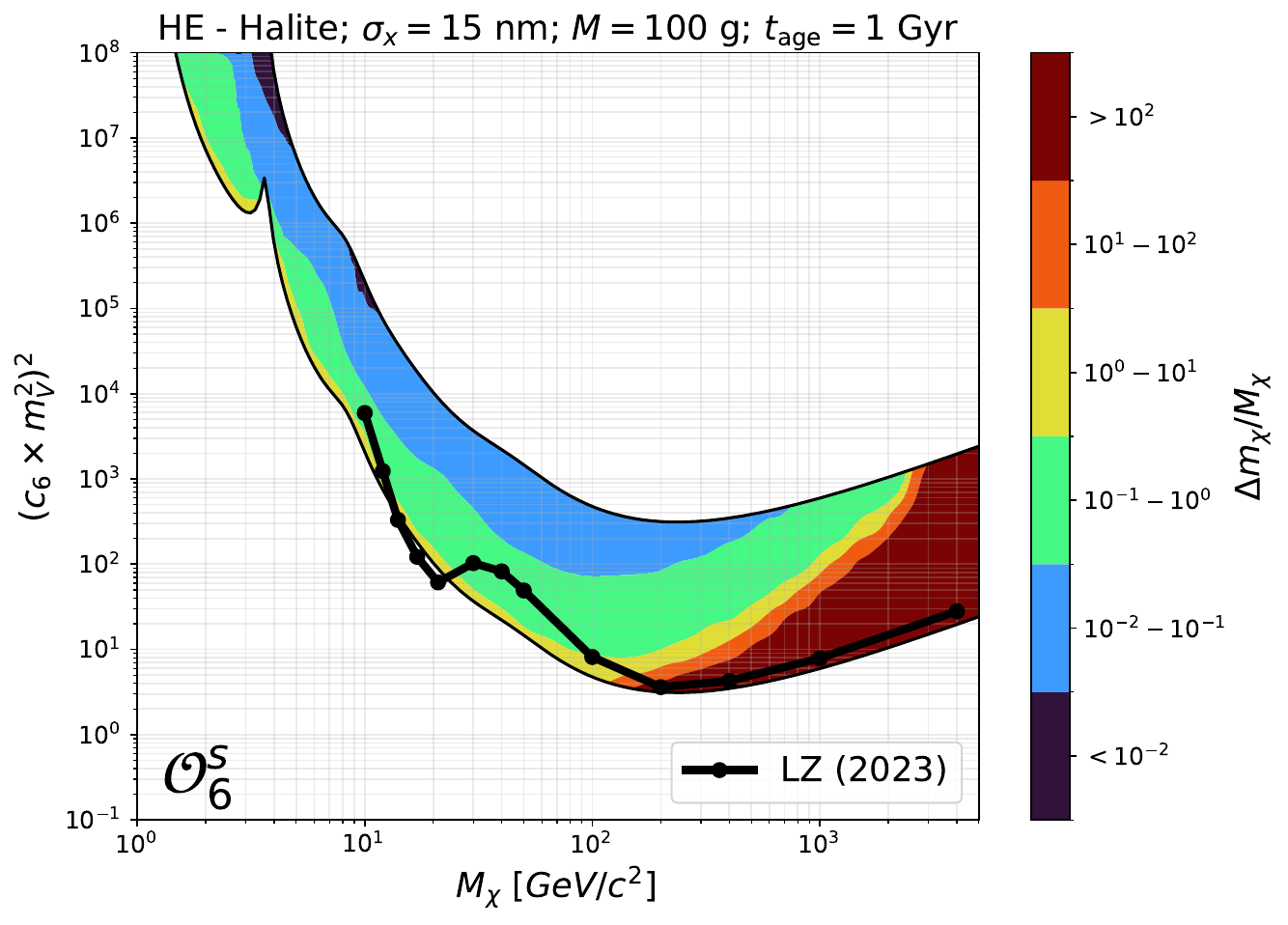}
    \end{subfigure}

    \vspace{0.1ex}
    
    \begin{subfigure}{0.35\textwidth}
    \includegraphics[width=\linewidth]{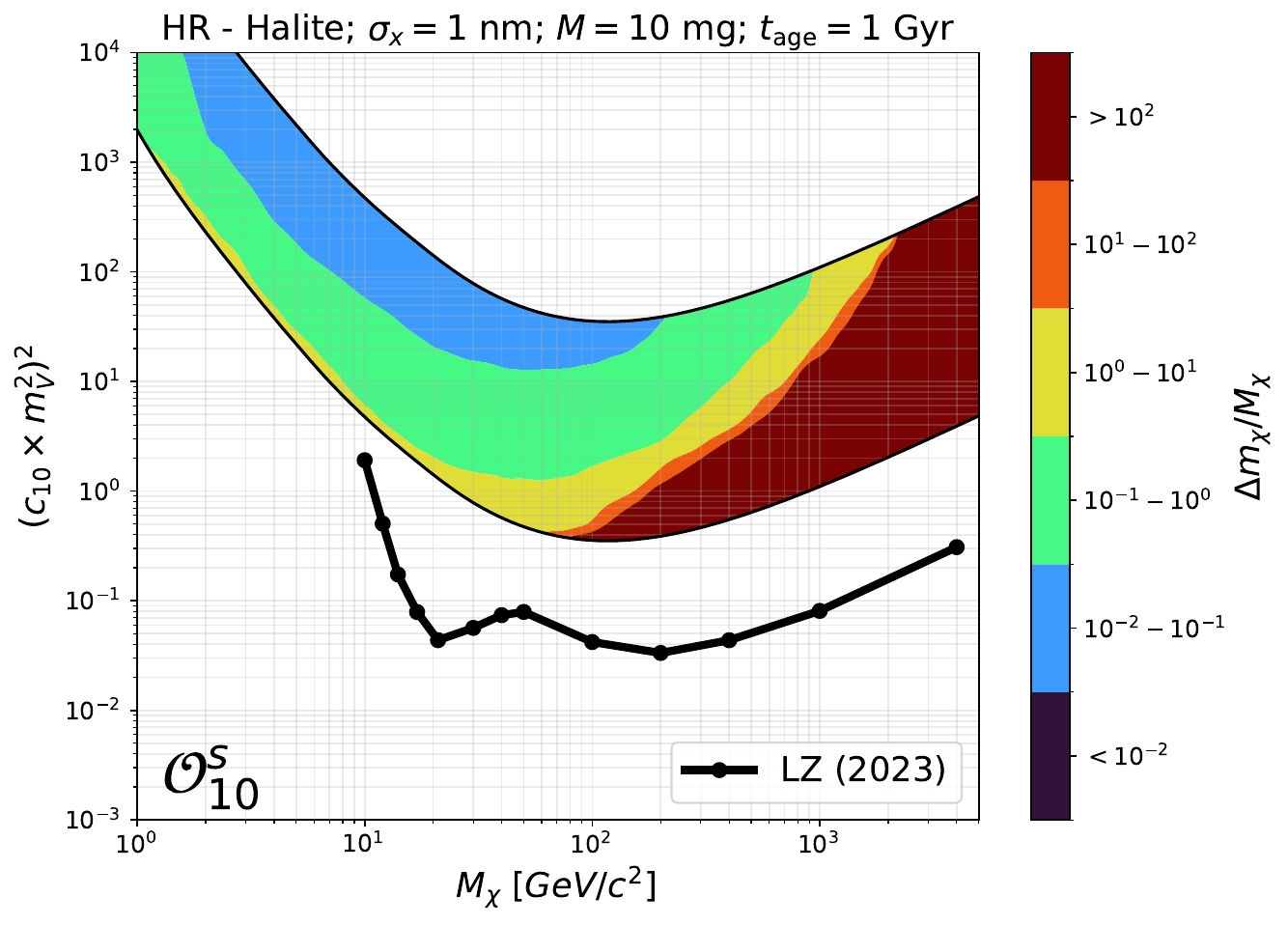}
    \end{subfigure}
    \begin{subfigure}{0.35\textwidth}
    \includegraphics[width=\linewidth]{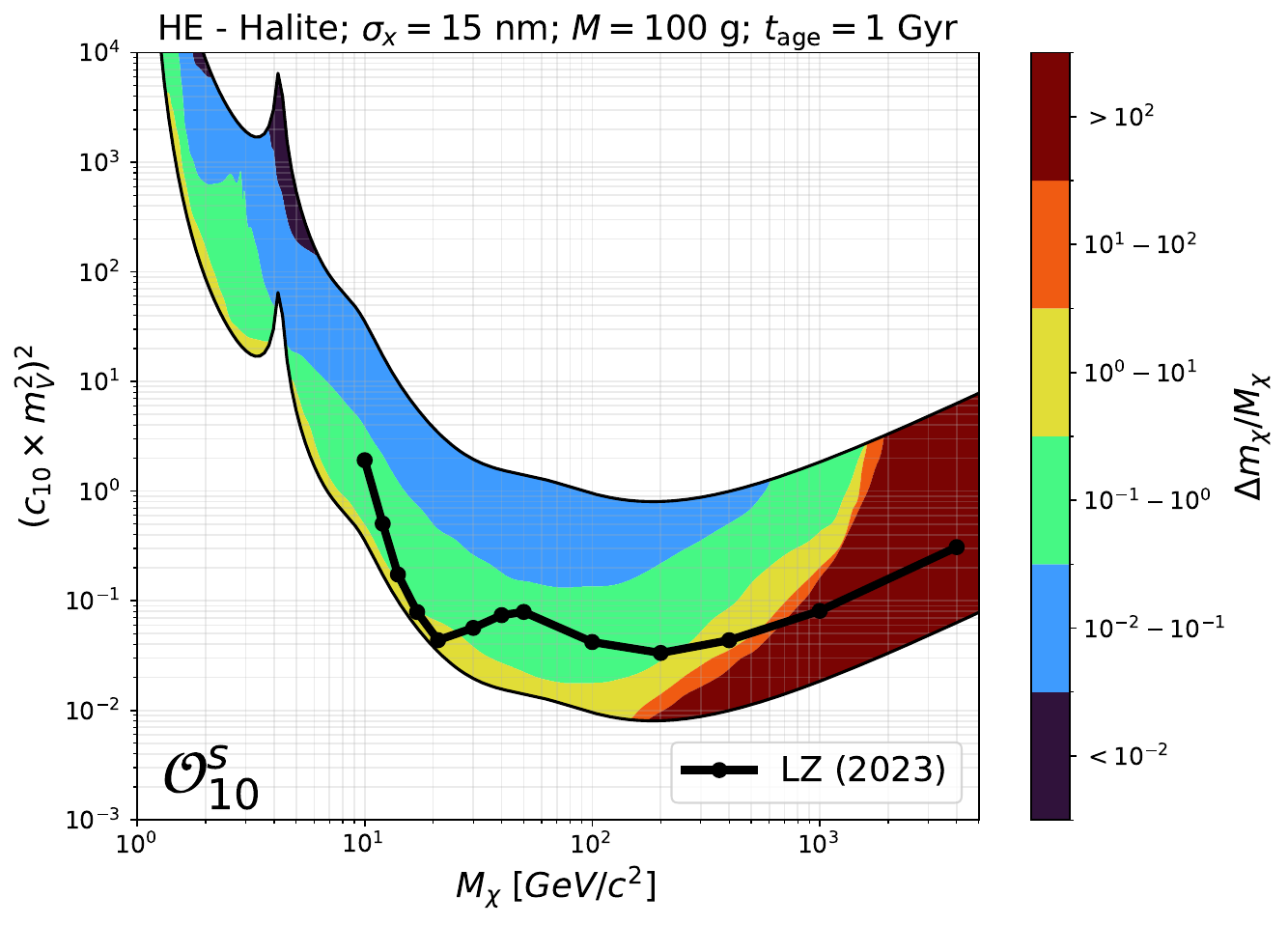}
    \end{subfigure}

    \vspace{0.1ex}

    \begin{subfigure}{0.35\textwidth}
    \includegraphics[width=\linewidth]{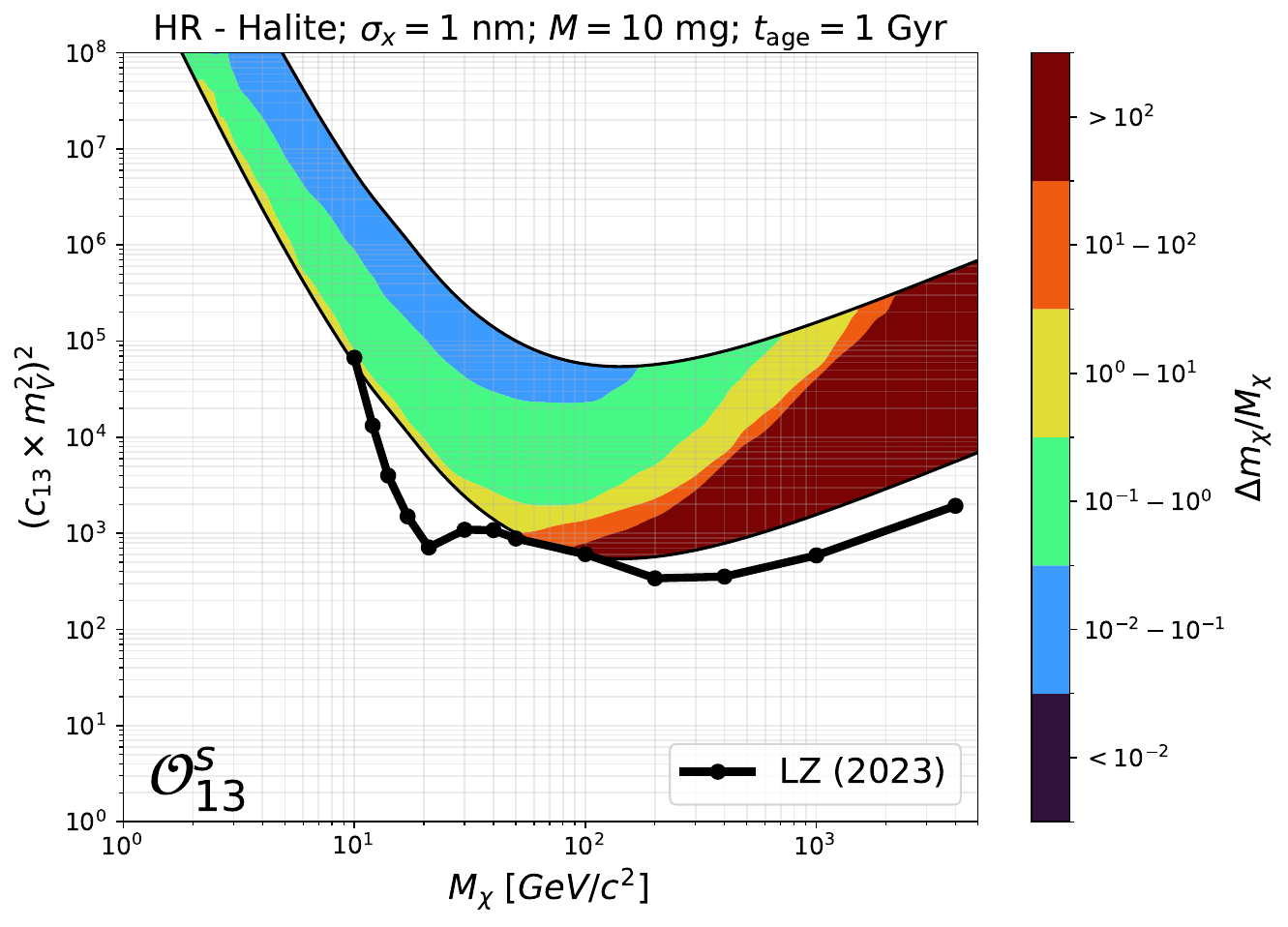}
    \end{subfigure}
    \begin{subfigure}{0.35\textwidth}
    \includegraphics[width=\linewidth]{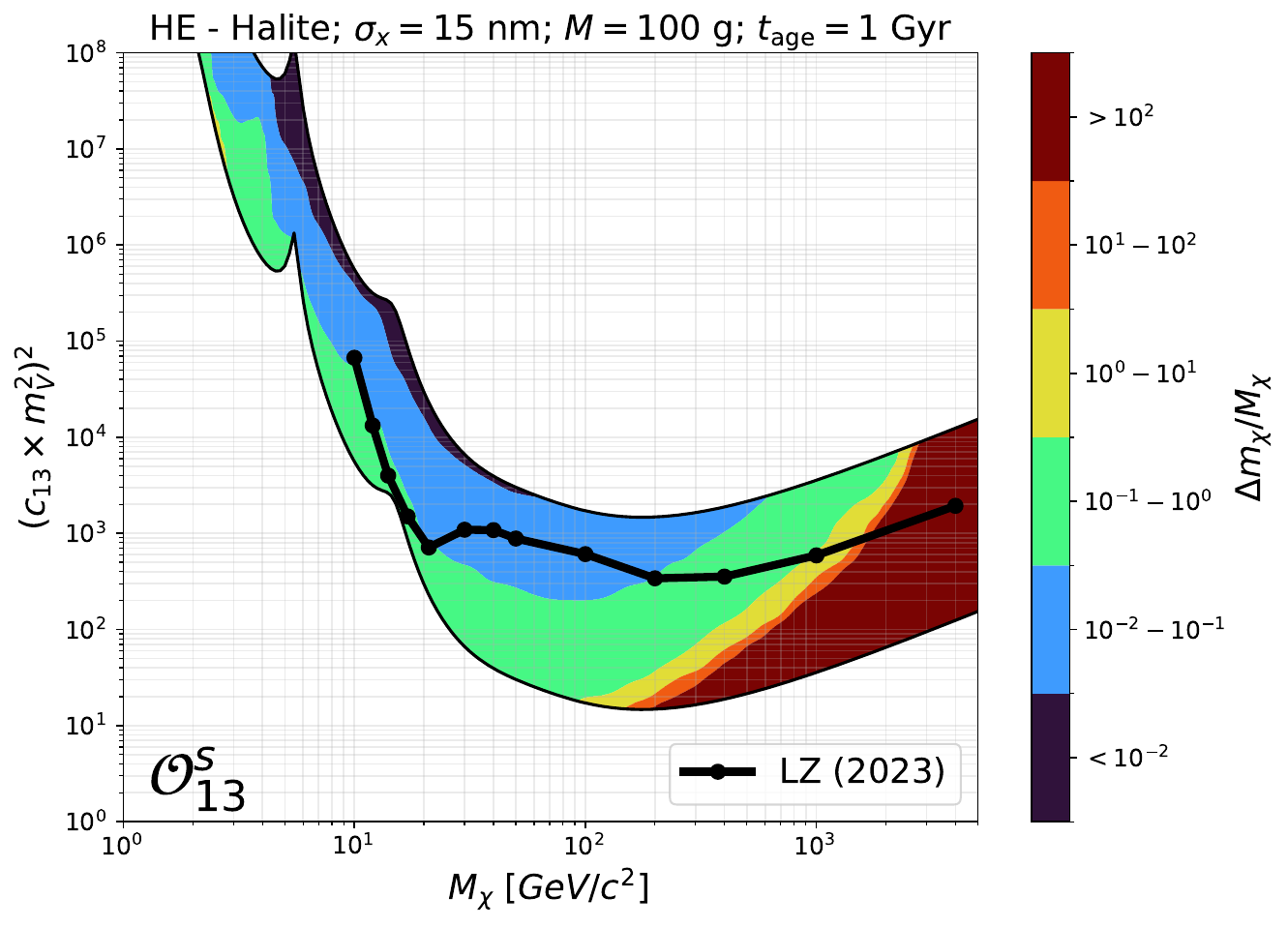}
    \end{subfigure}

    \caption{Projected constraints on the DM mass $M_{\chi}$ from a hypothetical signals generated by elastic interactions via the NREFT operators $\mathcal{O}_4^s$, $\mathcal{O}_6^s$, $\mathcal{O}_{10}^s$, and $\mathcal{O}_{13}^s$ in halite, with $C^{238}=10^{-11}\ \mathrm{g/g}$ and $t_{\rm age}=1\ \mathrm{Gyr}$. 
    Results are shown for the high-resolution (HR: $\sigma_x=1\ \mathrm{nm}$, $M=10\ \mathrm{mg}$, left) and high-exposure (HE: $\sigma_x=15\ \mathrm{nm}$, $M=100\ \mathrm{g}$, right) scenarios. 
    Colored regions indicate the fractional uncertainty $\Delta m_\chi / M_\chi$, where $\Delta m_\chi$ defines the range of masses around the true DM mass $M_\chi$ that cannot be excluded at $2\sigma$. The lower thin black curve without circular markers shows the projected 90\% C.L. exclusion limits for halite paleo-detectors from Ref.~\cite{Theodosopoulos:2026ehn}. Contours are restricted to the region between these limits and a factor of 100 above them (upper thin black curve without circular markers). Portions of the parameter space are already excluded by existing experiments, such as LUX–ZEPLIN~\cite{LZ:2023lvz} (thick black line with circular markers). Halite paleo-detectors are projected to reconstruct the DM mass with a relative uncertainty as small as $\Delta m_\chi/M_\chi \sim 10^{-2}$ for $M_\chi \simeq 1$--$10\ \mathrm{GeV}/c^2$, a mass range that is difficult to probe with conventional direct-detection experiments. In the HE scenario, mass reconstruction remains possible over the range $M_\chi \simeq 40$--$1000\ \mathrm{GeV}/c^2$ with a relative uncertainty $\Delta m_\chi/M_\chi \gtrsim 0.1$ for the operators $\mathcal{O}_{4}^s$, $\mathcal{O}_{10}^s$ and $\mathcal{O}_{13}^s$.}

    \label{fig:mass_reconstruction_O4}
\end{figure}

\begin{figure}
    \captionsetup{justification=raggedright,singlelinecheck=false}
    \centering
    \begin{subfigure}{0.46\textwidth}
    \includegraphics[width=\linewidth]{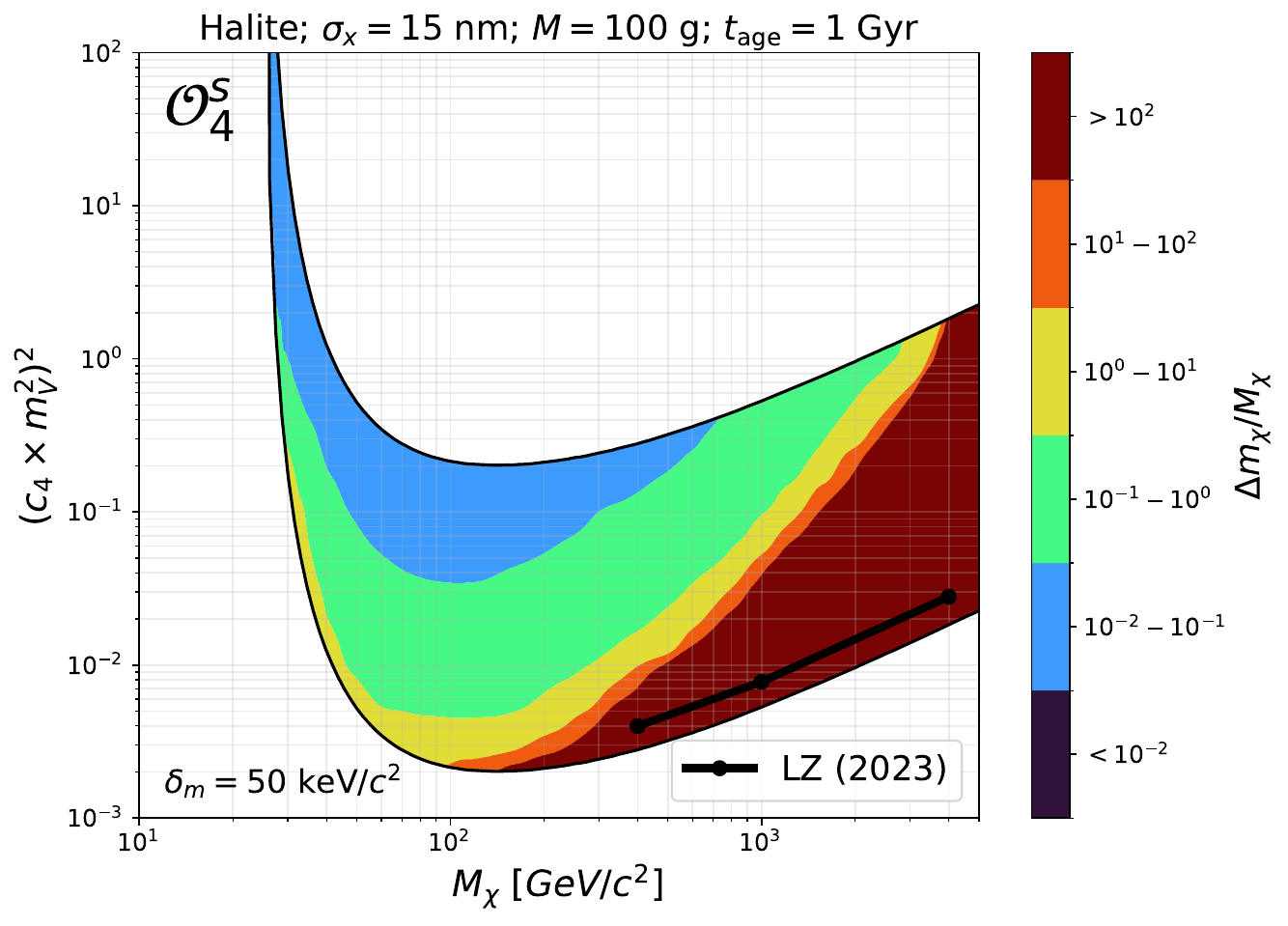}
    \end{subfigure}
    \begin{subfigure}{0.46\textwidth}
    \includegraphics[width=\linewidth]{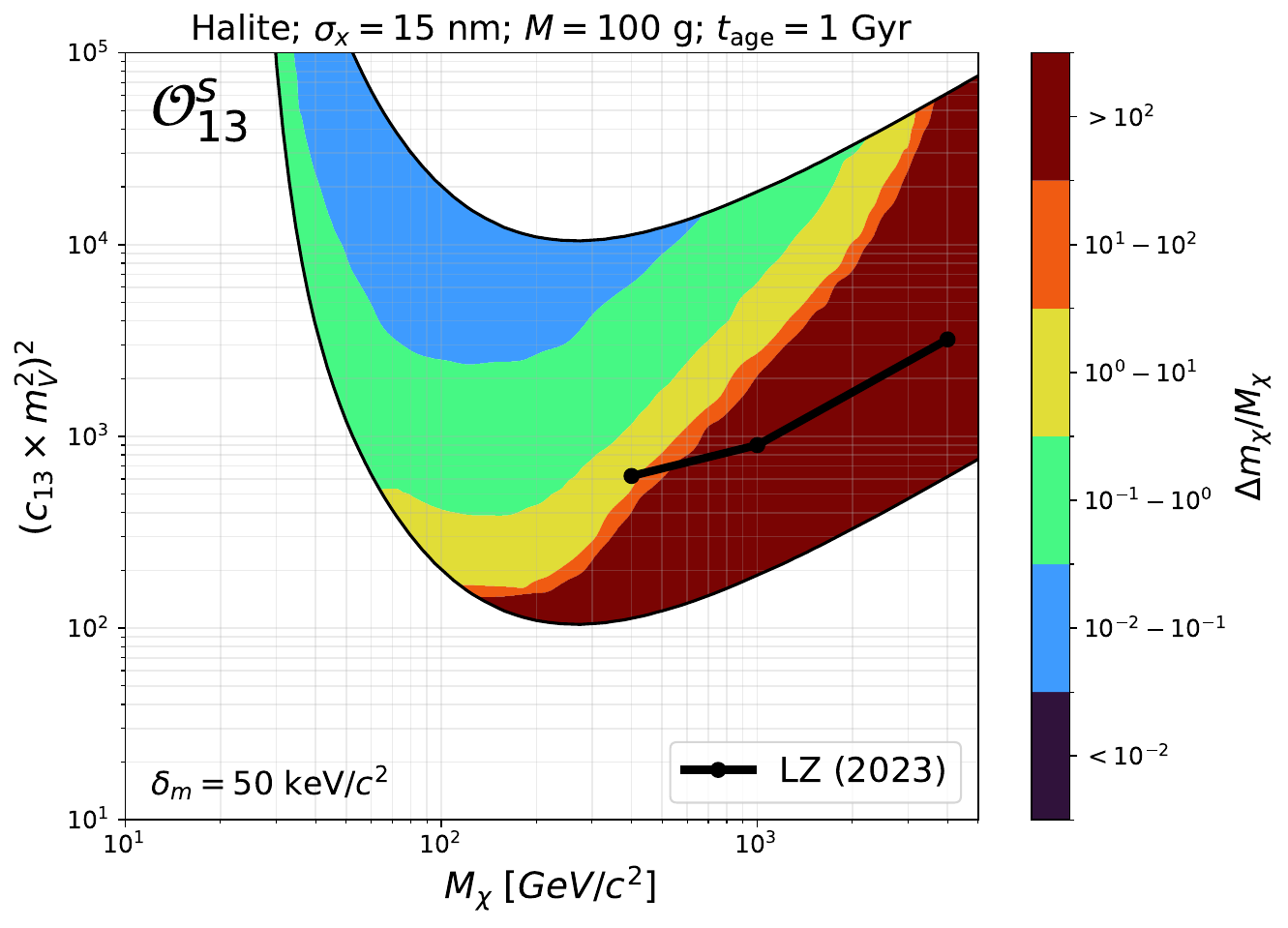}
    \end{subfigure}

    \caption{Projected constraints on the DM mass $M_{\chi}$ from a hypothetical signal generated by inelastic interactions via the $\mathcal{O}_{4}^s$ operator (left) and $\mathcal{O}_{13}^s$ (right), with $C^{238}=10^{-11}\ \mathrm{g/g}$ and $t_{\rm age}=1\ \mathrm{Gyr}$, for halite as the target mineral.
    The horizontal axis gives the true DM mass $M_\chi$, while the vertical axis gives the squared dimensionless isoscalar coupling $(c_j m_V^2)^2$ for the operator shown in each panel, with $m_V=246.2\ \mathrm{GeV}$.
    Results are shown for a mass-splitting $\delta_m=50\ \mathrm{keV}/c^2$ in the high-exposure (HE: $\sigma_x=15\ \mathrm{nm}$, $M=100\ \mathrm{g}$, right) scenario. The colored regions indicate the fractional uncertainty $\Delta m_\chi / M_\chi$, where $\Delta m_\chi$ defines the range of masses around $M_\chi$ that cannot be excluded at $2\sigma$. The lower thin black curve without circular markers shows the projected 90\% C.L. exclusion limits for halite paleo-detectors from Ref.~\cite{Theodosopoulos:2026ehn}. Contours are restricted to the region between these limits and a factor of 100 above them (upper thin black curve without circular markers). Portions of the parameter space are already excluded by existing experiments, such as LUX–ZEPLIN~\cite{LZ:2023lvz} (thick black line with circular markers). Halite paleo-detectors are projected to reconstruct $M_\chi$ with relative uncertainties down to $\Delta m_{\chi}/m_{\chi}\sim10^{-2}$ for DM masses between $30$ and $400\,\mathrm{GeV}/c^2$, a regime that is difficult to probe with conventional direct-detection experiments.}

    \label{fig:mass_reconstruction_Halite_inelastic}
\end{figure}

We next investigate the reconstruction of the DM mass in halite paleo-detectors. Fig.~\ref{fig:mass_reconstruction_O4} presents the projected reconstructed mass intervals for elastic scattering mediated by the operators
$\mathcal{O}_4^s$, $\mathcal{O}_6^s$, $\mathcal{O}_{10}^s$, and $\mathcal{O}_{13}^s$.
The overall behavior closely resembles that observed in the SI analysis for gypsum. 
The remaining spin-dependent operators accessible to halite paleo-detectors, $\mathcal{O}_{7}^s$, $\mathcal{O}_{9}^s$, and $\mathcal{O}_{14}^s$, yield mass-reconstruction projections similar to those shown in Fig.~\ref{fig:mass_reconstruction_O4} and are therefore omitted.
In the HR scenario, halite paleo-detectors are projected to reconstruct DM masses in the range $M_\chi\sim1$--$10\,\mathrm{GeV}/c^2$,
while in the HE scenario the reconstructed mass intervals remain relatively constrained up to $M_\chi\sim10^3\,\mathrm{GeV}/c^2$ for the operators $\mathcal{O}_{4}^s$, $\mathcal{O}_{10}^s$, and $\mathcal{O}_{13}^s$.
As discussed in the main text, the reconstructed mass intervals broaden substantially at very large DM masses because the recoil spectrum becomes only weakly dependent on $m_\chi$ through the reduced WIMP--nucleus mass. Fig.~\ref{fig:mass_reconstruction_Halite_inelastic} shows the corresponding projections for inelastic scattering mediated by
$\mathcal{O}_4^s$ and $\mathcal{O}_{13}^s$ assuming $\delta_m=50\,\mathrm{keV}/c^2$.
For DM masses between roughly $30$ and $400\,\mathrm{GeV}/c^2$, halite paleo-detectors are projected to reconstruct the DM mass with relative uncertainties down to $\Delta m_\chi/M_\chi\sim10^{-2}$, similarly to the gypsum SI analysis.
We omit mass-reconstruction projections for the operators $\mathcal{O}_6^s$ and $\mathcal{O}_{10}^s$,
since halite paleo-detectors are insensitive to these operators for inelastic scattering at $M_\chi \gtrsim 400\ \mathrm{GeV}/c^2$ (see TFKS), while at lower masses the reconstructed mass intervals for these operators follow trends similar to those obtained for $\mathcal{O}_4^s$.

\begin{figure}
    \captionsetup{justification=raggedright,singlelinecheck=false}
    \centering
    \begin{subfigure}{0.46\textwidth}
    \includegraphics[width=\linewidth]{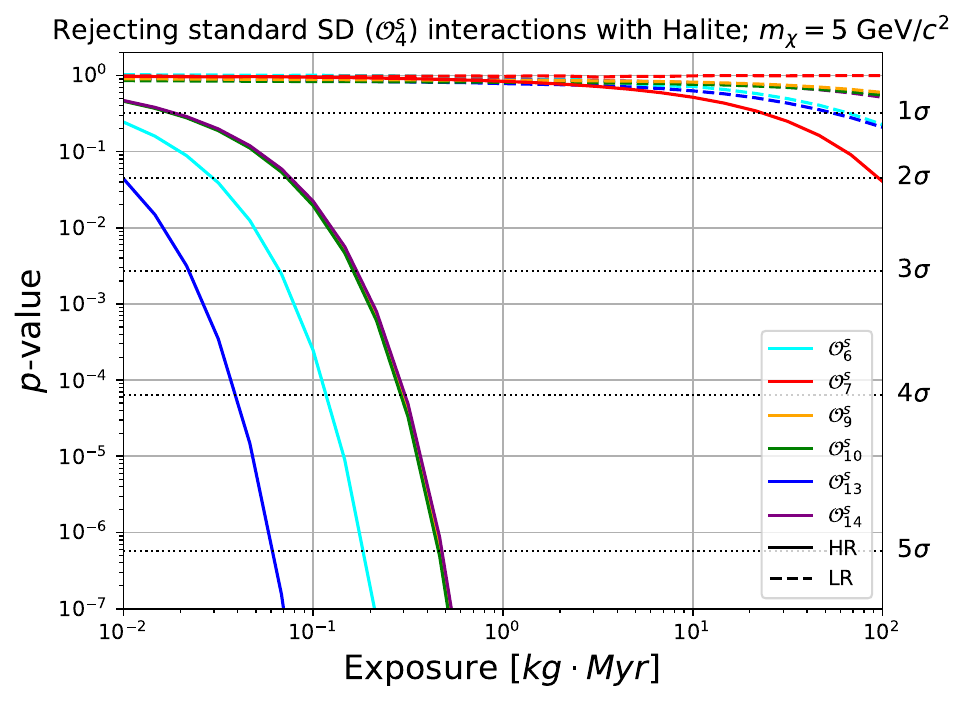}
    \end{subfigure}
    \begin{subfigure}{0.465\textwidth}
    \includegraphics[width=\linewidth]{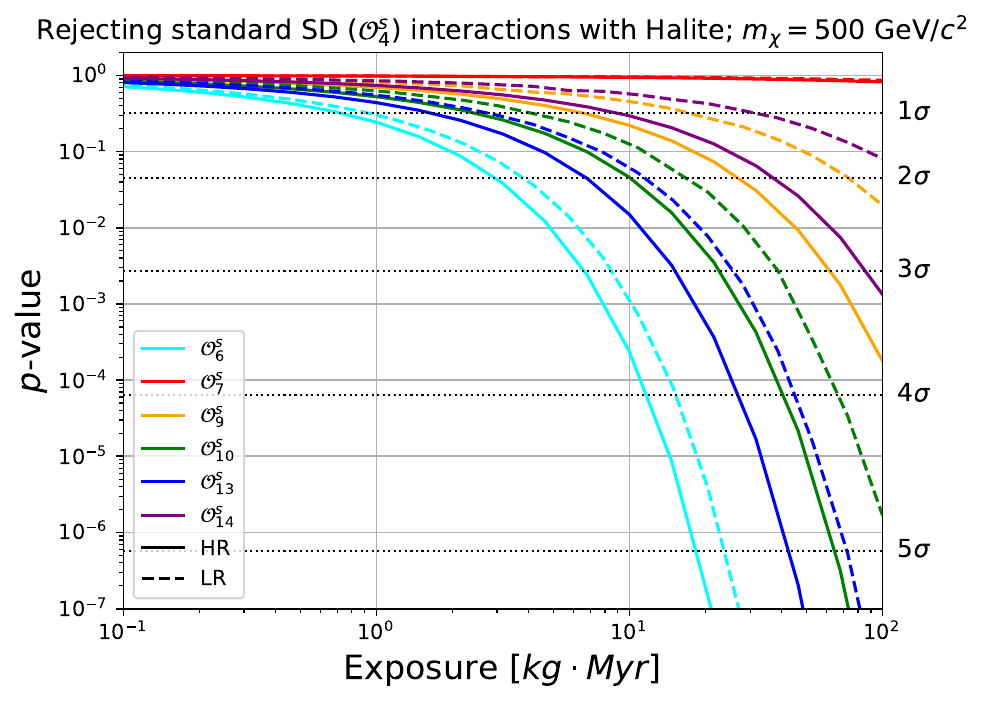}
    \end{subfigure}

    \caption{Projected significance for rejecting the elastic standard SD ($\mathcal{O}_4^s$) interaction hypothesis  as a function of exposure, for halite as the target mineral. The $p$-value and corresponding significance (in units of $\sigma$) are shown assuming a DM signal generated by non-standard NREFT operators ($\mathcal{O}^{s}_{6}$, $\mathcal{O}^{s}_{7}$, $\mathcal{O}^{s}_{9}$, $\mathcal{O}^{s}_{10}$, $\mathcal{O}^{s}_{13}$, and $\mathcal{O}^{s}_{14}$) in halite. The left (right) panel corresponds to $m_{\chi}=5\ \mathrm{GeV}/c^2$ ($500\ \mathrm{GeV}/c^2$). Solid and dashed curves denote high-resolution (HR, $\sigma_x = 1\ \mathrm{nm}$) and low-resolution (LR, $\sigma_x = 15\ \mathrm{nm}$) scenarios, respectively. The track-length spectra used in the calculations are normalized such that $R = \int dx_T\, dR/dx_T=6\times10^4\ \mathrm{kg}^{-1}\mathrm{Myr}^{-1}$ (left) and $10^3\ \mathrm{kg}^{-1}\mathrm{Myr}^{-1}$ (right) events per unit exposure, compatible with projected sensitivities for halite paleo-detectors~\cite{Theodosopoulos:2026ehn}. The standard SD operator $\mathcal{O}_4^s$ can be rejected with higher confidence levels for scattering via $\mathcal{O}_{6}^s$ and $\mathcal{O}_{13}^s$; whereas it cannot be rejected in the LR scenario for a light DM particle ($m_{\chi}=5\ \mathrm{GeV}/c^2$), nor for signals generated by $\mathcal{O}_7^s$ at either DM mass.}

    \label{fig:halite_comp}
\end{figure}

\begin{figure}
    \captionsetup{justification=raggedright,singlelinecheck=false}
    \centering
    \begin{subfigure}{0.46\textwidth}
    \includegraphics[width=\linewidth]{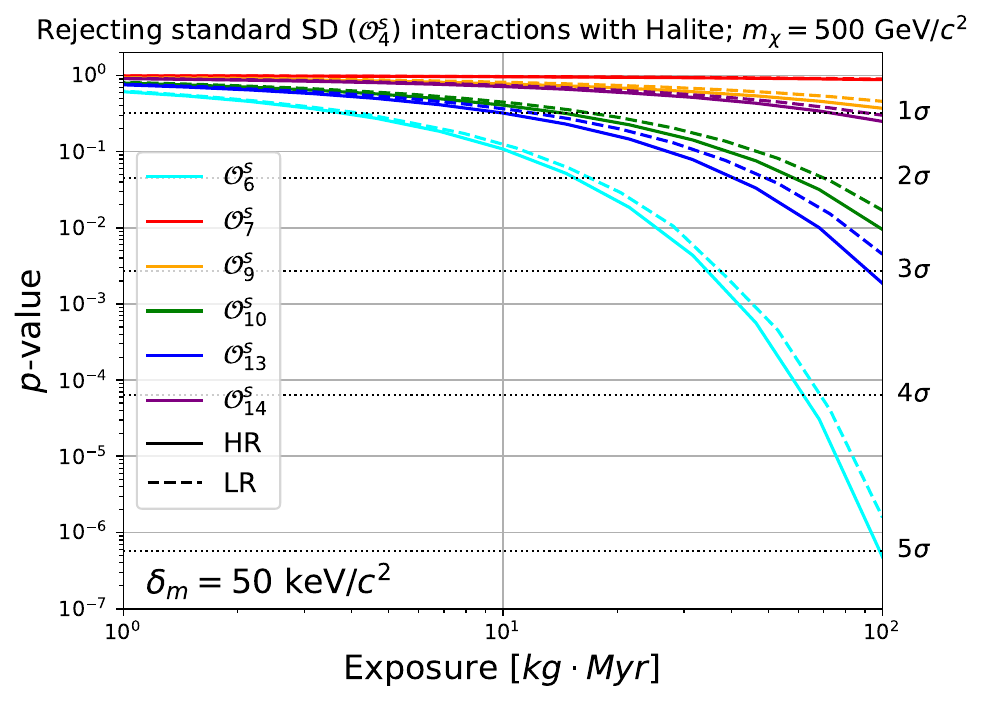}
    \end{subfigure}
    \begin{subfigure}{0.465\textwidth}
    \includegraphics[width=\linewidth]{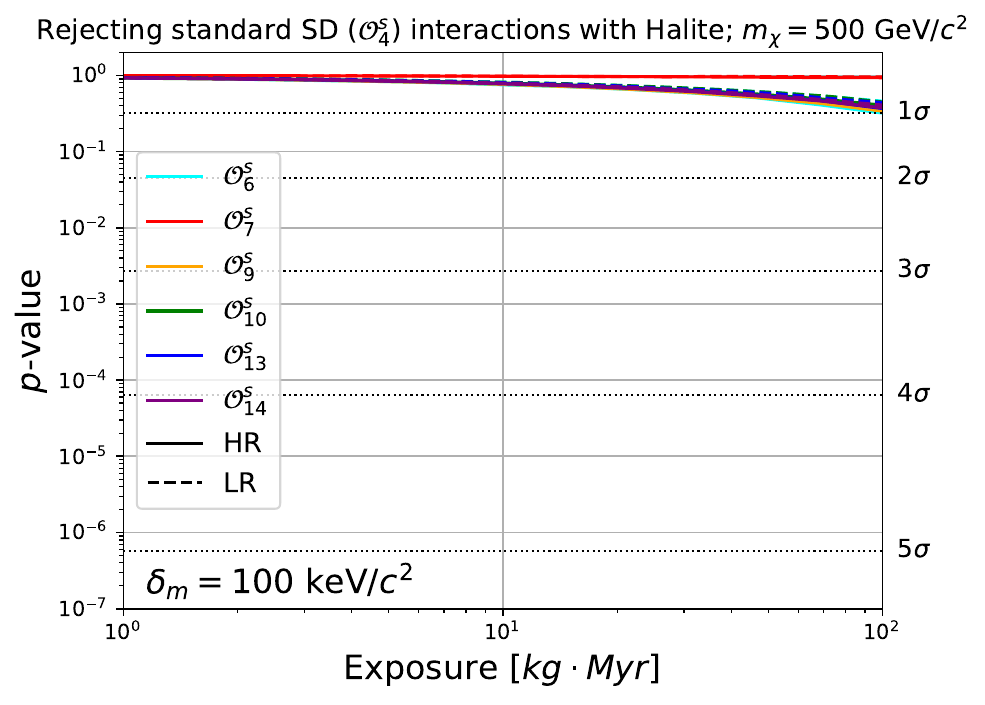}
    \end{subfigure}

    \caption{Projected significance for rejecting the inelastic standard SD ($\mathcal{O}_4^s$) interaction hypothesis as a function of exposure, for halite as the target mineral. The $p$-value and corresponding significance (in units of $\sigma$) are shown assuming a DM signal generated by non-standard NREFT operators ($\mathcal{O}^{s}_{6}$, $\mathcal{O}^{s}_{7}$, $\mathcal{O}^{s}_{9}$, $\mathcal{O}^{s}_{10}$, $\mathcal{O}^{s}_{13}$, and $\mathcal{O}^{s}_{14}$) in halite. The left (right) panel corresponds to $\delta_{m}=50\ \mathrm{keV}$ ($100\ \mathrm{keV}$). Solid and dashed curves denote high-resolution (HR, $\sigma_x = 1\ \mathrm{nm}$) and low-resolution (LR, $\sigma_x = 15\ \mathrm{nm}$) scenarios, respectively. The DM mass is set to $m_{\chi} = 500\ \mathrm{GeV}/c^2$. The track-length spectra used in the calculations are normalized such that $R = \int dx_T\, dR/dx_T=500\ \mathrm{kg}^{-1}\mathrm{Myr}^{-1}$ (left) and $150\ \mathrm{kg}^{-1}\mathrm{Myr}^{-1}$ (right) events per unit exposure, compatible with projected sensitivities for halite paleo-detectors~\cite{Theodosopoulos:2026ehn}. The standard SD operator $\mathcal{O}_4^s$ can be rejected with lower confidence levels for inelastic scattering, especially as the mass-splitting $\delta_m$ approaches the maximum values paleo-detectors can probe: $\delta_{m}=100\ \mathrm{keV}$.}

    \label{fig:halite_inelastic}
\end{figure}

We also study the ability of halite paleo-detectors to discriminate between standard SD interactions and non-standard interaction hypotheses. Figs.~\ref{fig:halite_comp} and \ref{fig:halite_inelastic} show the projected significance for rejecting the standard SD interaction hypothesis as a function of exposure for elastic and inelastic scattering, respectively. In the elastic case, the largest rejection significances are obtained for signals generated by $\mathcal{O}_6^s$ and $\mathcal{O}_{13}^s$,
consistent with the pronounced differences between their spectra and that of
$\mathcal{O}_4^s$.
By contrast, $\mathcal{O}_7^s$ cannot be distinguished from the standard SD interaction hypothesis because the corresponding track-length spectra are nearly identical. As in the gypsum analysis, the discrimination power is significantly reduced for light DM particles in the LR scenario because of the limited sensitivity to short damage tracks. In the inelastic case, the rejection significance decreases as the mass splitting increases, reflecting the reduced sensitivity of paleo-detectors to inelastic interactions.

\begin{figure}
    \captionsetup{justification=raggedright,singlelinecheck=false}
    \centering
    \begin{subfigure}{0.36\textwidth}
    \includegraphics[width=\linewidth]{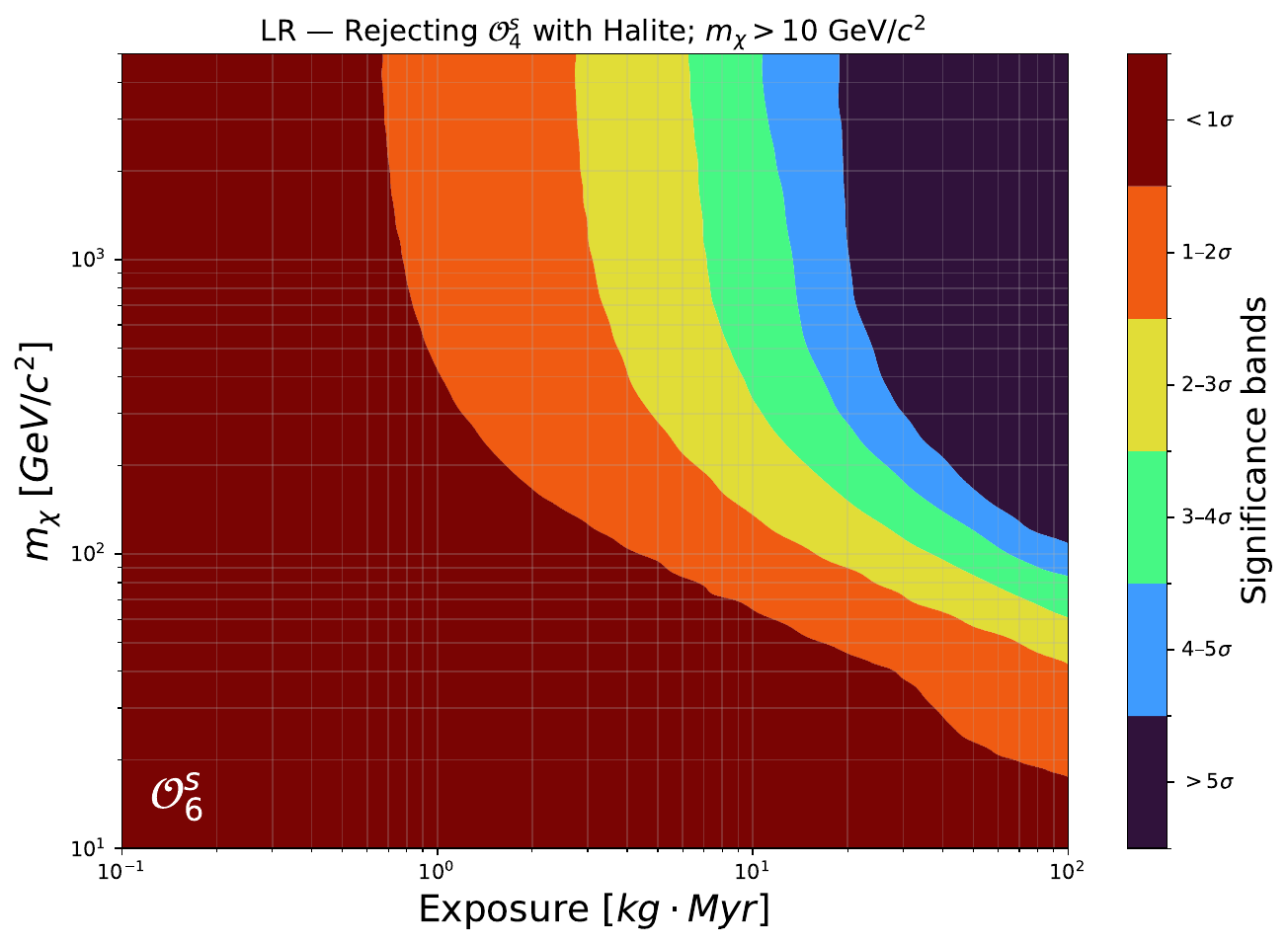}
    \end{subfigure}
    \begin{subfigure}{0.36\textwidth}
    \includegraphics[width=\linewidth]{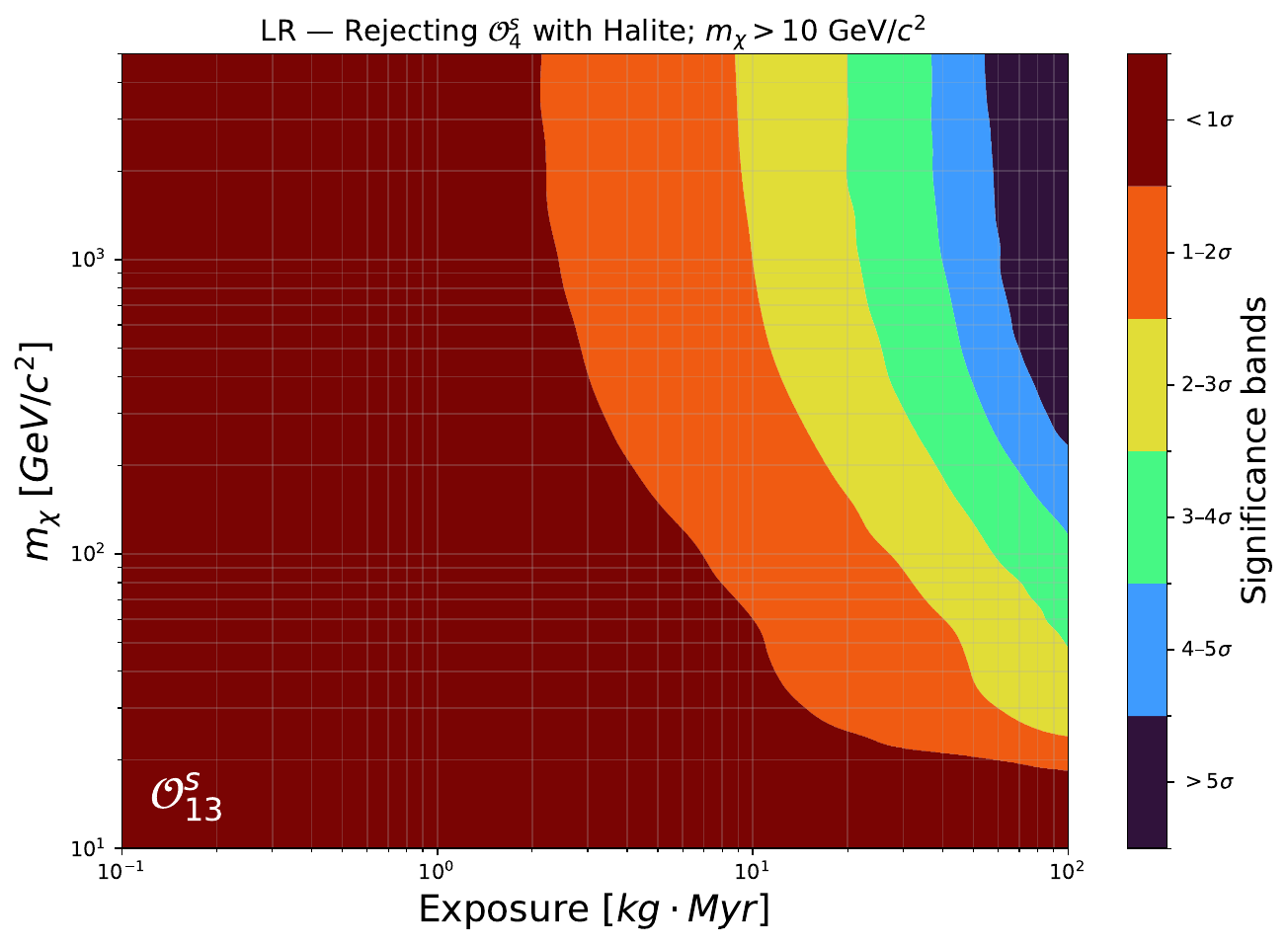}
    \end{subfigure}
    
    \vspace{0.1ex}

    \begin{subfigure}{0.36\textwidth}
    \includegraphics[width=\linewidth]{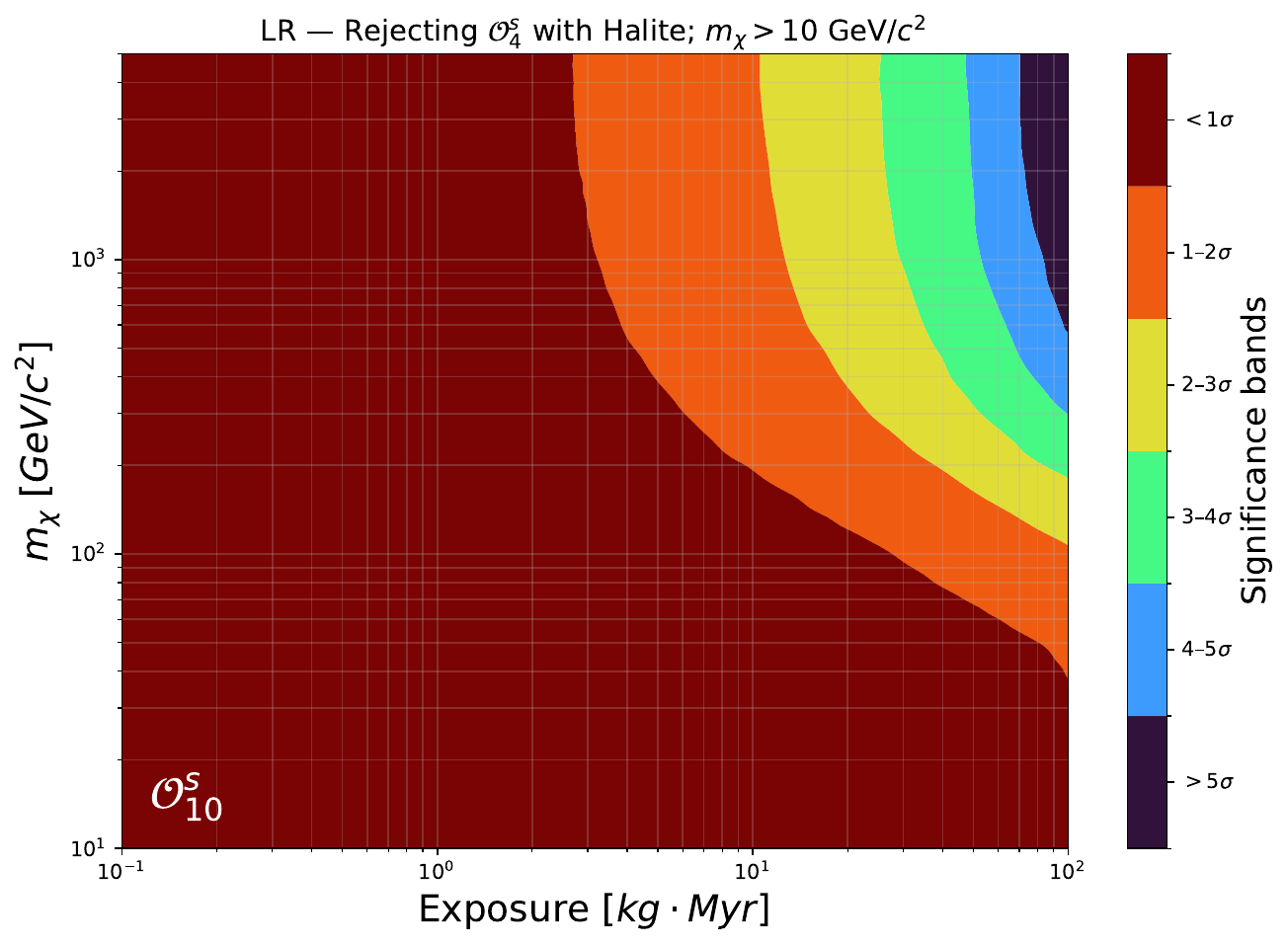}
    \end{subfigure}
    \begin{subfigure}{0.36\textwidth}
    \includegraphics[width=\linewidth]{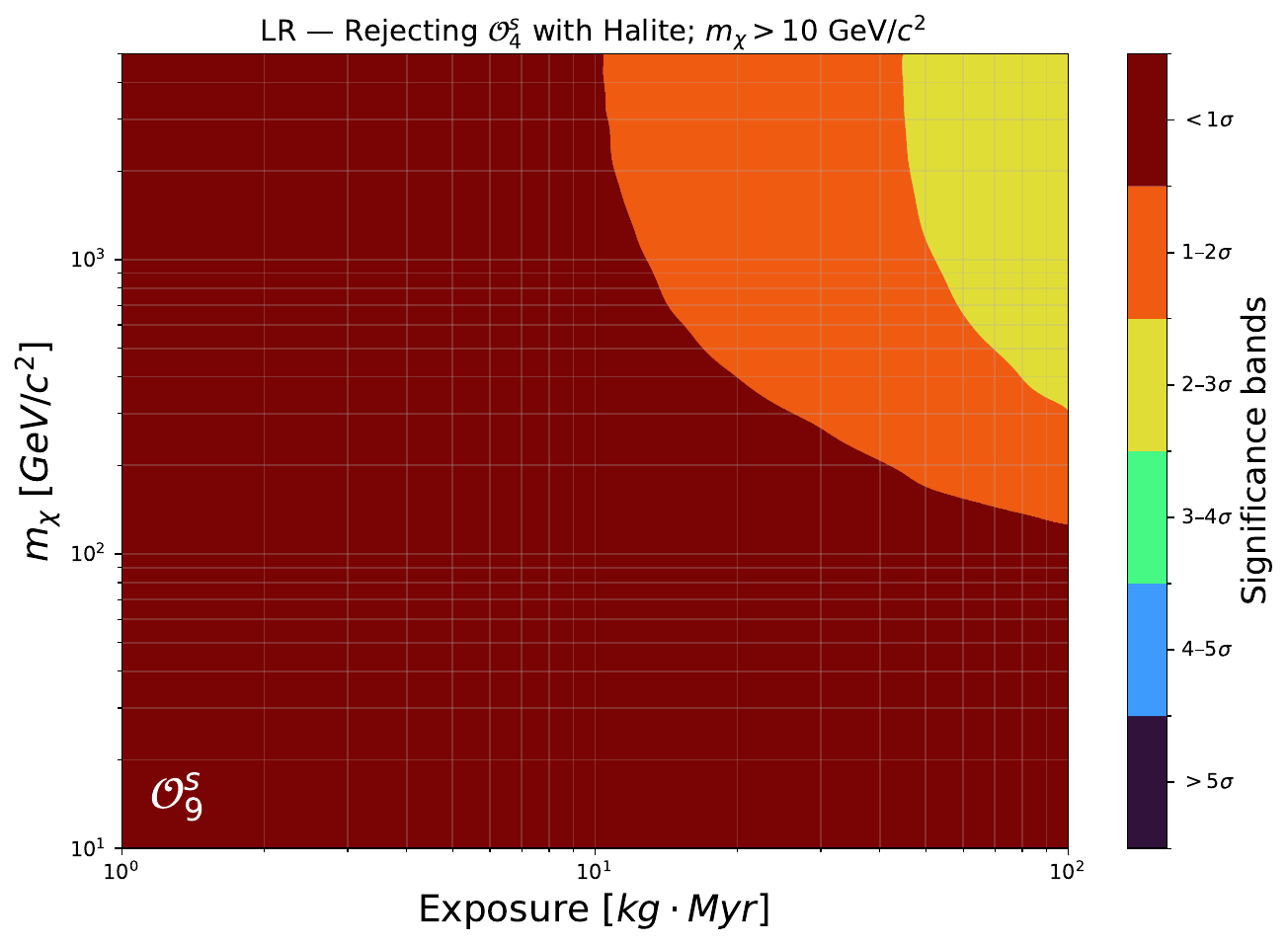}
    \end{subfigure}
    \caption{Projected confidence levels for rejecting the standard SD-only ($\mathcal{O}_4^s$) elastic interaction hypothesis, assuming a DM signal generated entirely by $\mathcal{O}_{6}^{s}$ and $\mathcal{O}_{13}^{s}$ (top panels), and by $\mathcal{O}_{10}^{s}$ and $\mathcal{O}_{9}^{s}$ (bottom panels), for halite as the target mineral. Contours corresponding to $1\sigma$–$5\sigma$ significance are shown as a function of DM mass and exposure for the low-resolution (LR, $\sigma_x=15\ \mathrm{nm}$) scenario. The track-length spectra used in the calculations are normalized such that $R = \int dx_T\, dR/dx_T=10^3\ \mathrm{kg}^{-1}\mathrm{Myr}^{-1}$ events per unit exposure, compatible with projected sensitivities for halite paleo-detectors~\cite{Theodosopoulos:2026ehn}. The standard SD operator ($\mathcal{O}_4^s$) cannot be rejected for $m_{\chi}\lesssim20\,\mathrm{GeV}/c^2$ for signals generated by $\mathcal{O}_6^s$ or $\mathcal{O}_{13}^s$, and for $m_{\chi}\lesssim100\,\mathrm{GeV}/c^2$ for signals generated by $\mathcal{O}_{9}^s$.}

    \label{fig:halite_contour}
\end{figure}

\begin{figure}
    \captionsetup{justification=raggedright,singlelinecheck=false}
    \centering
    \begin{subfigure}{0.36\textwidth}
    \includegraphics[width=\linewidth]{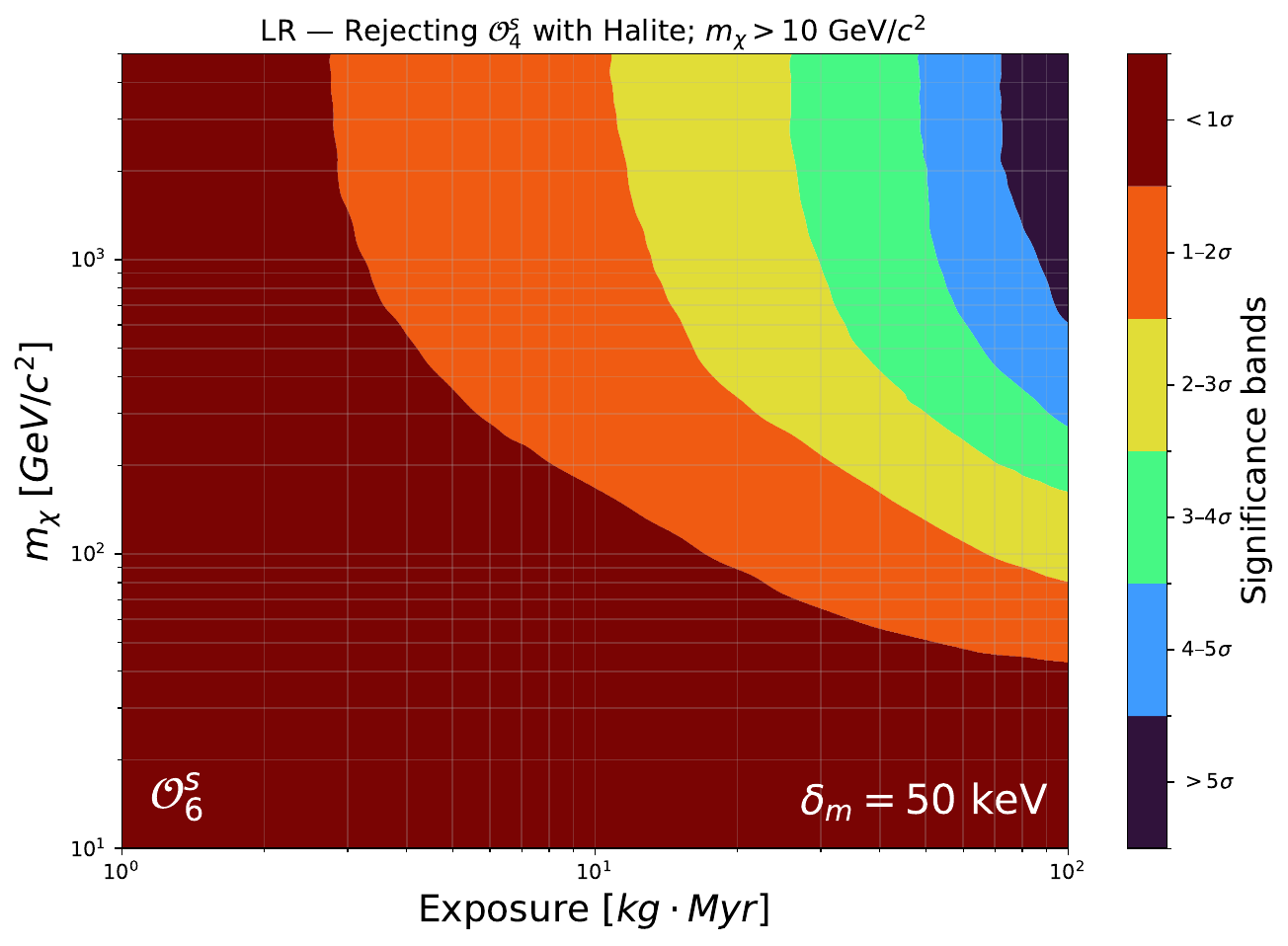}
    \end{subfigure}
    \begin{subfigure}{0.36\textwidth}
    \includegraphics[width=\linewidth]{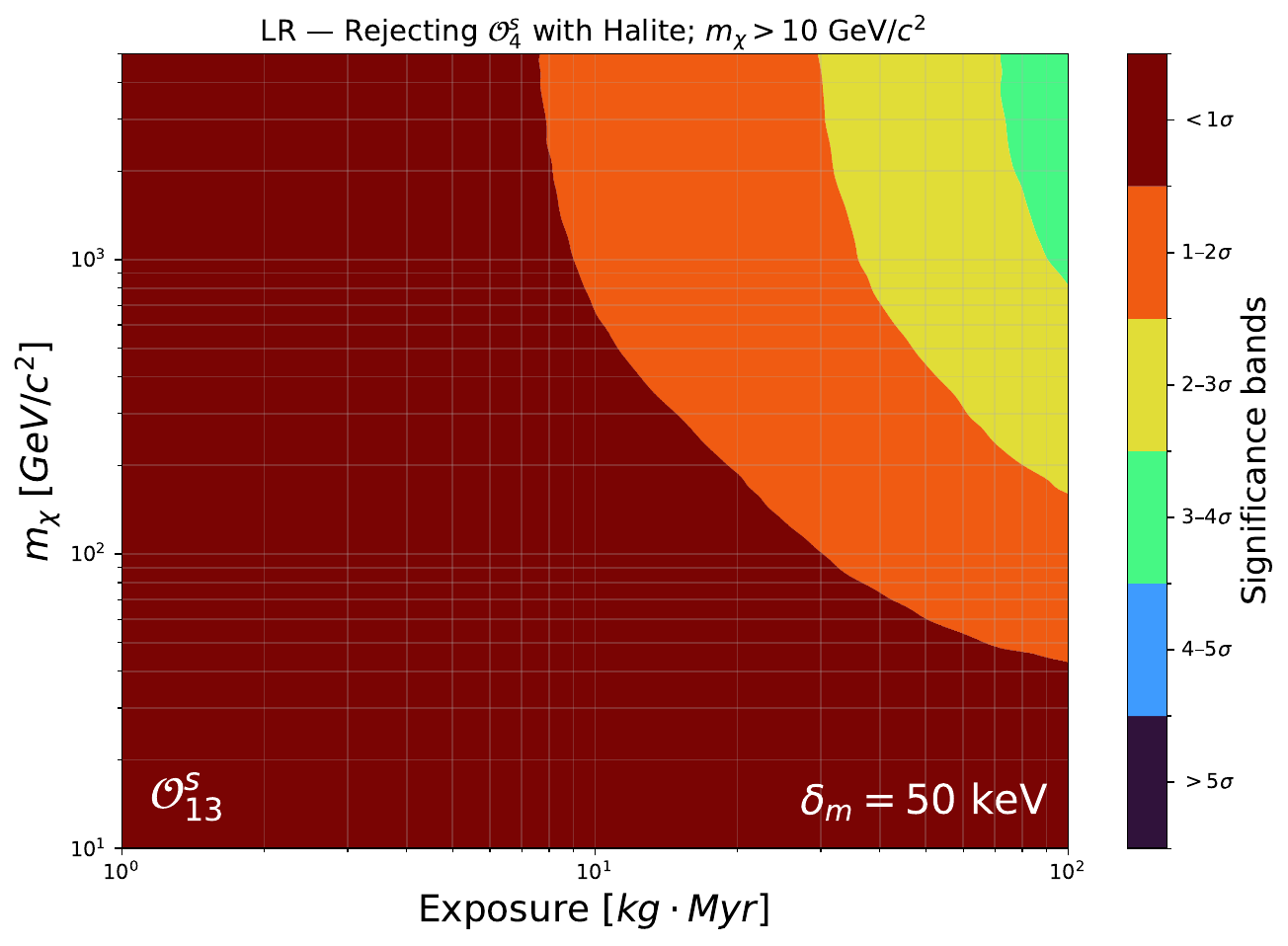}
    \end{subfigure}
    \caption{Projected confidence levels for rejecting the standard SD-only ($\mathcal{O}_4^s$) inelastic interaction hypothesis, assuming a DM signal generated entirely by $\mathcal{O}_{6}^{s}$ (left) and $\mathcal{O}_{13}^{s}$ (right), for halite as the target mineral and a mass splitting $\delta_m=50\,\mathrm{keV}/c^2$. Contours corresponding to $1\sigma$--$5\sigma$ significance are shown as a function of DM mass and exposure for the low-resolution (LR, $\sigma_x=15\,\mathrm{nm}$) scenario. The track-length spectra used in the calculations are normalized such that $R = \int dx_T\, dR/dx_T=500\ \mathrm{kg}^{-1}\mathrm{Myr}^{-1}$ events per unit exposure, compatible with projected sensitivities for halite paleo-detectors~\cite{Theodosopoulos:2026ehn}. For inelastic scattering, the standard SD operator cannot be rejected for $m_{\chi}\lesssim40\ \mathrm{GeV}/c^2$.}

    \label{fig:halite_contour_inelastic}
\end{figure}

Finally, Figs.~\ref{fig:halite_contour} and \ref{fig:halite_contour_inelastic} present contour plots of the projected confidence levels for rejecting the standard SD interaction hypothesis as a function of DM mass and exposure. The qualitative behavior closely mirrors the SI contour analysis presented in the main text. 
Operators such as $\mathcal{O}_6^s$ and $\mathcal{O}_{13}^s$ yield the highest rejection confidence levels because of the associated track-length spectra that differ significantly from that predicted by the standard SD interaction. For elastic scattering, the contours for $\mathcal{O}_{10}^s$ and $\mathcal{O}_{9}^s$ in Fig.~\ref{fig:halite_contour} are progressively shifted toward larger exposures, indicating reduced discrimination power. 
The sensitivity of paleo-detectors to inelastic WIMP--nucleon scattering is lower than in the elastic case (see TFKS). Consequently, the contours for $\mathcal{O}_6^s$ and $\mathcal{O}_{13}^s$ in Fig.~\ref{fig:halite_contour_inelastic} are shifted toward larger exposures relative to the corresponding contours in the elastic-scattering case. The inelastic-scattering contours for $\mathcal{O}_{10}^s$ and $\mathcal{O}_{9}^s$ exhibit qualitatively similar behavior and are therefore not shown. 
The discrimination power decreases for smaller DM masses and larger mass splittings, and becomes substantially reduced for
$m_\chi\lesssim20\,\mathrm{GeV}/c^2$ in the elastic case and
$m_\chi\lesssim40\,\mathrm{GeV}/c^2$ for inelastic scattering with $\delta_m\sim50\,\mathrm{keV}/c^2$.
Overall, the halite analysis confirms that paleo-detectors can provide statistically significant discrimination between different SD interaction hypotheses over a broad region of parameter space.

\bibliography{refs}

@article{Fitzpatrick:2012ix,
    author = "Fitzpatrick, A. Liam and Haxton, Wick and Katz, Emanuel and Lubbers, Nicholas and Xu, Yiming",
    title = "{The Effective Field Theory of Dark Matter Direct Detection}",
    eprint = "1203.3542",
    archivePrefix = "arXiv",
    primaryClass = "hep-ph",
    doi = "10.1088/1475-7516/2013/02/004",
    journal = "JCAP",
    volume = "02",
    pages = "004",
    year = "2013"
}

@article{Cerdeno:2024uqt,
    author = "Cerdeno, David and de los Rios, Martin and Perez, Andres D.",
    title = "{Bayesian technique to combine independently-trained machine-learning models applied to direct dark matter detection}",
    eprint = "2407.21008",
    archivePrefix = "arXiv",
    primaryClass = "hep-ph",
    reportNumber = "IFT-UAM/CSIC-24-116",
    doi = "10.1088/1475-7516/2025/01/038",
    journal = "JCAP",
    volume = "01",
    pages = "038",
    year = "2025"
}

@inproceedings{Miller:2021hys,
    author = "Miller, Benjamin Kurt and Cole, Alex and Forr{\'e}, Patrick and Louppe, Gilles and Weniger, Christoph",
    title = "{Truncated Marginal Neural Ratio Estimation}",
    booktitle = "{35th Conference on Neural Information Processing Systems}",
    eprint = "2107.01214",
    archivePrefix = "arXiv",
    primaryClass = "stat.ML",
    doi = "10.5281/zenodo.5043706",
    month = "7",
    year = "2021"
}

@misc{Hedges:2026pgf,
    author = "Hedges, Samuel and Huber, Patrick",
    title = "{Calorimetric approach to paleo-detection of dark matter}",
    eprint = "2605.13659",
    archivePrefix = "arXiv",
    primaryClass = "hep-ph",
    month = "5",
    year = "2026"
}

@article{Fieguth:2018vob,
    author = "Fieguth, A. and Hoferichter, M. and Klos, P. and Men{\'e}ndez, J. and Schwenk, A. and Weinheimer, C.",
    title = "{Discriminating WIMP-nucleus response functions in present and future XENON-like direct detection experiments}",
    eprint = "1802.04294",
    archivePrefix = "arXiv",
    primaryClass = "hep-ph",
    reportNumber = "INT-PUB-18-006",
    doi = "10.1103/PhysRevD.97.103532",
    journal = "Phys. Rev. D",
    volume = "97",
    number = "10",
    pages = "103532",
    year = "2018"
}

@article{Anand:2013yka,
    author = "Anand, Nikhil and Fitzpatrick, A. Liam and Haxton, W. C.",
    title = "{Weakly interacting massive particle-nucleus elastic scattering response}",
    eprint = "1308.6288",
    archivePrefix = "arXiv",
    primaryClass = "hep-ph",
    doi = "10.1103/PhysRevC.89.065501",
    journal = "Phys. Rev. C",
    volume = "89",
    number = "6",
    pages = "065501",
    year = "2014"
}

@article{Baxter:2021pqo,
    author = "Baxter, D. and others",
    title = "{Recommended conventions for reporting results from direct dark matter searches}",
    eprint = "2105.00599",
    archivePrefix = "arXiv",
    primaryClass = "hep-ex",
    doi = "10.1140/epjc/s10052-021-09655-y",
    journal = "Eur. Phys. J. C",
    volume = "81",
    number = "10",
    pages = "907",
    year = "2021"
}

@article{Read:2014qva,
    author = "Read, J. I.",
    title = "{The Local Dark Matter Density}",
    eprint = "1404.1938",
    archivePrefix = "arXiv",
    primaryClass = "astro-ph.GA",
    reportNumber = "JPHYSG-100038.R1",
    doi = "10.1088/0954-3899/41/6/063101",
    journal = "J. Phys. G",
    volume = "41",
    pages = "063101",
    year = "2014"
}

@article{10.1111/j.1365-2966.2010.16253.x,
    author = {Schönrich, Ralph and Binney, James and Dehnen, Walter},
    title = {Local kinematics and the local standard of rest},
    journal = {Monthly Notices of the Royal Astronomical Society},
    volume = {403},
    number = {4},
    pages = {1829-1833},
    year = {2010},
    month = {04},
    issn = {0035-8711},
    doi = {10.1111/j.1365-2966.2010.16253.x},
    url = {https://doi.org/10.1111/j.1365-2966.2010.16253.x},
}

@article{Sch_nrich_2012,
   title={Galactic rotation and solar motion from stellar kinematics},
   volume={427},
   ISSN={1365-2966},
   url={http://dx.doi.org/10.1111/j.1365-2966.2012.21631.x},
   DOI={10.1111/j.1365-2966.2012.21631.x},
   number={1},
   journal={Monthly Notices of the Royal Astronomical Society},
   publisher={Oxford University Press (OUP)},
   author={Schönrich, Ralph},
   year={2012},
   month=oct, pages={274–287} }

@article{Bland_Hawthorn_2016,
   title={The Galaxy in Context: Structural, Kinematic, and Integrated Properties},
   volume={54},
   ISSN={1545-4282},
   url={http://dx.doi.org/10.1146/annurev-astro-081915-023441},
   DOI={10.1146/annurev-astro-081915-023441},
   number={1},
   journal={Annual Review of Astronomy and Astrophysics},
   publisher={Annual Reviews},
   author={Bland-Hawthorn, Joss and Gerhard, Ortwin},
   year={2016},
   month=sep, pages={529–596} }

@article{2021,
   title={Improved GRAVITY astrometric accuracy from modeling optical aberrations},
   volume={647},
   ISSN={1432-0746},
   url={http://dx.doi.org/10.1051/0004-6361/202040208},
   DOI={10.1051/0004-6361/202040208},
   journal={Astron.; Astrophys.},
   publisher={EDP Sciences},
   author={Abuter, R. and others},
   year={2021},
   month=mar, pages={A59} }

@article{10.1111/j.1365-2966.2007.11964.x,
    author = {Smith, Martin C. and others},
    title = {The RAVE survey: constraining the local Galactic escape speed},
    journal = {Monthly Notices of the Royal Astronomical Society},
    volume = {379},
    number = {2},
    pages = {755-772},
    year = {2007},
    month = {07},
    issn = {0035-8711},
    doi = {10.1111/j.1365-2966.2007.11964.x},
    url = {https://doi.org/10.1111/j.1365-2966.2007.11964.x},
    eprint = {https://academic.oup.com/mnras/article-pdf/379/2/755/3399611/mnras0379-0755.pdf},
}

@article{Jeong:2021bpl,
    author = "Jeong, Injun and Kang, Sunghyun and Scopel, Stefano and Tomar, Gaurav",
    title = "{WimPyDD: An object{\textendash}oriented Python code for the calculation of WIMP direct detection signals}",
    eprint = "2106.06207",
    archivePrefix = "arXiv",
    primaryClass = "hep-ph",
    reportNumber = "CQUeST-2021-0663, TUM-HEP 1343/21",
    doi = "10.1016/j.cpc.2022.108342",
    journal = "Comput. Phys. Commun.",
    volume = "276",
    pages = "108342",
    year = "2022"
}

@article{Gorton:2022eed,
    author = "Gorton, Oliver C. and Johnson, Calvin W. and Jiao, Changfeng and Nikoleyczik, Jonathan",
    title = "{dmscatter: A fast program for WIMP-nucleus scattering}",
    eprint = "2209.09187",
    archivePrefix = "arXiv",
    primaryClass = "nucl-th",
    doi = "10.1016/j.cpc.2022.108597",
    journal = "Comput. Phys. Commun.",
    volume = "284",
    pages = "108597",
    year = "2023"
}

@article{ZIEGLER20101818,
title = {SRIM – The stopping and range of ions in matter (2010)},
journal = {Nuclear Instruments and Methods in Physics Research Section B: Beam Interactions with Materials and Atoms},
volume = {268},
number = {11},
pages = {1818-1823},
year = {2010},
note = {19th International Conference on Ion Beam Analysis},
issn = {0168-583X},
doi = {https://doi.org/10.1016/j.nimb.2010.02.091},
url = {https://www.sciencedirect.com/science/article/pii/S0168583X10001862},
author = {James F. Ziegler and M.D. Ziegler and J.P. Biersack}
}

@article{Drukier:2018pdy,
    author = "Drukier, Andrzej K. and Baum, Sebastian and Freese, Katherine and G{\'o}rski, Maciej and Stengel, Patrick",
    title = "{Paleo-detectors: Searching for Dark Matter with Ancient Minerals}",
    eprint = "1811.06844",
    archivePrefix = "arXiv",
    primaryClass = "astro-ph.CO",
    reportNumber = "NORDITA-2018-117, LCTP-18-25",
    doi = "10.1103/PhysRevD.99.043014",
    journal = "Phys. Rev. D",
    volume = "99",
    number = "4",
    pages = "043014",
    year = "2019"
}

@article{Fan:2010gt,
    author = "Fan, JiJi and Reece, Matthew and Wang, Lian-Tao",
    title = "{Non-relativistic effective theory of dark matter direct detection}",
    eprint = "1008.1591",
    archivePrefix = "arXiv",
    primaryClass = "hep-ph",
    doi = "10.1088/1475-7516/2010/11/042",
    journal = "JCAP",
    volume = "11",
    pages = "042",
    year = "2010"
}

@article{XENON:2017fdd,
    author = "Aprile, E. and others",
    collaboration = "XENON",
    title = "{Effective field theory search for high-energy nuclear recoils using the XENON100 dark matter detector}",
    eprint = "1705.02614",
    archivePrefix = "arXiv",
    primaryClass = "astro-ph.CO",
    doi = "10.1103/PhysRevD.96.042004",
    journal = "Phys. Rev. D",
    volume = "96",
    number = "4",
    pages = "042004",
    year = "2017"
}

@article{LZ:2023lvz,
    author = "Aalbers, J. and others",
    collaboration = "LZ",
    title = "{First constraints on WIMP-nucleon effective field theory couplings in an extended energy region from LUX-ZEPLIN}",
    eprint = "2312.02030",
    archivePrefix = "arXiv",
    primaryClass = "hep-ex",
    doi = "10.1103/PhysRevD.109.092003",
    journal = "Phys. Rev. D",
    volume = "109",
    number = "9",
    pages = "092003",
    year = "2024"
}

@article{PandaX-II:2018woa,
    author = "Xia, Jingkai and others",
    collaboration = "PandaX-II",
    title = "{PandaX-II Constraints on Spin-Dependent WIMP-Nucleon Effective Interactions}",
    eprint = "1807.01936",
    archivePrefix = "arXiv",
    primaryClass = "hep-ex",
    doi = "10.1016/j.physletb.2019.02.043",
    journal = "Phys. Lett. B",
    volume = "792",
    pages = "193--198",
    year = "2019"
}

@incollection{HILL201265,
title = {Chapter 2 - Scanning Helium Ion Microscopy},
editor = {Peter W. Hawkes},
series = {Advances in Imaging and Electron Physics},
publisher = {Elsevier},
volume = {170},
pages = {65-148},
year = {2012},
booktitle = {Advances in Imaging and Electron Physics},
issn = {1076-5670},
doi = {https://doi.org/10.1016/B978-0-12-394396-5.00002-6},
url = {https://www.sciencedirect.com/science/article/pii/B9780123943965000026},
author = {Ray Hill and John A. Notte and Larry Scipioni}
}

@article{VANGASTEL20122104,
title = {Subsurface analysis of semiconductor structures with helium ion microscopy},
journal = {Microelectronics Reliability},
volume = {52},
number = {9},
pages = {2104-2109},
year = {2012},
note = {Special issue 23rd european symposium on the reliability of electron devices, failure physics and analysis},
issn = {0026-2714},
doi = {https://doi.org/10.1016/j.microrel.2012.06.130},
url = {https://www.sciencedirect.com/science/article/pii/S0026271412003277},
author = {Raoul {van Gastel} and Gregor Hlawacek and Harold J.W. Zandvliet and Bene Poelsema}
}

@article{Joens2013,
  author    = {Matthew S. Joens and others},
  title     = {Helium Ion Microscopy (HIM) for the imaging of biological samples at sub-nanometer resolution},
  journal   = {Scientific Reports},
  year      = {2013},
  volume    = {3},
  number    = {1},
  pages     = {3514},
  doi       = {10.1038/srep03514},
  url       = {https://doi.org/10.1038/srep03514},
  issn      = {2045-2322}
}

@article{ECHLIN20151,
title = {The TriBeam system: Femtosecond laser ablation in situ SEM},
journal = {Materials Characterization},
volume = {100},
pages = {1-12},
year = {2015},
issn = {1044-5803},
doi = {https://doi.org/10.1016/j.matchar.2014.10.023},
url = {https://www.sciencedirect.com/science/article/pii/S104458031400326X},
author = {McLean P. Echlin and Marcus Straw and Steven Randolph and Jorge Filevich and Tresa M. Pollock}
}

@article{PFEIFENBERGER2017109,
title = {The use of femtosecond laser ablation as a novel tool for rapid micro-mechanical sample preparation},
journal = {Materials and Design},
volume = {121},
pages = {109-118},
year = {2017},
issn = {0264-1275},
doi = {https://doi.org/10.1016/j.matdes.2017.02.012},
url = {https://www.sciencedirect.com/science/article/pii/S0264127517301417},
author = {Manuel J. Pfeifenberger and Melanie Mangang and Stefan Wurster and Jens Reiser and Anton Hohenwarter and Wilhelm Pfleging and Daniel Kiener and Reinhard Pippan}
}

@article{10.1116/1.5047806,
    author = {Randolph, Steven Jeffrey and Filevich, Jorge and Botman, Aurelien and Gannon, Renae and Rue, Chad and Straw, Marcus},
    title = {In situ femtosecond pulse laser ablation for large volume 3D analysis in scanning electron microscope systems},
    journal = {Journal of Vacuum Science and Technology B},
    volume = {36},
    number = {6},
    pages = {06JB01},
    year = {2018},
    month = {09},
    issn = {2166-2746},
    doi = {10.1116/1.5047806},
    url = {https://doi.org/10.1116/1.5047806}
}

@article{RODRIGUEZ2014150,
title = {SAXS and TEM investigation of ion tracks in neodymium-doped yttrium aluminium garnet},
journal = {Nuclear Instruments and Methods in Physics Research Section B: Beam Interactions with Materials and Atoms},
volume = {326},
pages = {150-153},
year = {2014},
note = {17th International Conference on Radiation Effects in Insulators (REI)},
issn = {0168-583X},
doi = {https://doi.org/10.1016/j.nimb.2013.10.076},
url = {https://www.sciencedirect.com/science/article/pii/S0168583X14001013},
author = {M.D. Rodriguez and W.X. Li and F. Chen and C. Trautmann and T. Bierschenk and B. Afra and D. Schauries and R.C. Ewing and S.T. Mudie and P. Kluth}
}

@article{Schaff2015,
  author  = {Florian Schaff and Martin Bech and Paul Zaslansky and Christoph Jud and Marianne Liebi and Manuel Guizar-Sicairos and Franz Pfeiffer},
  title   = {Six-dimensional real and reciprocal space small-angle X-ray scattering tomography},
  journal = {Nature},
  year    = {2015},
  volume  = {527},
  number  = {7578},
  pages   = {353--356},
  doi     = {10.1038/nature16060},
  url     = {https://doi.org/10.1038/nature16060}
}

@article{Holler2014,
  author  = {M. Holler and A. Diaz and M. Guizar-Sicairos and P. Karvinen and Elina Färm and Emma Härkönen and Mikko Ritala and A. Menzel and J. Raabe and O. Bunk},
  title   = {X-ray ptychographic computed tomography at 16 nm isotropic 3D resolution},
  journal = {Scientific Reports},
  year    = {2014},
  volume  = {4},
  number  = {1},
  pages   = {3857},
  doi     = {10.1038/srep03857},
  url     = {https://doi.org/10.1038/srep03857}
}

@article{Baum:2018tfw,
    author = "Baum, Sebastian and Drukier, Andrzej K. and Freese, Katherine and G{\'o}rski, Maciej and Stengel, Patrick",
    title = "{Searching for Dark Matter with Paleo-Detectors}",
    eprint = "1806.05991",
    archivePrefix = "arXiv",
    primaryClass = "astro-ph.CO",
    reportNumber = "NORDITA-2018-043, LCTP-18-15",
    doi = "10.1016/j.physletb.2020.135325",
    journal = "Phys. Lett. B",
    volume = "803",
    pages = "135325",
    year = "2020"
}

@article{Edwards:2018hcf,
    author = "Edwards, Thomas D. P. and Kavanagh, Bradley J. and Weniger, Christoph and Baum, Sebastian and Drukier, Andrzej K. and Freese, Katherine and G{\'o}rski, Maciej and Stengel, Patrick",
    title = "{Digging for dark matter: Spectral analysis and discovery potential of paleo-detectors}",
    eprint = "1811.10549",
    archivePrefix = "arXiv",
    primaryClass = "hep-ph",
    reportNumber = "NORDITA-2018-119; LCTP-18-26",
    doi = "10.1103/PhysRevD.99.043541",
    journal = "Phys. Rev. D",
    volume = "99",
    number = "4",
    pages = "043541",
    year = "2019"
}

@article{Baum:2019fqm,
    author = "Baum, Sebastian and Edwards, Thomas D. P. and Kavanagh, Bradley J. and Stengel, Patrick and Drukier, Andrzej K. and Freese, Katherine and G{\'o}rski, Maciej and Weniger, Christoph",
    title = "{Paleodetectors for Galactic supernova neutrinos}",
    eprint = "1906.05800",
    archivePrefix = "arXiv",
    primaryClass = "astro-ph.GA",
    reportNumber = "NORDITA-2019-060; LCTP-19-12",
    doi = "10.1103/PhysRevD.101.103017",
    journal = "Phys. Rev. D",
    volume = "101",
    number = "10",
    pages = "103017",
    year = "2020"
}

@article{Jordan:2020gxx,
    author = "Jordan, Johnathon R. and Baum, Sebastian and Stengel, Patrick and Ferrari, Alfredo and Morone, Maria Cristina and Sala, Paola and Spitz, Joshua",
    title = "{Measuring Changes in the Atmospheric Neutrino Rate Over Gigayear Timescales}",
    eprint = "2004.08394",
    archivePrefix = "arXiv",
    primaryClass = "hep-ph",
    doi = "10.1103/PhysRevLett.125.231802",
    journal = "Phys. Rev. Lett.",
    volume = "125",
    number = "23",
    pages = "231802",
    year = "2020"
}

@article{Baum:2021jak,
    author = "Baum, Sebastian and Edwards, Thomas D. P. and Freese, Katherine and Stengel, Patrick",
    title = "{New Projections for Dark Matter Searches with Paleo-Detectors}",
    eprint = "2106.06559",
    archivePrefix = "arXiv",
    primaryClass = "astro-ph.CO",
    doi = "10.3390/instruments5020021",
    journal = "Instruments",
    volume = "5",
    number = "2",
    pages = "21",
    year = "2021"
}

@article{Baum:2023cct,
    author = "Baum, Sebastian and others",
    title = "{Mineral detection of neutrinos and dark matter. A whitepaper}",
    eprint = "2301.07118",
    archivePrefix = "arXiv",
    primaryClass = "astro-ph.IM",
    reportNumber = "FERMILAB-PUB-23-501-SQMS-V",
    doi = "10.1016/j.dark.2023.101245",
    journal = "Phys. Dark Univ.",
    volume = "41",
    pages = "101245",
    year = "2023"
}

@misc{Baum:2024eyr,
    author = "Baum, Sebastian and others",
    title = "{Mineral Detection of Neutrinos and Dark Matter 2024. Proceedings}",
    eprint = "2405.01626",
    archivePrefix = "arXiv",
    primaryClass = "astro-ph.CO",
    month = "5",
    year = "2024"
}

@misc{Hirose:2025jht,
    author = "Hirose, Shigenobu and others",
    title = "{Mineral Detection of Neutrinos and Dark Matter 2025 Proceedings}",
    eprint = "2508.20482",
    archivePrefix = "arXiv",
    primaryClass = "physics.ins-det",
    month = "8",
    year = "2025"
}

@article{Kavanagh:2015jma,
    author = "Kavanagh, Bradley J.",
    title = "{New directional signatures from the nonrelativistic effective field theory of dark matter}",
    eprint = "1505.07406",
    archivePrefix = "arXiv",
    primaryClass = "hep-ph",
    reportNumber = "SACLAY-T15-093",
    doi = "10.1103/PhysRevD.92.023513",
    journal = "Phys. Rev. D",
    volume = "92",
    number = "2",
    pages = "023513",
    year = "2015"
}

@article{Cowan:2010js,
    author = "Cowan, Glen and Cranmer, Kyle and Gross, Eilam and Vitells, Ofer",
    title = "{Asymptotic formulae for likelihood-based tests of new physics}",
    eprint = "1007.1727",
    archivePrefix = "arXiv",
    primaryClass = "physics.data-an",
    doi = "10.1140/epjc/s10052-011-1554-0",
    journal = "Eur. Phys. J. C",
    volume = "71",
    pages = "1554",
    year = "2011",
    note = "[Erratum: Eur.Phys.J.C 73, 2501 (2013)]"
}

@article{Billard_2012,
   title={Assessing the discovery potential of directional detection of dark matter},
   volume={85},
   ISSN={1550-2368},
   url={http://dx.doi.org/10.1103/PhysRevD.85.035006},
   DOI={10.1103/physrevd.85.035006},
   number={3},
   journal={Physical Review D},
   publisher={American Physical Society (APS)},
   author={Billard, J. and Mayet, F. and Santos, D.},
   year={2012},
   month=feb }

@article{Conrad:2014nna,
    author = "Conrad, Jan",
    title = "{Statistical Issues in Astrophysical Searches for Particle Dark Matter}",
    eprint = "1407.6617",
    archivePrefix = "arXiv",
    primaryClass = "astro-ph.CO",
    doi = "10.1016/j.astropartphys.2014.09.003",
    journal = "Astropart. Phys.",
    volume = "62",
    pages = "165--177",
    year = "2015"
}

@article{Wilks:1938dza,
    author = "Wilks, S. S.",
    title = "{The Large-Sample Distribution of the Likelihood Ratio for Testing Composite Hypotheses}",
    doi = "10.1214/aoms/1177732360",
    journal = "Annals Math. Statist.",
    volume = "9",
    number = "1",
    pages = "60--62",
    year = "1938"
}

@book{Gradstein2012,
  editor    = {F. M. Gradstein and J. G. Ogg and M. D. Schmitz and G. M. Ogg},
  title     = {The Geologic Time Scale},
  publisher = {Elsevier},
  address   = {Amsterdam, The Netherlands},
  year      = {2012}
}

@article{Gallagher,
author = {Gallagher, Kerry and Brown, Roderick and Johnson, Christopher},
year = {1998},
month = {05},
pages = {519-572},
title = {Fission Track Analysis and its Application to Geological Problems},
volume = {26},
journal = {Annual Review of Earth and Planetary Sciences},
doi = {10.1146/annurev.earth.26.1.519}
}

@book{vandenHaute1998,
  editor    = {P. van den Haute and F. de Corte},
  title     = {Advances in Fission-Track Geochronology},
  publisher = {Springer},
  address   = {Berlin/Heidelberg, Germany},
  year      = {1998}
}

@article{OHare:2020lva,
    author = "O'Hare, Ciaran A. J.",
    title = "{Can we overcome the neutrino floor at high masses?}",
    eprint = "2002.07499",
    archivePrefix = "arXiv",
    primaryClass = "astro-ph.CO",
    reportNumber = "CPPC-2020-13",
    doi = "10.1103/PhysRevD.102.063024",
    journal = "Phys. Rev. D",
    volume = "102",
    number = "6",
    pages = "063024",
    year = "2020"
}

@misc{SuperCDMS:2022crd,
    author = "Albakry, M. F. and others",
    collaboration = "SuperCDMS",
    title = "{Effective Field Theory Analysis of CDMSlite Run 2 Data}",
    eprint = "2205.11683",
    archivePrefix = "arXiv",
    primaryClass = "astro-ph.CO",
    month = "5",
    year = "2022"
}

@misc{Edwards:2017kqw,
    author = "Edwards, Thomas D. P. and Weniger, Christoph",
    title = "{swordfish: Efficient Forecasting of New Physics Searches without Monte Carlo}",
    eprint = "1712.05401",
    archivePrefix = "arXiv",
    primaryClass = "hep-ph",
    month = "12",
    year = "2017"
}

@article{Edwards:2018lsl,
    author = "Edwards, Thomas D. P. and Kavanagh, Bradley J. and Weniger, Christoph",
    title = "{Assessing Near-Future Direct Dark Matter Searches with Benchmark-Free Forecasting}",
    eprint = "1805.04117",
    archivePrefix = "arXiv",
    primaryClass = "hep-ph",
    doi = "10.1103/PhysRevLett.121.181101",
    journal = "Phys. Rev. Lett.",
    volume = "121",
    number = "18",
    pages = "181101",
    year = "2018"
}

@article{Drukier:1984vhf,
    author = "Drukier, A. and Stodolsky, Leo",
    editor = "Tran Thanh Van, J.",
    title = "{Principles and Applications of a Neutral Current Detector for Neutrino Physics and Astronomy}",
    reportNumber = "MPI-PAE/PTh 36/82",
    doi = "10.1103/PhysRevD.30.2295",
    journal = "Phys. Rev. D",
    volume = "30",
    pages = "2295",
    year = "1984"
}

@article{Goodman:1984dc,
    author = "Goodman, Mark W. and Witten, Edward",
    editor = "Srednicki, M. A.",
    title = "{Detectability of Certain Dark Matter Candidates}",
    reportNumber = "Print-85-0030 (PRINCETON)",
    doi = "10.1103/PhysRevD.31.3059",
    journal = "Phys. Rev. D",
    volume = "31",
    pages = "3059",
    year = "1985"
}

@article{Drukier:1986tm,
    author = "Drukier, A. K. and Freese, Katherine and Spergel, D. N.",
    title = "{Detecting Cold Dark Matter Candidates}",
    doi = "10.1103/PhysRevD.33.3495",
    journal = "Phys. Rev. D",
    volume = "33",
    pages = "3495--3508",
    year = "1986"
}

@article{LZ:2022lsv,
    author = "Aalbers, J. and others",
    collaboration = "LZ",
    title = "{First Dark Matter Search Results from the LUX-ZEPLIN (LZ) Experiment}",
    eprint = "2207.03764",
    archivePrefix = "arXiv",
    primaryClass = "hep-ex",
    doi = "10.1103/PhysRevLett.131.041002",
    journal = "Phys. Rev. Lett.",
    volume = "131",
    number = "4",
    pages = "041002",
    year = "2023"
}

@article{DarkSide-50:2022qzh,
    author = "Agnes, P. and others",
    collaboration = "DarkSide-50",
    title = "{Search for low-mass dark matter WIMPs with 12~ton-day exposure of DarkSide-50}",
    eprint = "2207.11966",
    archivePrefix = "arXiv",
    primaryClass = "hep-ex",
    reportNumber = "FERMILAB-PUB-22-589-ND-PPD-SCD",
    doi = "10.1103/PhysRevD.107.063001",
    journal = "Phys. Rev. D",
    volume = "107",
    number = "6",
    pages = "063001",
    year = "2023"
}

@article{XENON:2023cxc,
    author = "Aprile, E. and others",
    collaboration = "XENON",
    title = "{First Dark Matter Search with Nuclear Recoils from the XENONnT Experiment}",
    eprint = "2303.14729",
    archivePrefix = "arXiv",
    primaryClass = "hep-ex",
    doi = "10.1103/PhysRevLett.131.041003",
    journal = "Phys. Rev. Lett.",
    volume = "131",
    number = "4",
    pages = "041003",
    year = "2023"
}

@article{PandaX:2024qfu,
    author = "Bo, Zihao and others",
    collaboration = "PandaX",
    title = "{Dark Matter Search Results from 1.54{\,}{\,}Tonne{\textperiodcentered}Year Exposure of PandaX-4T}",
    eprint = "2408.00664",
    archivePrefix = "arXiv",
    primaryClass = "hep-ex",
    doi = "10.1103/PhysRevLett.134.011805",
    journal = "Phys. Rev. Lett.",
    volume = "134",
    number = "1",
    pages = "011805",
    year = "2025"
}

@article{PhysRevLett.74.4133,
  title = {Limits on Dark Matter Using Ancient Mica},
  author = {Snowden-Ifft, D. P. and Freeman, E. S. and Price, P. B.},
  journal = {Phys. Rev. Lett.},
  volume = {74},
  issue = {21},
  pages = {4133--4136},
  numpages = {0},
  year = {1995},
  month = {May},
  publisher = {American Physical Society},
  doi = {10.1103/PhysRevLett.74.4133},
  url = {https://link.aps.org/doi/10.1103/PhysRevLett.74.4133}
}

@article{Collar:1995aw,
    author = "Collar, Juan I.",
    title = "{Comments on 'limits on dark matter using ancient mica'}",
    eprint = "astro-ph/9511055",
    archivePrefix = "arXiv",
    reportNumber = "JICUSC-95-7",
    doi = "10.1103/PhysRevLett.76.331",
    journal = "Phys. Rev. Lett.",
    volume = "76",
    pages = "331",
    year = "1996"
}

@article{Snowden-Ifft:1996dug,
    author = "Snowden-Ifft, D. P. and Freeman, E. S. and Price, P. B.",
    title = "{Snowden-Ifft, Freemen, and Price Reply (A Reply to the Comment by Juan I Collar)}",
    doi = "10.1103/PhysRevLett.76.332",
    journal = "Phys. Rev. Lett.",
    volume = "76",
    pages = "332",
    year = "1996"
}

@article{SinghSidhu:2019znk,
    author = "Singh Sidhu, Jagjit and Starkman, Glenn and Harvey, Ralph",
    title = "{Counter-top search for macroscopic dark matter}",
    eprint = "1905.10025",
    archivePrefix = "arXiv",
    primaryClass = "astro-ph.HE",
    doi = "10.1103/PhysRevD.100.103015",
    journal = "Phys. Rev. D",
    volume = "100",
    number = "10",
    pages = "103015",
    year = "2019"
}

@article{Ebadi:2021cte,
    author = "Ebadi, Reza and others",
    title = "{Ultraheavy dark matter search with electron microscopy of geological quartz}",
    eprint = "2105.03998",
    archivePrefix = "arXiv",
    primaryClass = "hep-ph",
    doi = "10.1103/PhysRevD.104.015041",
    journal = "Phys. Rev. D",
    volume = "104",
    number = "1",
    pages = "015041",
    year = "2021"
}

@article{Acevedo:2021tbl,
    author = "Acevedo, Javier F. and Bramante, Joseph and Goodman, Alan",
    title = "{Old rocks, new limits: excavated ancient mica searches for dark matter}",
    eprint = "2105.06473",
    archivePrefix = "arXiv",
    primaryClass = "hep-ph",
    doi = "10.1088/1475-7516/2023/11/085",
    journal = "JCAP",
    volume = "11",
    pages = "085",
    year = "2023"
}

@article{Fung:2025cub,
    author = "Fung, Audrey and Lucas, Thalles and Balogh, Levente and Leybourne, Matthew and Vincent, Aaron C.",
    title = "{Refining the sensitivity of new physics searches with ancient minerals}",
    eprint = "2504.08885",
    archivePrefix = "arXiv",
    primaryClass = "hep-ph",
    doi = "10.1103/clvr-mhx5",
    journal = "Phys. Rev. D",
    volume = "112",
    number = "4",
    pages = "043040",
    year = "2025"
}

@article{Tucker-Smith:2001myb,
    author = "Tucker-Smith, David and Weiner, Neal",
    title = "{Inelastic dark matter}",
    eprint = "hep-ph/0101138",
    archivePrefix = "arXiv",
    reportNumber = "UCB-PTH-00-43, LBNL-47234, UW-PT-00-17",
    doi = "10.1103/PhysRevD.64.043502",
    journal = "Phys. Rev. D",
    volume = "64",
    pages = "043502",
    year = "2001"
}

@article{Barello:2014uda,
    author = "Barello, G. and Chang, Spencer and Newby, Christopher A.",
    title = "{A Model Independent Approach to Inelastic Dark Matter Scattering}",
    eprint = "1409.0536",
    archivePrefix = "arXiv",
    primaryClass = "hep-ph",
    doi = "10.1103/PhysRevD.90.094027",
    journal = "Phys. Rev. D",
    volume = "90",
    number = "9",
    pages = "094027",
    year = "2014"
}

@article{Mei:2005gm,
    author = "Mei, Dongming and Hime, A.",
    title = "{Muon-induced background study for underground laboratories}",
    eprint = "astro-ph/0512125",
    archivePrefix = "arXiv",
    reportNumber = "LA-UP-05-3704",
    doi = "10.1103/PhysRevD.73.053004",
    journal = "Phys. Rev. D",
    volume = "73",
    pages = "053004",
    year = "2006",
    note = "[Erratum: Phys.Rev.D 109, 019901 (2024)]"
}

@article{Tapia-Arellano:2021cml,
    author = "Tapia-Arellano, Natalia and Horiuchi, Shunsaku",
    title = "{Measuring solar neutrinos over gigayear timescales with paleo detectors}",
    eprint = "2102.01755",
    archivePrefix = "arXiv",
    primaryClass = "hep-ph",
    doi = "10.1103/PhysRevD.103.123016",
    journal = "Phys. Rev. D",
    volume = "103",
    number = "12",
    pages = "123016",
    year = "2021"
}

@article{Cappellaro:2003eg,
    author = "Cappellaro, Enrico and Barbon, Roberto and Turatto, Massimo",
    editor = "Marcaide, Juan-Mar{\'\i}a and Weiler, Kurt W.",
    title = "{Supernova statistics}",
    eprint = "astro-ph/0310859",
    archivePrefix = "arXiv",
    doi = "10.1007/3-540-26633-X_48",
    journal = "Springer Proc. Phys.",
    volume = "99",
    pages = "347--354",
    year = "2005"
}

@article{Diehl:2006cf,
    author = "Diehl, Roland and others",
    title = "{Radioactive Al-26 and massive stars in the galaxy}",
    eprint = "astro-ph/0601015",
    archivePrefix = "arXiv",
    doi = "10.1038/nature04364",
    journal = "Nature",
    volume = "439",
    pages = "45--47",
    year = "2006"
}

@misc{Strumia:2006db,
    author = "Strumia, Alessandro and Vissani, Francesco",
    title = "{Neutrino masses and mixings and...}",
    eprint = "hep-ph/0606054",
    archivePrefix = "arXiv",
    reportNumber = "IFUP-TH-2004-1",
    month = "6",
    year = "2006"
}

@article{Leaman_2011,
   title={Nearby supernova rates from the Lick Observatory Supernova Search - I. The methods and data base: Nearby SN rates from LOSS - I},
   volume={412},
   ISSN={0035-8711},
   url={http://dx.doi.org/10.1111/j.1365-2966.2011.18158.x},
   DOI={10.1111/j.1365-2966.2011.18158.x},
   number={3},
   journal={Monthly Notices of the Royal Astronomical Society},
   publisher={Oxford University Press (OUP)},
   author={Leaman, Jesse and Li, Weidong and Chornock, Ryan and Filippenko, Alexei V.},
   year={2011},
   month=mar, pages={1419–1440} }

@article{Botticella_2012,
   title={A comparison between star formation rate diagnostics and rate of core collapse supernovae within 11 Mpc},
   volume={537},
   ISSN={1432-0746},
   url={http://dx.doi.org/10.1051/0004-6361/201117343},
   DOI={10.1051/0004-6361/201117343},
   journal={Astronomy and amp; Astrophysics},
   publisher={EDP Sciences},
   author={Botticella, M. T. and Smartt, S. J. and Kennicutt, R. C. and Cappellaro, E. and Sereno, M. and Lee, J. C.},
   year={2012},
   month=jan, pages={A132} }

@article{Adams:2013ana,
    author = "Adams, Scott M. and Kochanek, C. S. and Beacom, John F. and Vagins, Mark R. and Stanek, K. Z.",
    title = "{Observing the Next Galactic Supernova}",
    eprint = "1306.0559",
    archivePrefix = "arXiv",
    primaryClass = "astro-ph.HE",
    doi = "10.1088/0004-637X/778/2/164",
    journal = "Astrophys. J.",
    volume = "778",
    pages = "164",
    year = "2013"
}

@techreport{sources4a1999,
  title        = {SOURCES 4A: A Code for Calculating (Alpha, n), Spontaneous Fission, and Delayed Neutron Sources and Spectra},
  institution  = {Los Alamos National Laboratory},
  address      = {Los Alamos, NM, USA},
  number       = {LA-13639-MS},
  year         = {1999},
  note         = {Technical Report}
}

@article{KONING20122841,
title = {Modern Nuclear Data Evaluation with the TALYS Code System},
journal = {Nuclear Data Sheets},
volume = {113},
number = {12},
pages = {2841-2934},
year = {2012},
note = {Special Issue on Nuclear Reaction Data},
issn = {0090-3752},
doi = {https://doi.org/10.1016/j.nds.2012.11.002},
url = {https://www.sciencedirect.com/science/article/pii/S0090375212000889},
author = {A.J. Koning and D. Rochman}
}

@article{SOPPERA2014294,
title = {JANIS 4: An Improved Version of the NEA Java-based Nuclear Data Information System},
journal = {Nuclear Data Sheets},
volume = {120},
pages = {294-296},
year = {2014},
issn = {0090-3752},
doi = {https://doi.org/10.1016/j.nds.2014.07.071},
url = {https://www.sciencedirect.com/science/article/pii/S0090375214005237},
author = {N. Soppera and M. Bossant and E. Dupont}
}

@misc{Theodosopoulos:2026ehn,
    author = "Theodosopoulos, Dionysios P. and Freese, Katherine and Kelso, Chris and Stengel, Patrick",
    title = "{Projected Sensitivity of Paleo-Detectors to Dark Matter Effective Interactions with Nuclei}",
    eprint = "2603.13629",
    archivePrefix = "arXiv",
    primaryClass = "astro-ph.CO",
    month = "3",
    year = "2026"
}

@misc{XLZD:2024nsu,
    author = "Aalbers, J. and others",
    collaboration = "XLZD",
    title = "{The XLZD Design Book: Towards the Next-Generation Liquid Xenon Observatory for Dark Matter and Neutrino Physics}",
    eprint = "2410.17137",
    archivePrefix = "arXiv",
    primaryClass = "hep-ex",
    month = "10",
    year = "2024"
}

@article{DarkSide-20k:2017zyg,
    author = "Aalseth, C. E. and others",
    collaboration = "DarkSide-20k",
    title = "{DarkSide-20k: A 20 tonne two-phase LAr TPC for direct dark matter detection at LNGS}",
    eprint = "1707.08145",
    archivePrefix = "arXiv",
    primaryClass = "physics.ins-det",
    reportNumber = "FERMILAB-PUB-17-298-PPD",
    doi = "10.1140/epjp/i2018-11973-4",
    journal = "Eur. Phys. J. Plus",
    volume = "133",
    pages = "131",
    year = "2018"
}

@article{SuperCDMS:2016wui,
    author = "Agnese, R. and others",
    collaboration = "SuperCDMS",
    title = "{Projected Sensitivity of the SuperCDMS SNOLAB experiment}",
    eprint = "1610.00006",
    archivePrefix = "arXiv",
    primaryClass = "physics.ins-det",
    reportNumber = "FERMILAB-PUB-16-467-AE",
    doi = "10.1103/PhysRevD.95.082002",
    journal = "Phys. Rev. D",
    volume = "95",
    number = "8",
    pages = "082002",
    year = "2017"
}

@article{Green:2007rb,
    author = "Green, Anne M.",
    title = "{Determining the WIMP mass using direct detection experiments}",
    eprint = "hep-ph/0703217",
    archivePrefix = "arXiv",
    doi = "10.1088/1475-7516/2007/08/022",
    journal = "JCAP",
    volume = "08",
    pages = "022",
    year = "2007"
}

@article{Green:2008rd,
    author = "Green, Anne M.",
    title = "{Determining the WIMP mass from a single direct detection experiment, a more detailed study}",
    eprint = "0805.1704",
    archivePrefix = "arXiv",
    primaryClass = "hep-ph",
    doi = "10.1088/1475-7516/2008/07/005",
    journal = "JCAP",
    volume = "07",
    pages = "005",
    year = "2008"
}

@article{Bozorgnia:2018jep,
    author = "Bozorgnia, Nassim and Cerde{\~n}o, David G. and Cheek, Andrew and Penning, Bjoern",
    title = "{Opening the energy window on direct dark matter detection}",
    eprint = "1810.05576",
    archivePrefix = "arXiv",
    primaryClass = "hep-ph",
    reportNumber = "IPPP/18/92",
    doi = "10.1088/1475-7516/2018/12/013",
    journal = "JCAP",
    volume = "12",
    pages = "013",
    year = "2018"
}

@article{Brenner:2022qku,
    author = "Brenner, Anja and Herrera, Gonzalo and Ibarra, Alejandro and Kang, Sunghyun and Scopel, Stefano and Tomar, Gaurav",
    title = "{Complementarity of experiments in probing the non-relativistic effective theory of dark matter-nucleon interactions}",
    eprint = "2203.04210",
    archivePrefix = "arXiv",
    primaryClass = "hep-ph",
    doi = "10.1088/1475-7516/2022/06/026",
    journal = "JCAP",
    volume = "06",
    number = "06",
    pages = "026",
    year = "2022"
}

@misc{Calabrese-Day:2026soq,
    author = "Calabrese-Day, Andrew and LaVoie-Ingram, Emilie and Ream, Kathryn and Ross, Hannah and Spitz, Joshua and Stengel, Patrick and Sun, Kai and Takla, Alexander",
    title = "{Toward Neutrino and Dark Matter Detection with Ancient Minerals: TEM Study of Heavy-Ion Tracks in Olivine}",
    eprint = "2604.09732",
    archivePrefix = "arXiv",
    primaryClass = "physics.ins-det",
    month = "4",
    year = "2026"
}

@article{Galelli:2025gss,
    author = "Galelli, Claudio and Caccianiga, Lorenzo and Apollonio, Lorenzo and Magnani, Paolo and Breton, Vincent",
    title = "{A volcanic chronosequence as a time-resolved paleo-detector array to study the cosmic-ray flux in the late Pleistocene and Holocene}",
    eprint = "2510.23126",
    archivePrefix = "arXiv",
    primaryClass = "astro-ph.HE",
    doi = "10.1088/1475-7516/2026/04/023",
    journal = "JCAP",
    volume = "04",
    pages = "023",
    year = "2026"
}

\end{document}